\documentclass[onecolumn,showkeys,aps,prd]{revtex4}
\usepackage{tikz}
\usepackage{amsfonts}
\usepackage{latexsym}
\usepackage{amsmath,amssymb}
\usetikzlibrary{positioning}
\usepackage{appendix}

\newcommand{\ds}{\displaystyle}
\newcommand{\del}{\partial}

\newcommand{\D}{\mathrm{d}}
\newcommand{\pb}[2]{\{\,  {#1} \,,\, {#2} \,\}  }

\usepackage{ntheorem}

\newcommand{\killing}[2]{ {\langle #1 , #2 \rangle } }

\newcommand{\realni}{\ensuremath{\mathbb{R}}}

\newcommand{\cA}{{\cal A}}
\newcommand{\cD}{{\cal D}}
\newcommand{\cF}{{\cal F}}
\newcommand{\cG}{{\cal G}}
\newcommand{\cH}{{\cal H}}

\newcommand{\cM}{{\cal M}}

\newcommand{\cT}{{\cal T}}
\newcommand{\cS}{{\cal S}}

\newtheorem{Theorem}{Theorem}

\newtheorem*{theorem-non}{Definition}

\newtheorem{Fundamental Theorem}{Fundamental Theorem}

\newenvironment{Proof}[1][Proof]{\textbf{#1.} }{\ \rule{0.5em}{0.5em}}

\begin{document}

\title{Gauge symmetry of the $3BF$ theory for a generic Lie $3$-group}

\author{Tijana Radenkovi\'c}
 \email{rtijana@ipb.ac.rs}
\affiliation{Institute of Physics, University of Belgrade, Pregrevica 118, 11080 Belgrade, Serbia}

\author{Marko Vojinovi\'c}
 \email{vmarko@ipb.ac.rs}
\affiliation{Institute of Physics, University of Belgrade, Pregrevica 118, 11080 Belgrade, Serbia}

%\pacs{04.60.Pp}


\begin{abstract}
The higher category theory can be employed to generalize the $BF$ action to the so-called $3BF$ action, by passing from the notion of a gauge group to the notion of a gauge $3$-group. In this work we determine the full gauge symmetry of the $3BF$ action. To that end, the complete Hamiltonian analysis of the $3BF$ action for an arbitrary semistrict Lie $3$-group is performed, by using the Dirac procedure. The Hamiltonian analysis is the first step towards a canonical quantization of a $3BF$ theory. This is an important stepping-stone for the quantization of the complete Standard Model of elementary particles coupled to Einstein-Cartan gravity, formulated as a $3BF$ action with suitable simplicity constraints. We show that the resulting gauge symmetry group consists of the already familiar $G$-, $H$-, and $L$-gauge transformations, as well as additional $M$- and $N$-gauge transformations, which have not been discussed in the existing literature.
\end{abstract}

\maketitle

\section{\label{SecI}Introduction}

Among the most important open problems in contemporary theoretical physics is the problem of quantization of the gravitational field. Within the framework of Loop Quantum Gravity (LQG), one of the most prominent candidates for the quantum theory of gravity, the study of nonperturbative quantization has evolved in two directions: the canonical and the covariant approach. See \cite{RovelliBook, RovelliVidottoBook, Thiemann2007, zakopane} for an overview and a comprehensive introduction to the theory.

The \emph{covariant quantization} approach focuses on defining the gravitational path integral of the theory:
\begin{equation}\label{eq1}
    Z_{gr}=\int \cD g\, e^{iS_{gr}[g]}.
\end{equation}
In order to give the rigorous definition of the path integral, the classical action of the theory $S_{gr}$ is written as a sum of the topological $BF$ action, i.e., the action with no propagating degrees of freedom, and the part featuring the simplicity constraints, i.e., sum of products of Langrange multipliers and the corresponding simplicity constraints imposed on the variables of the topological part of the action. Next, one defines the path-integral of the topological theory given by the $BF$ action, using the Topological Quantum Field Theory (TQFT) formalism. Once a path-integral is defined for the topological sector, it is deformed into a non-topological theory, by imposing the simplicity constraints. This quantization technique is known as the \emph{spinfoam quantization} method.

The spinfoam quantization procedure has been successfully employed in various theories, including the three-dimensional topological Ponzano-Regge model of quantum gravity \cite{PonzanoRegge1968}, the four-dimensional topological Ooguri model \cite{Ooguri}, the Barrett-Crane model of gravity in four dimensions \cite{BarrettCrane,BarrettCrane1, crane2003}, and others. The most successful among these is the renowned EPRL/FK model \cite{EPRL,FK}, which had been specifically formulated to correspond to the quantum theory of gravity obtained by the \emph{canonical loop quantization}, where a state of the gravitational field is described by the so-called \emph{spin network}.

However, note that all mentioned models, formulated as constrained $BF$ actions, are theories of pure gravity, without matter fields.  Recently, as an endeavor to formulate a theory that unifies all the known interactions, one interesting new avenue of research has been opened, based on a categorical generalization of the $BF$ action in the context of Higher Gauge Theory (HGT) formalism \cite{BaezHuerta2011}. One novel candidate discussed in the literature \cite{Radenkovic2019}, uses the $3$-group structure to formulate the $3BF$ action as a categorical generalization of the $BF$ theory. Then, modifying the pure $3BF$ action by adding the appropriate simplicity constraints, one obtains the \emph{constrained} $3BF$ action, describing the theory of all the fields present in the Standard model coupled in a standard way to Einstein-Cartan gravity.

Once the appropriate classical theory has been constructed, one needs to quantize it by constructing a topological state sum $Z$ using the algebraic structure underlying the topological sector of the constrained $3BF$ action, i.e.\,, the underlying $2$-crossed module. This construction has been recently carried out in \cite{RadVoj2022}, where a triangulation independent state sum $Z$ of a topological higher gauge theory for an arbitrary $2$-crossed module and a $4$-dimensional closed and orientable spacetime manifold $\cM_4$ is defined. Once the topological state sum is formulated, one could proceed to modify the amplitudes of the state sum in order to impose the simplicity constraints and obtain the state sum describing the full theory. In this way one would finally arrive at the rigorous definition of a path integral given by the equation \eqref{eq1}.

In addition to the covariant approach, one can also study the constrained $3BF$ action, using the \emph{canonical quantization}. This approach focuses on defining the quantum theory via a triple $(\cH, \cA, W)$, i.e., the Hilbert space of states $\cH$, the algebra of observables $\cA$, and the dynamics $W$ given by the transition amplitudes. Specifically, in canonical LQG, the algebra of fields that are promoted to the quantum operators is chosen to be the algebra based on the holonomies of the gravitational connection. However, in the case of the $3BF$ theory, the notion of connection is generalized to the notion of $3$-connection, which makes its canonical quantization approach an interesting avenue of research. The first step toward the canonical quantization of the theory is the Hamiltonian analysis, resulting in the algebra of first-class and second-class constraints. The first-class constraints become conditions on the physical states determining the Hilbert space, while the Hamiltonian constraint determines the dynamics.

The results presented in this paper are the natural continuation of the results presented in \cite{Radenkovic2019}. The main result is the calculation of the full symmetry group of the pure $3BF$ action. To that end, the complete Hamiltonian analysis of the $3BF$ action for a semistrict Lie $3$-group is performed by using the Dirac procedure (see \cite{blagojevic2002gravitation} for an overview and a comprehensive introduction to the Hamiltonian analysis). It is a generalization of the Hamiltonian analysis of a $2BF$ action performed in \cite{MOV2016,MOV2019,Mikovic2015,MikovicOliveira2014}, and of the Hamiltonian analysis for the special case of a $2$-crossed module corresponding to the theory of scalar electrodynamics, carried out in \cite{Radenkovic2020}. The analysis of the Hamiltonian structure of the theory gives us the algebra of first-class and second-class constraints present in the theory. As usual, the first-class constraints generate gauge transformations, which do not change the physical state of the system. Using the Castellani's procedure, one can find the generator of the gauge transformations in the theory on a spatial hypersurface. Then, the results obtained by this method are generalized to the whole spacetime. The complete gauge symmetry, consisting of five types of finite gauge transformations, along with the proofs that they are indeed the gauge symmetries of $3BF$ action, is presented. With these results in hand, the structure of the full gauge symmetry group is analyzed, and its corresponding Lie algebra is determined. 

The obtained results give rise to a connection between the gauge symmetry group of the $3BF$ action, and its underlining $3$-group structure, establishing a \emph{duality} between the two. This analysis is an important step towards the study of the gauge symmetry group of the theory of gravity with matter, formulated as the constrained $3BF$ action \cite{Radenkovic2019}, as well as its canonical quantization. Furthermore, it is important for the overall understanding of the physical meaning of the $3$-group structure and its interpretation as the underlining symmetry of the pure $3BF$ action, which represents a basis for the constrained $3BF$ action describing the physical theory. 

The layout of the paper is as follows. In Section \ref{secII}, we give a brief overview of $BF$ and $2BF$ theories, and introduce the $3BF$ action. Section \ref{secIII} contains the Hamiltonian analysis for the $3BF$ theory. In subsection \ref{secIIIa}, the canonical structure of the theory is obtained, while in subsection \ref{secIIIb} the resulting first-class and second-class constraints present in the theory, as well as the algebra of constraints, are presented. In the subsection \ref{secIIIc} we analyze the Bianchi identities that the first-class constraints satisfy, which enforce restrictions in the sense of Hamiltonian analysis, and reduce the number of independent first-class constraints present in the theory. We then proceed with the counting of the physical degrees of freedom. Finally, this section concludes with the subsection \ref{secIIId} where we construct the generator of the gauge symmetries for the topological theory, based on the calculations done in section \ref{secIIIb}.

Section \ref{secIV} contains the main results of our paper and is devoted to the analysis of the symmetries of the $3BF$ action. Having results of the subsection \ref{secIIId} in hand, we find the form variations of all variables and their canonical momenta, and use that result to determine all gauge transformations of the theory. This is done in four steps.
The subsection \ref{secIVa} deals with the gauge group $G$, and the corresponding $G$-gauge transformations. In subsection \ref{secIVb} we discuss the gauge group $\tilde{H}_L$ which consists of the familiar $H$-gauge and $L$-gauge transformations (familiar from \cite{Wang2014}), while the subsection \ref{secIVc} examines the novel $M$-gauge and $N$-gauge transformations which also arise in the theory. The results of the subsections \ref{secIVa}, \ref{secIVb}, and \ref{secIVc} are summarized in subsection \ref{secIVd}, where the complete structure of the symmetry group is presented, including its Lie algebra. Our concluding remarks are given in section \ref{secV}, containing a summary and a discussion of the obtained results, as well as possible future lines of investigation. The Appendices contain various technical details concerning $3$-groups, additional relations of the constraint algebra, the computation of the generator of gauge symmetries, form-variations of all fields and momenta, and some other technical details.

Our notation and conventions are as follows. Spacetime indices, denoted by the mid-alphabet Greek letters $\mu,\nu,\dots$, are raised and lowered by the spacetime metric $g_{\mu\nu}$. The spatial part of these is denoted with lo\-wer\-case mid-alphabet Latin indices $i,j,\dots$, and the time component is denoted with $0$. The indices that are counting the generators of groups $G$, $H$, and $L$ are denoted with initial Greek letters $\alpha, \beta, \dots$, lowercase initial Latin letters $a, b, c,\dots$, and uppercase Latin indices $A,B,C,\dots$, respectively. The antisymmetrization over two indices is denoted as $A_{[a_1|a_2\dots a_{n-1}|a_n]}=\frac{1}{2}\left(A_{a_1a_2\dots a_{n-1}a_n}-A_{a_{n}a_2\dots a_{n-1}a_1}\right)$, while the total antisymmetrization is denoted as $A_{[a_1\dots a_n]}=\frac{1}{n!}\sum_{\sigma\in S_n}(-1)^{\mathrm{sgn}(\sigma)}A_{a_{\sigma(1)}\dots a_{\sigma(n)}}$. Likewise, the symmetrization over two indices is denoted as $A_{(a_1|a_2\dots a_{n-1}|a_n)}=\frac{1}{2}\left(A_{a_1a_2\dots a_{n-1}a_n}+A_{a_{n}a_2\dots a_{n-1}a_1}\right)$, while the total symmetrization is denoted as $A_{(a_1\dots a_n)}=\frac{1}{n!}\sum_{\sigma\in S_n}A_{a_{\sigma(1)}\dots a_{\sigma(n)}}$. We work in the natural system of units, defined by $c=\hbar=1$ and $G = l_p^2$, where $l_p$ is the Planck length. All additional notation and conventions used throughout the paper are explicitly defined in the text where they appear.

%%%%%%%%%%%%%%%%%%%%%%%%%%%%%%%%%%%%%%%%%%%%%%%%%%%%%%%%%%%%%%%%%%%%%%%%%%%%%%%%%%%%%%%%%%%%%%%%%%%%%%%%%%%%%%%%%%%%%%%%%%%%%%
\section{\label{secII}The $3BF$ theory}

Given a Lie group $G$ and its corresponding Lie algebra $\mathfrak{g}$, one can introduce the so-called $BF$ action as
\begin{equation}\label{eq:bf}
    S_{BF} =\int_{\cM_4} \langle B \wedge F \rangle_\mathfrak{g}\,,
\end{equation}
where $F\equiv \D \alpha+ \alpha\wedge \alpha$ is the curvature $2$-form for the algebra-valued connection $1$-form $\alpha \in \cA^1(\cM_4\,, \mathfrak{g})$ on a trivial principal $G$-bundle over a $4$-dimensional compact and orientable spacetime manifold $\cM_4$, and $B \in \cA^2(\cM_4\,, \mathfrak{g})$ is a Lagrange multiplier $2$-form. The $\langle \_\,,\_\rangle{}_{\mathfrak{g}}$ denotes the $G$-invariant bilinear symmetric nondegenerate form on $\mathfrak{g}$. For more details see \cite{BFgravity2016, plebanski1977, baez2000}.

Varying the action (\ref{eq:bf}) with respect to the Lagrange multiplier $B$ and the connection $\alpha$, one obtains the equations of motion,
\begin{equation}
    F=0\,,\quad \quad \nabla B \equiv \D B + \alpha \wedge B  =0\,.
\end{equation}
These equations of motion imply that $\alpha$ is a flat connection, while the Lagrange multiplier $B$ is a constant field. Therefore, the theory given by the $BF$ action has no local propagating degrees of freedom, i.e., the theory is topological. 

Within the framework of Higher Gauge Theory, one can define the categorical generalization of the $BF$ action to the so-called $2BF$ action, by passing from the notion of a gauge group to the notion of a gauge $2$-group, see \cite{GirelliPfeifferPopescu2008,FariaMartinsMikovic2011,MikovicVojinovic2012}. In the category theory, a $2$-group is defined as a $2$-category consisting of only one object, where all the morphisms and $2$-morphisms are invertible. It has been shown that every strict $2$-group is equivalent to a crossed module $(H \stackrel{\del}{\to}G \,, \rhd)$, where $G$ and $H$ are groups, $\delta$ is a homomorphism from $H$ to $G$, while $\rhd:G\times H \to H$ is an action of $G$ on $H$. Given a crossed-module $(H \stackrel{\del}{\to}G \,, \rhd)$, one can introduce a generalization of the $BF$ action, the so-called $2BF$ action \cite{GirelliPfeifferPopescu2008,FariaMartinsMikovic2011}:
\begin{equation}\label{eq:bfcg}
S_{2BF} =\int_{\cM_4} \langle B \wedge \cF \rangle_{\mathfrak{ g}} +  \langle C \wedge \cG \rangle_{\mathfrak{h}} \,,
\end{equation}
where the $2$-form $B \in \cA^2(\cM_4\,, \mathfrak{g})$ and the $1$-form $C\in \cA^1(\cM_4\,,\mathfrak{h})$ are Lagrange multipliers, and $\mathfrak{h}$ is a Lie algebra of the Lie group $H$. The variables $\cF\in \cA^2(\cM_4\,,\mathfrak{g})$ and $\cG\in \cA^3(\cM_4\,,\mathfrak{h})$ define the {\em fake $2$-curvature} $(\cF,\cG)$ for the $2$-connection $(\alpha\,,\beta)$ on a trivial principal $2$-bundle over a $4$-dimensional compact and oriented spacetime manifold $\cM_4$. See \cite{Baez2004} for a rigorous definition. Here the $2$-connection $(\alpha\,,\beta)$ is given by $\mathfrak{g}$-valued $1$-form $\alpha \in \cA^1(\cM_4\,,\mathfrak{g})$ and an $\mathfrak{h}$-valued $2$-form $\beta \in \cA^2(\cM_4\,,\mathfrak{h})$:
\begin{equation}\label{eq:krivine}
    \cF=\D \alpha+ \alpha \wedge \alpha - \partial\beta\,, \quad \quad \cG= \D\beta+\alpha\wedge^\rhd \beta\,.
\end{equation} 
The $2$-curvature $(\cF, \cG)$ is called \emph{fake}, because of the additional term $\partial \beta$, see \cite{BaezHuerta2011}. Also, $\langle \_\,,\_\rangle{}_{\mathfrak{g}}$ and $\langle \_\,,\_\rangle{}_{\mathfrak{h}}$ denote the $G$-invariant bilinear symmetric nondegenerate forms for the algebras $\mathfrak{g}$ and $\mathfrak{h}$, respectively. See \cite{GirelliPfeifferPopescu2008,FariaMartinsMikovic2011} for review and references. Varying the $2BF$ action (\ref{eq:bfcg}) with respect to variables $B$ and $C$ one obtains the equations of motion
\begin{equation}\label{eq:jedn2bf}
    \cF=0\,, \quad \quad \cG=0\,,
\end{equation}
while varying with respect to $\alpha$ and $\beta$ one obtains
\begin{gather}
\D B_\alpha- {f_{\alpha \beta}}^\gamma B_\gamma \wedge  \alpha^\beta - {\rhd_{\alpha a}}^b C_b\wedge \beta^a =0\,, \\ 
\D C_a - {\partial_a}^\alpha B_\alpha + {\rhd_{\alpha a}}^b C_b \wedge \alpha^\alpha =0\,.
\end{gather}
Here, the coefficiants ${f_{\alpha \beta}}^\gamma$ are the structure constants of the algebra $\mathfrak{g}$, ${\rhd_{\alpha a}}^b$ are the coefficients of the action $\rhd$ of the algebra $\mathfrak{g}$ on $\mathfrak{h}$, while $\partial_a{}^\alpha$ are the coefficients of the map $\partial$, given in the bases of algebras $\mathfrak{g}$ and $\mathfrak{h}$ (see the equations \eqref{eq:basis}--\eqref{eq12} below). Similarly to the case of the $BF$ action, the $2BF$ action defines a topological theory, i.e., a theory with no propagating degrees of freedom, see \cite{MikovicOliveira2014,MOV2016}.

Continuing the categorical generalization one step further, one can generalize the notion of a $2$-group to the notion of a $3$-group. Similarly to the definition of a group and a $2$-group within the category theory formalism, a $3$-group is defined as a $3$-category with only one object, where all morphisms, $2$-morphisms, and $3$-morphisms are invertible. Moreover, analogously as a strict $2$-group is equivalent to a crossed-module, it has been proved that a semistrict $3$-group is equivalent to a $2$-crossed module \cite{martins2011}. 

A Lie $2$-crossed module, denoted as $ (L\stackrel{\delta}{\to} H \stackrel{\partial}{\to}G\,, \rhd\,, \{\_\,,\_\}{}_{\mathrm{pf}})$ (see Appendix \ref{AppA} for the precise definition), is an algebraic structure specified by three Lie groups $G$, $H$, and $L$, together with the homomorphisms $ \delta: L\to H$ and $ \partial: H\to G$, an action $\rhd$ of the group $G$ on all three groups, and a $G$-equivariant map, called the Peiffer lifting:
\begin{displaymath}
\{ \_ \,, \_ \}{}_{\mathrm{pf}} : H\times H \to L\,.
\end{displaymath}
In order for this structure to be a $3$-group, the structure constants of algebras $\mathfrak{g}$, $\mathfrak{h}$, and $\mathfrak{l}$, together with the maps $\del$ and $\delta$, the action $\rhd$, and the Peiffer lifting, must satisfy certain axioms, see \cite{Radenkovic2019}. Here $ \mathfrak {g} $, $ \mathfrak{h}$, and $ \mathfrak{l} $ denote the Lie algebras corresponding to the Lie groups $ G $, $ H $, and $ L $.

Analogously to the definition of a $2$-connection given in \cite{Baez2004}, one can define a $3$-connection as follows. Given a $2$-crossed module and a $4$-dimensional compact and orientable spacetime manifold $\cM_4$, one can introduce a trivial principal $3$-bundle using the $2$-crossed module as a fiber over the base manifold $\cM_4$. See \cite{martins2011, Wang2014} for the precise definition of a corresponding $3$-holonomy. This gives rise to a $3$-connection, which can be represented as an ordered triple $(\alpha, \beta, \gamma) $, where $\alpha$, $\beta$, and $\gamma$ are algebra-valued differential forms, $ \alpha \in \cA ^ 1 (\cM_4, \mathfrak {g}) $, $ \beta \in \cA ^ 2 (\cM_4, \mathfrak {h}) $, and $ \gamma \in \cA ^ 3 (\cM_4, \mathfrak {l}) $. The corresponding fake $ 3 $-curvature $ (\cal F, G, H) $ is defined as:
\begin{equation}\label{eq:3krivine}
\begin{array}{c}
\ds  \cF = \D \alpha+\alpha \wedge \alpha - \partial \beta \,, \quad \quad \cG = \D \beta + \alpha \wedge^\rhd \beta - \delta \gamma\,, \vphantom{\ds\int} \\
\ds  \cH = \D \gamma + \alpha\wedge^\rhd \gamma + \{\beta \wedge \beta\}{}_{\mathrm{pf}} \,.
\end{array}
\end{equation}
Similarly as in the case of the $2BF$ theory, the $3$-curvature $(\cF, \cG, \cH)$ is called \emph{fake}, because of the additional terms $\partial \beta$, $\delta \gamma$, and $\{\beta \wedge \beta\}{}_{\mathrm{pf}}$.
Fixing the bases in algebras $\mathfrak{g}$, $\mathfrak{h}$, and $\mathfrak{l}$ as $\tau_\alpha \in \mathfrak{g}$, $t_a \in \mathfrak{h}$, and $T_A \in \mathfrak{l}$, one defines the structure constants
\begin{equation}\label{eq:basis}
     [\tau_\alpha,\tau_\beta]={f_{\alpha\beta}}^\gamma \, \tau_\gamma\,, \quad \quad [t_a,t_b]={f_{ab}}^c \, t_c\,, \quad \quad [T_A,T_B]={f_{AB}}^C \, T_C\,,  
\end{equation}
maps $ \partial: H\to G$ and $\delta: L\to H$ as 
\begin{equation}
    \partial (t_a)={\partial_a}^\alpha \, \tau_\alpha\,, \quad \quad \quad \delta (T_A)={\delta_A}^a \, t_a\,,
\end{equation}
and an action of $\mathfrak{g}$ on the generators of $\mathfrak{g}$,  $\mathfrak{h}$, and $\mathfrak{l}$ as
\begin{equation}\label{eq12}
        \tau_\alpha \rhd \tau_\beta = {f_{\alpha\beta}}^\gamma\, \tau_\gamma\,, \quad \quad \tau_\alpha \rhd t_a={\rhd_{\alpha a}}^b\, t_b\,,\quad \quad\tau_\alpha \rhd T_A={\rhd_{\alpha A}}^B \,T_B \,,
\end{equation}
respectively. To define the Peiffer lifting in a basis, one specifies the coefficients ${X_{ab}}^A$:
\begin{equation}
    \{t_a,\,t_b\}{}_{\mathrm{pf}}={X_{ab}}^A T_A\,.
\end{equation}
Writing the curvature in the bases of the corresponding algebras and differential forms
\begin{equation*}
{\cal F}=\frac{1}{2}{{\cal F}^\alpha{}_{\mu\nu}} \tau_\alpha \D x^\mu \wedge \D x^\nu\,, \quad {\cal G}=\frac{1}{3!}{{\cal G}^a{}_{\mu \nu \rho}} t_a \D x^\mu \wedge \D x^\nu \wedge \D x^\rho\,, \quad {\cal H}=\frac{1}{4!}{\cal H }^A{}_{\mu\nu\rho\sigma} T_A\D x^\mu \wedge \D x^\nu \wedge \D x^\rho \wedge \D x^\sigma\,,
\end{equation*}
one obtains the corresponding components:
\begin{equation} 
\begin{array}{lcl}
\label{eq:krivinekoeficijenti3BF}
{{\cal F}^\alpha{}_{\mu \nu}}&=&\partial_{\mu} {\alpha^\alpha{}_{\nu}}-\partial_{\nu}{\alpha^\alpha{}_{\mu}} + {f_{\beta\gamma}}^\alpha {\alpha}^\beta{}_{\mu}{\alpha}^\gamma{}_{\nu} -\beta^a{}_{\mu\nu}\partial_a{}^\alpha\,,\vphantom{\int\ds}\\
{{\cal G}^a{}_{\mu \nu\rho}}&=&\partial_{\mu}\beta^a{}_{\nu \rho}+\partial_{\nu}\beta^a{}_{\rho \mu}+\partial_{\rho}{\beta^a{}_{\mu \nu}} +{\alpha}^\alpha{}_{\mu}\beta^b{}_{\nu \rho}{\rhd_{\alpha b}}^a+{\alpha}^\alpha{}_{\nu}\beta^b{}_{\rho \mu}{\rhd_{\alpha b}}^a+{\alpha}^\alpha{}_{\rho}\beta^b{}_{\mu \nu}{\rhd_{\alpha b}}^a-\gamma^A{}_{\mu\nu\rho}{}\delta_A{}^a\,,\vphantom{\int\ds}\\
{{\mathcal{H}}^A{}_{\mu \nu \rho \sigma}}&=&\partial_{\mu} {\gamma^A{}_{\nu \rho \sigma}}-\partial_{\nu} {\gamma^A{}_{\rho \sigma\mu}}+\partial_{\rho} {\gamma^A{}_{ \sigma\mu\nu}}-\partial_{\sigma} {\gamma^A{}_{\mu\nu \rho}}\vphantom{\int\ds}\\&&+2\beta^a{}_{\mu \nu} \beta^b{}_{\rho\sigma} X_{(ab)}{}^A-2\beta^a{}_{\mu\rho}\beta^b{}_{\nu\sigma} X_{(ab)}{}^A+2\beta^a{}_{\mu\sigma}\beta^b{}_{\nu\rho} X_{(ab)}{}^A\vphantom{\int\ds}\\&&+{\alpha}^\alpha{}_{\mu}{\gamma^B{}_{\nu\rho\sigma}}{\rhd_{\alpha B}}^A-{\alpha}^\alpha{}_{\nu}{\gamma^B{}_{\rho\sigma\mu}}{\rhd_{\alpha B}}^A+{\alpha}^\alpha{}_{\rho}{\gamma^B{}_{\sigma\mu\nu}}{\rhd_{\alpha B}}^A-{\alpha}^\alpha{}_{\sigma}{\gamma^B{}_{\mu\nu\rho}}{\rhd_{\alpha B}}^A\,.\vphantom{\int\ds}    
\end{array}
\end{equation}
Then, similarly to the construction of $ BF $ and $ 2BF $ actions, one can define the gauge invariant topological $ 3BF $ action, with the underlying structure of a $3$-group. For the $4$-dimensional compact and orientable manifold $ \mathcal{M}_4 $ and the $2$-crossed module $ (L \stackrel {\delta} { \to} H \stackrel {\partial} {\to} G \,, \rhd \,, \{\_ \,, \_ \}{}_{\mathrm{pf}}) $, that gives rise to $3$-curvature (\ref{eq:3krivine}), one defines the $3BF$ action as
\begin{equation}\label{eq:bfcgdh}
S_{3BF} =\int_{\mathcal{M}_4} \langle B \wedge  {\cal F} \rangle_{\mathfrak{ g}} +  \langle C \wedge  {\cal G} \rangle_{\mathfrak{h}} + \langle D \wedge {\cal H} \rangle_{\mathfrak{l}} \,, 
\end{equation}
where $B \in \cA^2(\cM_4,\mathfrak{g})$, $C \in \cA^1(\cM_4,\mathfrak{h})$, and $D \in \cA^0(\cM_4,\mathfrak{l})$ are Lagrange multipliers. The forms $\killing{\_}{\_}_{\mathfrak{g}}$, $\killing{\_}{\_}_{\mathfrak{h}}$, and $\killing{\_}{\_}_{\mathfrak{l}}$ are $G$-invariant bilinear symmetric nondegenerate forms on $\mathfrak{g}$,  $\mathfrak{h}$, and $\mathfrak{l}$, respectively. Note that in the case of a semisimple Lie algebra, a natural choice for this bilinear form is the Killing form. However, one can also choose it differently, and moreover for a solvable Lie algebra one can introduce a non-trivial bilinear form, despite the fact that the Killing form is degenerate in this case. Fixing the basis in algebras $\mathfrak{g}$, $\mathfrak{h}$, and $\mathfrak{l}$, as defined in (\ref{eq:basis}), the forms $\killing{\_}{\_}_{\mathfrak{g}}$, $\killing{\_}{\_}_{\mathfrak{h}}$, and $\killing{\_}{\_}_{\mathfrak{l}}$ map pairs of basis vectors of algebras $\mathfrak{g}$, $\mathfrak{h}$, and $\mathfrak{l}$, to the metrics on their vector spaces, $g_{\alpha\beta}$, $g_{ab}$, and $g_{AB}$:
\begin{equation}
    \killing{\tau_\alpha}{\tau_\beta}_{\mathfrak{g}}=g_{\alpha\beta}\,,\quad \quad \killing{t_a}{t_b}_{\mathfrak{h}}=g_{ab}\,,\quad\quad \killing{T_A}{T_B}_{\mathfrak{l}}=g_{AB}\,.
\end{equation}
As the symmetric maps are nondegenerate, the inverse metrics $g^{\alpha\beta}$, $g^{ab}$, and $g^{AB}$ are well defined, and are used to raise and lower indices of the corresponding algebras.

Varying the action (\ref{eq:bfcgdh}) with respect to Lagrange multipliers $B^{\alpha}$, $C^a$, and  $D^A$ one obtains the equations of motion
\begin{equation}\label{eq:jed3bf}
\mathcal{F}^{\alpha} = 0 \,, \quad \quad \mathcal{G}^a = 0\,, \quad \quad \mathcal{H}^A = 0 \,,
\end{equation}
while varying with respect to the $3$-connection variables $\alpha^{\alpha}$, $\beta^a$, and $\gamma^A$ one gets:
\begin{gather}
\D B_\alpha- {f_{\alpha \beta}}^\gamma B_\gamma \wedge  \alpha^\beta - {\rhd_{\alpha a}}^b C_b\wedge \beta^a+\rhd_{\alpha B}{}^A D_A\wedge\gamma^B=0\,, \\ 
\D C_a - {\partial_a}^\alpha B_\alpha + {\rhd_{\alpha a}}^b C_b \wedge \alpha^\alpha + 2X_{(ab)}{}^AD_A\wedge \beta^b=0\,,\\
\D D_A - \rhd_{ \alpha A}{}^B D_B\wedge \alpha^\alpha+\delta_A {}^a C_a=0\,.
\end{gather}
For further details see \cite{martins2011,Wolf2014,Wang2014} for the definition of the $3$-group, and \cite{Radenkovic2019} for the definition of the pure $3BF$ action. 

Choosing the convenient underlying $2$-crossed module structure and imposing the appropriate simplicity constraints onto the degrees of freedom present in the $3BF$ action, one can obtain the non-trivial classical dynamics of the gravitational and matter fields. A reader interested in the construction of the constrained $2BF$ actions describing the Yang-Mills field and Einstein-Cartan gravity, and $3BF$ actions describing the Klein-Gordon, Dirac, Weyl and Majorana fields coupled to gravity in the standard way, is referred to \cite{MikovicVojinovic2012,Radenkovic2019}. One can also introduce higher dimensional, $nBF$ actions, see for example \cite{MV2020}. Various properties of these models have been studied in \cite{MV2012, MV2013, MV2015}. Naturally, if one is interested in theories defined on a $4$-dimensional spacetime manifold, there is an upper limit on the order of the differential forms one can use to construct a $n$-connection, and in $4$ dimensions that is $n=3$.
%%%%%%%%%%%%%%%%%%%%%%%%%%%%%%%%%%%%%%%%%%%%%%%%%%%%%%%%%%%%%%%%%%%%%%%%%%%%%%%%%%%%%%%%%%%%%%%%%%%%%%%%%%%%%%%%%%%%%%%%%%%%%%%%%%%%%%%%%%%%%%%%%%%%%
\section{Hamiltonian analysis of the $3BF$ theory}\label{secIII}
In this section, the canonical structure of the theory is presented, with the resulting first-class and second-class constraints present in the theory. The algebra of Poisson brackets between all, the first-class and the second-class constraints, is obtained.  We will use this result to calculate the total number of degrees of freedom in the theory, and in order to do that, we will have to analyse the Bianchi identities that the first-class constraints satisfy, which enforce restrictions in the sense of Hamiltonian analysis. They reduce the number of independent first-class constraints present in the theory, thus increasing the number of degrees of freedom. We will obtain that the pure $3BF$ theory is topological, i.e., there are no local propagating degrees of freedom. Finally, we will finish this section with the construction of the generator of gauge symmetries of the $3BF$ action, which is used to calculate the form-variations of all the variables and their canonical momenta. This result will be crucial for finding the gauge symmetries of $3BF$ action, which will be a topic of section \ref{secIV}. 
\subsection{Canonical structure and Hamiltonian}\label{secIIIa}
Assuming that the spacetime manifold $\cM_4 $ is globally hyperbolic, the Lagrangian on a spatial foliation $ \Sigma_3 $ of spacetime $\cM_4$ corresponding to the  $ 3BF $ action (\ref{eq:bfcgdh}) is given as:
\begin{equation}\label{eq:lagranžijan3BF}
    L_{3BF}= \int_{{\Sigma}_3} \D^3 \vec{x} \,\epsilon^{\mu\nu\rho\sigma}\big(\frac{1}{4}\,B^\alpha{}_{\mu\nu}\,\cF^\beta{}_{\rho\sigma}\, g_{\alpha\beta}+\frac{1}{3!}\, C^a{}_\mu\,
\cG^b{}_{\nu\rho\sigma}\, g_{ab}+ \frac{1}{4!} D^A \cH^B{}_{\mu\nu\rho\sigma} g_{AB}\big)\,.
\end{equation}
For the Lagrangian (\ref{eq:lagranžijan3BF}), the canonical momenta corresponding to all variables $B^\alpha{}_{\mu\nu}$, $\alpha^\alpha{}_\mu$, $C^a{}_\mu$, $\beta^a{}_{\mu\nu}$, $D^A$, and $\gamma^A{}_{\mu\nu\rho}$ are:
\begin{equation}\label{eq:3bfmomenti}
\begin{array}{lclcl}
\pi(B){_{\alpha}{}^{\mu\nu}} & = &  \ds\frac{\delta L}{\delta \partial_0 B{^{\alpha}{}_{\mu\nu}} }& = & 0\,, \vphantom{\ds\int}\\
\pi(\alpha){_{\alpha}{}^{\mu}} & = & \ds \frac{\delta L}{\delta \partial_0 \alpha{^{\alpha}{}_{\mu}}} & = & \ds \frac{1}{2}\epsilon^{0\mu\nu\rho} B_{\alpha\nu\rho}\,,\vphantom{\ds\int}\\
\pi(C){_a{}^\mu} & = &\ds \frac{\delta L}{\delta \partial_0 C{^a{}_\mu}} & = & 0 \,,\vphantom{\ds\int}\\
\pi(\beta){_a{}^{\mu\nu}} & = &\ds \frac{\delta L}{\delta \partial_0 \beta{^a{}_{\mu\nu}}} & = & - \epsilon^{0\mu\nu\rho} C_{a\rho}\,,\vphantom{\ds\int}\\
\pi(D){_A} & = & \ds\frac{\delta L}{\delta \partial_0 D{^{A}} } & = & 0\,,\vphantom{\ds\int}\\
\pi(\gamma){_A{}^{\mu\nu\rho}} & = & \ds\frac{\delta L}{\delta \partial_0 \gamma{}^A{}_{\mu\nu\rho}} & = & \epsilon^{0\mu\nu\rho} D_A\,.\vphantom{\ds\int}
\end{array}
\end{equation}
These momenta give rise to the six primary constraints of the theory, since none of them can be inverted for the time derivatives of the variables,
\begin{equation}
\begin{array}{lcl}
P(B){_{\alpha}{}^{\mu\nu} }& \equiv &\ds \pi(B){_{\alpha}{}^{\mu\nu}} \approx 0\,, \vphantom{\ds\int}\\
P(\alpha){_{\alpha}{}^{\mu}} & \equiv &\ds \pi(\alpha){_{\alpha}{}^{\mu}} -  \frac{1}{2}\epsilon^{0\mu\nu\rho} B_{\alpha\nu\rho} \approx 0\,,\vphantom{\ds\int} \\
P(C){_a{}^{\mu}} & \equiv & \ds\pi(C){_a{}^{\mu}}  \approx 0\,, \vphantom{\ds\int}\\
P(\beta){_a{}^{\mu\nu}} & \equiv & \ds\pi(\beta){_a{}^{\mu\nu}} + \epsilon^{0\mu\nu\rho}C_{a\rho} \approx 0\,, \vphantom{\ds\int}\\
P(D){_{A}{}}& \equiv & \ds\pi(D){_{A}{}} \approx 0\,, \vphantom{\ds\int}\\  
P(\gamma){_A{}^{\mu\nu\rho}} & \equiv & \ds\pi(\gamma)_A{}^{\mu\nu\rho} -
\epsilon^{0\mu\nu\rho} D_A \approx 0\,. \vphantom{\ds\int}\\
\end{array}
\end{equation}

Employing the following fundamental Poisson brackets,
\begin{equation}
\begin{array}{lcl}
\pb{B{^{\alpha}{}_{\mu\nu}}(\vec{x})}{\pi(B){_{\beta}{}^{\rho\sigma}(\vec{y})}} & = &  2\delta^\alpha_{\beta}  \delta^{\rho}_{[ \mu|} \delta^{\sigma}_{|\nu]} \,\delta^{(3)}(\vec{x}-\vec{y})\,, \vphantom{\ds\int}\\
\pb{\alpha{^{\alpha}{}_{\mu}}(\vec{x})}{\pi(\alpha){_{\beta}{}^{\nu}}(\vec{y})} & = &  \delta^\alpha_{\beta}  \delta^{\nu}_{\mu} \,\delta^{(3)}(\vec{x}-\vec{y})\,,\vphantom{\ds\int} \\
\pb{C{^a{}_{\mu}}(\vec{x})}{\pi(C){_b{}^{\nu}}(\vec{y})} & = & \delta^a_b \delta^{\nu}_{\mu} \,\delta^{(3)}(\vec{x}-\vec{y})\,, \vphantom{\ds\int}\\
\pb{\beta{^a{}_{\mu\nu}}(\vec{x})}{\pi(\beta){_b{}^{\rho\sigma}}(\vec{y})} & = & 2\delta^a_b \,\delta^{\rho}_{[\mu|} \delta^{\sigma}_{|\nu]} \,\delta^{(3)}(\vec{x}-\vec{y})\,, \vphantom{\ds\int}\\
\pb{D{^A{}}(\vec{x})}{\pi(D){_B{}}(\vec{y})} & = & \delta^A_B \,\delta^{(3)}(\vec{x}-\vec{y})\,, \vphantom{\ds\int}\\
\pb{\gamma{^A{}_{\mu\nu\rho}}(\vec{x})}{\pi(\gamma){_B{}^{\sigma\tau\xi}}(\vec{y})} & = & 3!\delta^A_B \,\delta^{\sigma}_{[\mu} \delta^{\tau}_{\nu} \delta^{\xi}_{\rho]}  \,\delta^{(3)}(\vec{x}-\vec{y})\,,\vphantom{\ds\int}
\end{array}
\end{equation}
one obtains the \emph{algebra of primary constraints}:
\begin{equation}\label{AlgebraPrimarnihVeza3bf}
\begin{array}{lcl}
\pb{P(B){_\alpha{}^{jk}}(\vec{x})}{P(\alpha){_{\beta}{}^i}(\vec{y})} & = & \epsilon^{0ijk}\,g_{\alpha\beta}(\vec{x}) \,\delta^{(3)}(\vec{x}-\vec{y})\,, \vphantom{\ds\int}\\
\pb{P(C){_a{}^{k}}(\vec{x})}{P(\beta){_b{}^{ij}}(\vec{y})} & = & - \epsilon^{0ijk}\, g_{ab}(\vec{x})\, \delta^{(3)}(\vec{x}-\vec{y})\,, \vphantom{\ds\int}\\
\pb{P(D){_A(\vec{x})}}{P(\gamma){_B{}^{ijk}}(\vec{y})} & = & \epsilon^{0ijk}\, g_{AB}(\vec{x})\, \delta^{(3)}(\vec{x}-\vec{y})\,. \vphantom{\ds\int}\\
\end{array}
\end{equation}
Note that all other Poisson brackets vanish.  The \emph{canonical, on-shell Hamiltonian} is given by the following expression:
\begin{equation}\label{Hcan3BF}
\begin{aligned}
H_c = \int_{\Sigma_3} & \D^3\vec{x} \bigg[\frac{1}{2} \pi(B){_{\alpha}{}^{\mu\nu}}\,\partial_0 B{^{\alpha}{}_{\mu\nu}} + \pi(\alpha){_{\alpha}{}^{\mu}} \,\partial_0 \alpha{^{\alpha}{}_{\mu}} + \pi(C){_a{}^{\mu}} \, \partial_0 C{^a{}_{\mu}}\\&+ \frac{1}{2} \pi(\beta){_a{}^{\mu\nu}}\, \partial_0 \beta{^a{}_{\mu\nu}} + \pi(D){_{A}} \, \partial_0 D{^{A}} +\frac{1}{3!} \pi(\gamma){_A{}^{\mu\nu\rho}} \, \partial_0 \gamma{^A{}_{\mu\nu\rho}}\bigg]-L\,.
\end{aligned}
\end{equation}
Employing the definition of the curvature components (\ref{eq:krivinekoeficijenti3BF}), the Hamiltonian (\ref{Hcan3BF}) can be written as the sum of terms that are equal to the product of the primary constraints and time derivatives of the variables, and the remainder. As the primary constraints are zero on-shell, the terms multiplying the time derivatives vanish, and the canonical Hamiltonian becomes:
\begin{equation}
\begin{aligned}
H_c = & - \int_{\Sigma_3} \D^3\vec{x}\, \epsilon^{0ijk} \bigg[ \frac{1}{2} B_{\alpha 0i}\, \cF{^{\alpha}{}_{jk}} + \frac{1}{6} C_{a 0} \,\cG{^a{}_{ijk}} + \beta{^a{}_{0i}} \bigg(\nabla_j C_{a k}-\frac{1}{2}\partial{_a{}^\alpha}B_{\alpha\,{jk}}+\beta{^b{}_{jk}}\,D_A\, X_{(ab)}{}^A\bigg)\\ & + \frac{1}{2} \alpha{^\alpha{}_0}\bigg( \nabla_i B{_{\alpha\,jk}} - C{_a{}_i}\,\rhd_{\alpha b}{}^a\, \beta{^b{}_{jk}}+\frac{1}{3}D_A \rhd_{\alpha B}{}^A\,\gamma^B {}_{ijk}\,\bigg)+ \frac{1}{2}\gamma{^A{_{0ij}}}\bigg(\nabla_k D_A+ C{_a{}_k} \,\delta_A{}^a \bigg) \bigg] \,. \vphantom{\ds\int}
\end{aligned}
\end{equation}
Adding to the canonical Hamiltonian the product of the Lagrange multipliers $ \lambda $ and the primary constraints, for every primary constraint, one gets the \emph{total, off-shell Hamiltonian}:
\begin{equation} \label{TotalniHamiltonijan3BF}
\begin{array}{ccl}
H_T & = & H_c \!+\!\ds  \int_{\Sigma_3}  \D^3\vec{x} \bigg[ \frac{1}{2} \lambda(B){^{\alpha}{}_{\mu\nu}} P(B){_{\alpha}{}^{\mu\nu}} + \lambda(\alpha){^{\alpha}{}_{\mu}} P(\alpha){_{\alpha}{}^{\mu}} + \lambda(C){^a{}_{\mu}} P(C){_a{}^{\mu}} \vphantom{\ds\int}\\ \ds
 & &\ds + \frac{1}{2} \lambda(\beta){^a{}_{\mu\nu}} P(\beta){_a{}^{\mu\nu}}+ \lambda(D){^A} P(D){_A}+\frac{1}{3!}\lambda(\gamma){^A{}_{\mu\nu\rho}} P(\gamma){_A{}^{\mu\nu\rho}} \bigg] \,. \vphantom{\ds\int}
\end{array}
\end{equation}

\subsection{Consistency conditions and algebra of constraints}\label{secIIIb}

In order for primary constraints to be preserved during the evolution of the system, they must satisfy the consistency conditions,
\begin{equation}
\dot{P} \equiv \pb{P}{H_T} \approx 0\,,\label{pcc3BF}
\end{equation}
for every primary constraint $P$. Imposing this condition on primary constraints ${P}(B)_{\alpha}{}^{0i}$, ${P}(\alpha){_{\alpha}{}^0}$, ${P}(C){_a{}^0}$, ${P}(\beta){_a{}^{0i}}$, and ${P}(\gamma){_{A}{}^{0ij}}$, one obtains the secondary constraints $\cS$,
\begin{equation} \label{SekundarneVeze3BF}
\begin{array}{lcl}
\cS(\cF){_\alpha}{}^{i} & \equiv & \ds \frac{1}{2}\epsilon^{0ijk} \cF{_{\alpha}{}_{jk}} \approx 0\,, \vphantom{\ds\int}\\
\cS(\nabla B)_{\alpha} & \equiv & \ds \frac{1}{2}\epsilon^{0ijk} \big( \nabla_{[i} B{_{\alpha\,jk]}} - C{_a{}_{[i}}\,\rhd_{\alpha b}{}^a\, \beta{^b{}_{jk]}}+\frac{1}{3}D_A \rhd_{\alpha B}{}^A\,\gamma^B {}_{ijk}\, \big) \approx 0 \,,\vphantom{\ds\int} \\
\cS(\cG)_a & \equiv & \ds \frac{1}{6}\epsilon^{0ijk}\cG{_a{}_{ijk}} \approx 0\,,\vphantom{\ds\int} \\
\cS(\nabla C){_a{}^{i}} & \equiv & \ds \epsilon^{0ijk}\big(\nabla_{[j|} C_{a |k]}-\frac{1}{2}\partial{_a{}^\alpha}B_{\alpha\,{jk}}+\beta{^b{}_{jk}}\,D_A\, X_{(ab)}{}^A\big) \approx 0\,,\vphantom{\ds\int} \\
\cS(\nabla D){_A{}^{ij}} & \equiv & \epsilon^{0ijk}\big( \nabla_k D_A+ C{_a{}_k} \,\delta_A{}^a  \big) \approx 0\,,\vphantom{\ds\int}
\end{array}
\end{equation}
while in the case of the constraints  ${P}(\alpha){_{\alpha}{}^k}$, ${P}(B){_{\alpha}{}^{jk}}$, ${P}(\beta){_a{}^{jk}}$, ${P}(C){_a{}^k}$, ${P}(\gamma){_A{}^{ijk}}$, and ${P}(D){_{A}}$ the corresponding consistency conditions determine the following Lagrange multipliers:
{\medmuskip=0mu
\thinmuskip=0mu
\thickmuskip=0mu
\begin{equation}
    \begin{array}{lcl}
        \lambda(B)_\alpha{}_{ij}&\approx&\ds\nabla_i B_\alpha{}_{0j}-\nabla_j B_\alpha{}_{0i}+C_a{}_0 \beta^b{}_{ij}\rhd_{\alpha}{}_b{}^a+ C{_{b}{}_{i}}\rhd_{\alpha}{}^b{}_a \beta{^{a}{}_{0j }}\vphantom{\ds\int}\\
        & & \ds-C{_{b}{}_{j}}\rhd_{\alpha}{}^b{}_a \beta{^{a}{}_{0i}}+g{}_{\beta}{}_\gamma{}^{\alpha} \alpha^\beta{}_0 B^\gamma{}_{ij}+D_B \gamma^A{}_{0ij}\rhd_\alpha{}^B{}_A \,,\vphantom{\ds\int}\\
        \lambda(\alpha){^{\alpha}{}_{i}} & \approx & \ds\nabla_i \alpha{^{\alpha}{}_0}+\partial{{}_a{}}^\alpha\beta{^a{}_{0i}}\,, \vphantom{\ds\int}\\
        \lambda(C)^a{}_i & \approx & \ds\nabla_i C^a{}_0+ C{^b{}_i} \rhd_{\alpha}{}^{a}{}_b \alpha{^\alpha{}_{0}}-2\beta_b{}_{0i}D_A X^{(ba)A}+B_\alpha{}_{0i}\partial^a{}^\alpha\,,\vphantom{\ds\int}\\
        \lambda(\beta){^a{}_{ij}} & \approx & \ds \nabla_{i} \beta{^a{}_{0j}} - \nabla_{j} \beta{^a{}_{0i}} - \beta{^b{}_{ij}}\rhd_{\alpha b}{}^a \alpha{^\alpha{}_0} + \gamma^A{}_{0ij}\delta_A{}^a \,, \vphantom{\ds\int}\\
        \lambda(D)_A & \approx & \ds\alpha^\alpha{}_0 D_B \rhd_{\alpha A}{}^B-C_a{}_0\delta_A{}^a\,,\vphantom{\ds\int}\\
        \lambda(\gamma){}^A{}_{ijk} & \approx  &\ds-2\beta^a{}_{0i}\beta^b{}_{jk}X_{(ab)}{}^A+2\beta^a{}_{0j}\beta^b{}_{ik}X_{(ab)}{}^A-2\beta^a{}_{0k}\beta^b{}_{ij}X_{(ab)}{}^A\vphantom{\ds\int}\\& &\ds-\alpha^\alpha{}_0
        \rhd_{\alpha B}{}^A \gamma^B{}_{ijk} + \nabla_i\gamma^A{}_{0jk}-\nabla_j\gamma^A{}_{0ik}+\nabla_k\gamma^A{}_{0ij}\,.\vphantom{\ds\int}\\
    \end{array}
\end{equation}}
Note that the rest of the Lagrange multipliers 
\begin{equation}
\lambda(B){^{\alpha}{}_{0i}}\,, \qquad
\lambda(\alpha){^{\alpha}{}_0}\,, \qquad
\lambda(C){^a{}_0}\,, \qquad 
\lambda(\beta){^a{}_{0i}}\,, \qquad
\lambda(\gamma){^A{}_{0ij}}\,,
\end{equation}
remain undetermined.

Further, as the secondary constraints must also be preserved during the evolution of the system, the consistency conditions of secondary constraints must be enforced. However, no tertiary constraints arise from these conditions (see equation (\ref{eq:secondclassconstwithHamiltonian}) in Appendix \ref{AppB}), leading the iterative procedure to an end. Finally, the total Hamiltonian can be written in the following form:
\begin{equation}\label{totalhamiltonian}
    \begin{aligned}
        H_T & = & \int_{\Sigma_3} \D^3{\vec{x}} \bigg[ \lambda(B){^{\alpha}{}_{0i}} \,\Phi(B){_{\alpha}{}^i} + \lambda(\alpha)^{\alpha} \,\Phi(\alpha)_{\alpha} + \lambda(C){^a{}_0} \,\Phi(C)_a + \lambda(\beta){^a{}_{0i}} \,\Phi(\beta){_a{}^{i}} +\frac{1}{2} \lambda(\gamma){^A{}_{0ij}}\Phi(\gamma)_A{}^{ij}\\
&   & \hphantom{\int}   - B_{\alpha 0i} \,\Phi(\cF)^{ai} - \alpha_{\alpha 0} \,\Phi(\nabla B)^{\alpha } - C_{a0} \,\Phi(\cG)^a - \beta_{a0i} \,\Phi(\nabla C)^{a i} - \frac{1}{2}\gamma_{A 0ij}\, \Phi(\nabla D)^{Aij} \bigg]\,,\\
    \end{aligned}
\end{equation}
where
\begin{equation}\label{eq:VezePrveKlase}
    \begin{array}{lcl}
\Phi(B){_{\alpha}{}^i} & = &P(B){_{\alpha}{}^{0i}}\,,\vphantom{\ds\int}\\
\Phi(\alpha){}_{\alpha} & = &P(\alpha){_{\alpha}{}^0}\,,\vphantom{\ds\int}\\
\Phi(C)_a & = &  P(C){_a{}^0}\,,\vphantom{\ds\int}\\
\Phi(\beta){_a{}^i} & = &P(\beta){_a{}^{0i}}\,,\vphantom{\ds\int}\\
\Phi(\gamma){_A{}^{ij}} & = &  P(\gamma){_A{}^{0ij}}\,,\vphantom{\ds\int}\\
\Phi(\cF)^{\alpha i} & = & \cS(\cF){^{\alpha i }} - \nabla_j P(B)^{\alpha ij}-P(C)_a{}^{i} \partial^{a\alpha}\,, \vphantom{\ds\int}\\
\Phi(\cG)_a & = & \ds\cS(\cG)_a + \nabla_i P(C)_{a}{}^{i}- \frac{1}{2} \beta_b{}_{ij}\rhd_{\alpha}{}^b{}_a P(B)^{\alpha ij}+P(D)^A\delta_A{}_a\,,\vphantom{\ds\int}\\
\Phi(\nabla C)_{a}{}^{i} & = & \ds \cS(\nabla C){_{a}{}^{i}} - \nabla_j P(\beta){}_a{}^{ij} + C_b{}_j \rhd{}_{\alpha}{}^b{}_a P(B)^{\alpha ij}\vphantom{\ds\int}\\& & \ds- \partial{_a{}^\alpha}P(\alpha){_{\alpha }{}^i}+2D_A X_{(ab)}{}^AP(C)^b{}^i+\beta^b{}_{jk}X_{(ab)}{}^AP(\gamma)_A{}^{ijk}\,,\vphantom{\ds\int}\\
\Phi(\nabla B)_{\alpha}& = & \ds\cS(\nabla B)_{\alpha} + \nabla_i P(\alpha)_{\alpha}{}^i -\frac{1}{2} f{}_{\alpha\gamma}{}^\beta B{_{\beta}{}_{ij}} P(B)^{\gamma}{}^{ij}\vphantom{\ds\int}\\
 & &-  \ds C{_{b}{}_i}\rhd_{\alpha a}{}^b P(C)^a{}^{i} - \frac{1}{2}\beta{_{b}{}_{ij}} \rhd_{\alpha a}{}^b P(\beta)^a{}^{ ij}\vphantom{\ds\int}\\& &-\ds P(D)^AD_B\rhd_{\alpha A}{}^B+\frac{1}{3!}P(\gamma)_A{}^{ijk}\gamma^B{}_{ijk}\rhd_{\alpha B}{}^A\,,\vphantom{\ds\int}\\
 \Phi(\nabla D)_A{}^{ij} & = & \ds \cS(\nabla D)_A{}^{ij}+\nabla_k P(\gamma){}_A{}^{ijk}-P(\beta){}_a{}^{ij}\delta_A{}^a-P(B)^\alpha{}^{ij}\rhd_\alpha{}^B{}_A D_B\,,\vphantom{\ds\int}\\
    \end{array}
\end{equation}
are the first-class constraints. The second-class constraints in the theory are:
\begin{equation}\label{eq:VezeDrugeKlase}
\begin{aligned}
\chi(B){_{\alpha}{}^{jk}}=P(B){_{\alpha}{}^{jk}}\,&,& \quad \chi(C){_a{}^i}=P(C){_a{}^i}\,&,& \quad \chi(D)_A=P(D)_A\,&,&\vphantom{\ds\int} \\ 
\chi(\alpha){_{\alpha}{}^i}=P(\alpha){_{\alpha}{}^i}\,&,& \quad  \chi(\beta){_a{}^{ij}}=P(\beta){_a{}^{ij}}\,&,& \quad \chi(\gamma){_A{}^{ijk}}=P(\gamma){_A{}^{ijk}}\,&.& \vphantom{\ds\int}
\end{aligned}
\end{equation}
The PB algebra of the first-class constraints is given by
\begin{equation} 
\begin{array}{lcl}
\pb{\Phi(\cF){^\alpha{}^{i}}(\vec{x})}{\Phi(\nabla B)_\beta(\vec{y})} & = & f{{}_{\beta\gamma}}{}^\alpha\,\Phi(\cF){^\gamma{}^{i}}(\vec{x})\,\delta^{(3)}(\vec{x}-\vec{y})\,, \vphantom{\ds\int}\\
\pb{\Phi(\nabla B)_\alpha(\vec{x})}{\Phi(\nabla B)_\beta(\vec{y})} & = & f{_{\alpha\beta}}{}^\gamma  \, \Phi(\nabla B)_\gamma (\vec{x})\,\delta^{(3)}(\vec{x}-\vec{y})\,,\vphantom{\ds\int}\\
\pb{\Phi(\cG)^a (\vec{x})}{\Phi(\nabla C)_{b}{}^{ i}(\vec{y})} & = &- \rhd{{}_{\alpha b}}{}^a \,\Phi(\cF)^{\alpha i}(\vec{x})\, \delta^{(3)}(\vec{x}-\vec{y}) \,,\vphantom{\ds\int}  \\
\pb{\Phi(\nabla C)_a{}^{ i} (\vec{x})}{\Phi(\nabla C)_{b}{}^{ j}(\vec{y})} & = &-2 X_{(ab)}{}^A \,\Phi(\nabla D)_{A}{}^{ij}(\vec{x})\, \delta^{(3)}(\vec{x}-\vec{y}) \,,\vphantom{\ds\int}  \\
\pb{\Phi(\cG)^a (\vec{x})}{\Phi(\nabla B)_{\alpha}(\vec{y})} & = &  \rhd_{\alpha b}{}^a \, \Phi(\cG)^{b}(\vec{x})\, \delta^{(3)}(\vec{x}-\vec{y}) \,, \vphantom{\ds\int} \\
\pb{\Phi(\nabla C){^a{}^i}(\vec{x})}{\Phi(\nabla B)_\alpha(\vec{y})} & = & \rhd{{}_{\alpha b}}{}^a\,\Phi(\nabla C){^b{}^i}(\vec{x})\,\delta^{(3)}(\vec{x}-\vec{y})\,,\vphantom{\ds\int}\\
\pb{\Phi(\nabla B)_{\alpha}(\vec{x})}{\Phi(\nabla D)_A{}^{ij}(\vec{y})}&=&
\rhd_\alpha{}_A{}^B\Phi(\nabla D){}_B{}^{ij}(\vec{x})\delta^{(3)}(\vec{x}-\vec{y})\,.\vphantom{\ds\int}
\end{array}
\end{equation}
The algebra between the first and the second class constraints is given in the Appendix \ref{AppB}, equation (\ref{eq:jednacinaPB1i2}). 

With the algebra of the constraints in hand, one can proceed to calculate the generator of gauge symmetries of the action. The generator will be used to calculate the form-variations of all the variables and their canonical momenta, which will help us find the finite gauge symmetries of the action. Additionally, we can determine the number of independent parameters of gauge transformations, since usually all the first class constraints generate unphysical transformations of dynamical variables, i.e., that to each parameter of the gauge symmetry there corresponds one first-class constraint. However, before we embark on the construction of the symmetry generator, we will devote some attention to the number of local propagating degrees of freedom in the theory, in order to determine if the $3BF$ action is topological or not.

\subsection{Number of degrees of freedom}\label{secIIIc}

In this subsection, we will show that the structure of the constraints implies that there are no local degrees of freedom in a $3BF$ theory. To that end, let us first specify all the Bianchi identities (BI) present in the theory. 

The $2$-form curvatures corresponding to $1$-forms $\alpha$ and $C$, given by
\begin{equation}
 F^\alpha = \D\alpha^\alpha + f{}_{\beta \gamma}{}^\alpha\,\alpha^\beta \wedge \alpha^\gamma \,,\qquad T^a= \D C^a + \rhd{}_{\alpha b}{}^a \, \alpha^\alpha \wedge C^b \,,
\end{equation}
satisfy the BI:
\begin{equation} 
\epsilon^{\lambda\mu\nu\rho} \,\nabla_\mu F^\alpha{}_{\nu\rho} = 0 \,,\label{bia2BF}
\end{equation}
\begin{equation}
\epsilon^{\lambda\mu\nu\rho}\left( \nabla_\mu T^a{}_{\nu\rho} -\,\rhd{}_{\alpha b}{}^a F^\alpha{}_{\mu\nu}C^b{}_\rho \right) = 0 \,.\label{bic2BF}
\end{equation}
Similarly, the $3$-form curvatures corresponding to $2$-forms $B$ and $\beta$, given by
\begin{equation}\label{bi2BF}
 S^\alpha = \D B^\alpha + f{}_{\beta\gamma}{}^\alpha\,\alpha^\beta \wedge B^\gamma \,, \qquad G^a = \D \beta^a + \rhd{}_{\alpha b}{}^a \,\alpha^\alpha \wedge \beta^\beta \,,
\end{equation}
satisfy the BI:
\begin{equation}
 \epsilon^{\lambda\mu\nu\rho}\left( \frac{2}{3} \nabla_\lambda \,S^\alpha{}_{\mu\nu\rho} -f{}_{\beta\gamma}{}^\alpha F^\beta{}_{\lambda\mu}\,B^\gamma{}_{\nu\rho}\right) = 0\,, \label{b2c2bf}
 \end{equation}
\begin{equation} 
\epsilon^{\lambda\mu\nu\rho}\left( \frac{2}{3} \nabla_\lambda \,G^a{}_{\mu\nu\rho} - \rhd{}_{\alpha b}{}^a \,F^\alpha{}_{\lambda\mu}\,\beta^b{}_{\nu\rho}\right) = 0 \,.\label{b2d}
\end{equation}
Finally, defining the $1$-form curvature for $D$, 
\begin{equation}
    Q^A = \D D^A + \rhd_{\alpha B}{}^A \alpha^\alpha \wedge D^B\,,
\end{equation}
one can write the corresponding BI for $Q^A$:
\begin{equation}\label{eq:BI3BF}
    \epsilon^{\lambda\mu\nu\rho}\left( \nabla_\nu Q^A{}_\rho - \frac{1}{2}\rhd_{\alpha B}{}^A F^\alpha{}_{\nu\rho} D^B \right) =0\,.
\end{equation}

These Bianchi identities play an important role in determining the number of degrees of freedom present in the theory.

As the general theory states, if there are $N$ fields in the theory, $F$ independent first-class constraints per space point, and $S$ independent second-class constraints per space point, the number of independent field components, i.e., the number of the physical degrees of freedom present in the theory, is given by:
\begin{equation}\label{n}
   n= N- F- \frac{S}{2}\,. 
\end{equation}
Let $p$ denote the dimensionality of the group $G$, $q$ the dimensionality of the group $H$, and $r$ the dimensionality of the group $L$. Determining the number of fields present in the $3BF$ theory, by counting the field components listed in Table \ref{tab:3bfN}, one obtains $N=10(p+q)+5r$.
\begin{table}[!ht]
    \centering
    \begin{tabular}{|c|c|c|c|c|c|} \hline
$\alpha^{\alpha}{}_{\mu}$ & $\beta^a{}_{\mu\nu}$ & $\gamma^A{}_{\mu\nu\rho}$ & $B^{\alpha}{}_{\mu\nu}$ & $C^a{}_{\mu}$ & $D^A$ \\ \hline
$4p$ & $6q$ & $4r$ & $6p$ & $4q$ & $r$ \\ \hline
    \end{tabular}
    \caption{The fields present in the $3BF$ theory.}
    \label{tab:3bfN}
\end{table}
Similarly, one determines the number of independent components of the second-class constraints by counting the components listed in Table \ref{tab:sc3BF} and obtains $S=6(p+q)+2r$.
\begin{table}[!ht]
    \centering
    \begin{tabular}{|c|c|c|c|c|c|} \hline
$\chi(B)_{\alpha}{}^{jk}$ & $\chi(C)_a{}^i$ & $\chi(D)_A$ & $\chi(\alpha)_{\alpha}{}^i$ & $\chi(\beta)_a{}^{ij}$ & $\chi(\gamma)_A{}^{ijk}$ \\ \hline
$3p$ & $3q$ & $r $& $3p$ & $3q$ & $r$\\ \hline
    \end{tabular}
    \caption{Second-class constraints in the $3BF$ theory.}
    \label{tab:sc3BF}
\end{table}
However, when counting the number of the first-class constraints $F$ one notes they are not all mutually independent. Namely, one can prove the following identities, as a consequence of the BI.

Taking the derivative of $\Phi(\cF)_\alpha{}^{i}$ one obtains
\begin{equation}\label{ra2bf}
\nabla_i \Phi(\cF)^\alpha{}^{i} + \partial_{a}{}^{\alpha} \Phi(\cG)^a
= \frac{1}{2}\epsilon^{0ijk}\nabla_i F^{\alpha}{}_{ jk} -\frac{1}{2}f_{\beta\gamma}{}^\alpha\cF^\beta{}_{ij}P(B)^{ij}  \,.\\
\end{equation}
This relation gives
\begin{equation}\label{eq:r1}
\nabla_i \Phi(\cF)^\alpha{}^{i} + \partial_{a}{}^{\alpha} \Phi(\cG)^a=0\,,
\end{equation}
since the first term on the right-hand side of (\ref{ra2bf}) is zero off-shell because $\epsilon ^{ijk}\, \nabla_i F^a{}_{jk} =0$ are the $\lambda=0$ components of BI (\ref{bia2BF}), and the second term on the right-hand side is also zero off-shell, since it is a product of two constraints:
\begin{equation}
    \frac{1}{2}f_{\beta\gamma}{}^\alpha\cF^\beta{}_{ij}P(B)^{ij} =\frac{1}{2}f_{\beta\gamma}{}^\alpha\epsilon_{0ijk}\cS(\cF)^\beta{}^{k}P(B)^{ij}=0\,.
\end{equation}
The relation (\ref{eq:r1}) means that $p$ components of the first-class constraints $\Phi(\cF)^\alpha{}^{i}$ and $\Phi(\cG)^a$ are not independent of the others. 
Furthermore, taking the derivative of $\Phi(\nabla C)_a{}^{i}$ one obtains
\begin{equation}
    \begin{aligned}\label{rb2bf} 
 &\phantom{\int\ds}\nabla_i \Phi(\nabla C)_a{}^{i} +\,C_{b i} \rhd{{}_\alpha{}_a{}^b} \Phi(\cF)^{\alpha i} + \partial_{a}{}^{\alpha} \Phi(\nabla B)_\alpha -\beta^b{}_{ij}X_{(ab)}{}^A\Phi(\nabla D)_A{}^{ij}-2D_AX_{(ab)}{}^A\,\Phi(\cG)^b\\&\phantom{\int\ds}=\frac{1}{2} \epsilon^{0ijk}\left( \nabla_i T_{a jk} -\rhd_{\alpha b}{}^a F^\alpha{}_{jk}C^b{}_i \right)-\frac{1}{2}\epsilon^{0ijk}\rhd_{\alpha a}{}^b\,P(B)^\alpha{}_{ij}\,S(\nabla C)_b{}_k+\epsilon^{0ijk}X_{(ab)}{}^A\,P(C)^b{}_i\,S(\nabla D)_A{}_{jk}\\&\quad \phantom{\int\ds}+\frac{1}{3}\epsilon^{0ijk}X_{(ab)}{}^A\,P(\gamma)^A{}_{ijk}\,S(\cG)^b+\frac{1}{2}\epsilon^{0ijk}\rhd_{\alpha a}{}^b\,P(\beta)^b{}_{ij}\,S(\cF)^\alpha{}_k. 
\end{aligned}
\end{equation}
Noting that the right-hand side of (\ref{rb2bf}) is zero off-shell as the $\lambda=0$ components of the BI (\ref{bic2BF}), and the remaining terms on the right-hand side are zero off-shell as products of two constraints, one obtains the following relation:
\begin{equation}\label{eq:r2}
\phantom{\int\ds}\nabla_i \Phi(\nabla C)_a{}^{i} +C_{b i} \rhd{{}_\alpha{}_a{}^b} \Phi(\cF)^{\alpha i} + \partial_{a}{}^{\alpha} \Phi(\nabla B)_\alpha -\beta^b{}_{ij}X_{(ab)}{}^A\Phi(\nabla D)_A{}^{ij}-2D_AX_{(ab)}{}^A\Phi(\beta)^b= 0\,.
\end{equation}
This relation means that $q$ components of the constraints $\Phi(\nabla C)_a{}^{i}$, $\Phi(\cF)^{\alpha i}$, $\Phi(\nabla B)_\alpha$, $\Phi(\nabla D)_A{}^{ij}$, and $\Phi(\beta)^b$, are not independent of the others, further lowering the number of the independent first-class constraints. Finally, the following relation is satisfied
\begin{equation}
\begin{aligned}
&\phantom{\int\ds}\nabla^j \Phi(\nabla D)^A{}_{ij} - \rhd_\alpha{}_B{}^A D^B \Phi(\cF){}_\alpha{}_{i}-\delta{}^A{}_a\Phi(\nabla C)^a{}_i\\&\phantom{\int\ds}=\epsilon_{0ijk}(\nabla^j Q_A{}^k+\frac{1}{2}\rhd_{\alpha A}{}^BF^\alpha{}^{jk}D_B)+\frac{1}{2} \epsilon^{0jkl}\rhd_{\alpha B}{}^A\, P(\gamma)^B{}_{ijk}\,S(\cF)_\alpha{}_l-\frac{1}{2}\epsilon^{0jkl}\rhd_{\alpha B}{}^A\,P(B)^\alpha{}_{ij}\,S(\nabla D)^B{}_{kl}\,.
\end{aligned}
\end{equation}
Since the first term on the right-hand side is precisely the $\lambda=0$ component of the BI (\ref{eq:BI3BF}), while the second and third terms are equal to zero as products of two constraints, this gives:
\begin{equation}\label{eq:r3}
\nabla^j \Phi(\nabla D)^A{}_{ij} - \rhd_{\alpha B}{}^A D^B \Phi(\cF){}_\alpha{}_{i}-\delta{}^A{}_a\Phi(\nabla C)^a{}_i=0\,.
\end{equation}
This relation suggests that $3r$ components of the primary constraints $\Phi(\nabla D)^A{}_{ij}$, $\Phi(\cF){}_\alpha{}_{i}$, and $\Phi(C)^a{}_i$ are not independent of the others. However, this is slightly misleading, since the covariant derivative of the BI (\ref{eq:BI3BF}) is automatically satisfied as a consequence of the BI (\ref{bia2BF}),

\begin{equation}
     \epsilon^{\lambda\mu\nu\rho} D^B \rhd_{\alpha B}{}^A \nabla _\mu F^\alpha{}_{\nu\rho}   =0\,,
\end{equation}
which means that there are in fact only $2r$ components of the constraint (\ref{eq:r3}). A formal proof of this statement would involve evaluating the Wronskian of all first-class constraints, and is out of the scope of this paper.

The number of independent components of first-class constraints is determined by counting the components listed in Table \ref {tab:3bffcc}, and then subtracting the number of independent relations (\ref{eq:r1}), (\ref{eq:r2}), and (\ref{eq:r3}).
\begin{table}[!ht]
    \centering
    \begin{tabular}{|c|c|c|c|c|c|c|c|c|c|c|} \hline
$\Phi(B)_{\alpha}{}^i$ & $\Phi(C)_a$ & $\Phi(\alpha)_{\alpha}$ & $\Phi(\beta)_a{}^i$  & $\Phi(\gamma)_A{}^{ij}$ & $\Phi(\cF)^{\alpha i}$ & $\Phi(\cG)^a$ & $\Phi( \nabla C)^{a i}$ & $\Phi(\nabla B)^{\alpha}$ & $\Phi(\nabla D)_A{}^{ij}$ \\ \hline
$3p$ & $q$ & $p$ & $3q$ & $3r$ & $3p-p$ & $q$ & $3q-q$ & $p$ & $3r-2r$ \\ \hline 
    \end{tabular}
    \caption{First-class constraints in the $3BF$ theory.}
    \label{tab:3bffcc}
\end{table}

Bearing the previous analysis in mind, one obtains the number of independent first-class constraints:
$$F=8(p+q)+6r - p - q -2r= 7(p+q)+ 4r \,.$$ 
Finally, using the definition (\ref{n}), the number of degrees of freedom in the $ 3BF $ theory is:
\begin{equation} \label{BrojFizickihStepeniSlobode3bf}
n = 10(p+q) + 5r - 7(p+q) - 4r - \frac{6(p+q)+2r}{2} = 0\,.
\end{equation}
Therefore, there are no local propagating degrees of freedom in a $3BF$ theory.
\subsection{Symmetry generator}\label{secIIId}

The unphysical transformations of dynamical variables are often referred to as gauge transformations. The gauge transformations are \emph{local}, meaning that the parameters of
the transformations are arbitrary functions of space and time. We shall now construct the generator of all gauge symmetries of the theory governed by the total Hamiltonian (\ref{totalhamiltonian}), using the Castellani's algorithm (see Chapter 5 in \cite{blagojevic2002gravitation} for a comprehensive overview of the procedure). The details of the construction are given in Appendix \ref{AppC}, and the following result is obtained
\begin{equation}\label{eq:generator3BF}
\begin{array}{ccl}
    G&=&\ds\int_{\Sigma_3}\D^3 \vec{x} \bigg( (\nabla_0 \epsilon_{\mathfrak{g}}{}^{\alpha})\,(\tilde{G}_1)_\alpha+\epsilon_{\mathfrak{g}}{}^\alpha\, (\tilde{G}_0)_\alpha+( \nabla_0 \epsilon_{\mathfrak{h}}{}^a{}_i)\,(\tilde{H}_1)_a{}^i+\epsilon_{\mathfrak{h}}{}^a{}_i\,(\tilde{H}_0)_a{}^i \\\ds& &\qquad\vphantom{\ds\int}\ds+\frac{1}{2}(\nabla_0\epsilon_{\mathfrak{l}}{}
   ^A{}_{ij})\,(\tilde{L}_1)_A{}^{ij}+ \frac{1}{2}\epsilon_{\mathfrak{l}}{}^A{}_{ij}\,(\tilde{L}_0)_A{}^{ij} \\ \ds &&\vphantom{\ds\int} \ds\qquad+(\nabla_0 \epsilon_{\mathfrak{m}}{}^{\alpha}{}_i)\,(\tilde{M}_1)_\alpha{}^i+\epsilon_{\mathfrak{m}}{}^{\alpha}{}_i\,(\tilde{M}_0)_\alpha{}^i+(\nabla_0\epsilon_{\mathfrak{n}}{}{}^{a})\,(\tilde{N}_1)_a+\epsilon_{\mathfrak{n}}{}^a(\tilde{N}_0)_a \bigg)\,,
\end{array}
\end{equation}
where 
\begin{equation}\label{eq:gtilda}
    \begin{array}{lcl}
    (\tilde{G}_1)_\alpha&=&-\Phi(\alpha)_\alpha\,,\vphantom{\ds\int}\\
    (\tilde{G}_0)_\alpha&=&-\big( f_{\alpha\gamma}{}^\beta B_{\beta 0i}\Phi(B)^\gamma{}^i+C_{a0}\rhd_{\alpha b}{}^a \Phi(C)^{b0}+\beta_{a0i}\rhd_{\alpha b}{}^a \Phi(\beta)^{b0i}-\frac{1}{2}\gamma^A{}_{0ij}\rhd_{ \alpha A}{}^B\Phi(\gamma)_B{}^{ij}-\Phi(\nabla B){}_{\alpha}\big)\,,\vphantom{\ds\int}\\
    (\tilde{H}_1)_a{}^i&=&-\Phi(\beta){}_a{}^i\,,\vphantom{\ds\int}\\
    (\tilde{H}_0)_a{}^i&=&C_{b0}\rhd_{\alpha a}{}^b\Phi(B)^{\alpha i}-2\beta^{b}{}_{0j}X_{(ab)}{}^A\Phi(\gamma)_A{}^{ij}+ \Phi(\nabla C)_a{}^i\,,\vphantom{\ds\int}\\
    (\tilde{L}_1)_a{}^{ij}&=&\Phi(\gamma)_A{}^{ij}\,,\vphantom{\ds\int}\\
    (\tilde{L}_0)_a{}^{ij}&=&-\Phi(\nabla D)_A{}^{ij}\,,\vphantom{\ds\int}\\
    (\tilde{M}_1)_\alpha{}^i&=&-\Phi(B){}_{\alpha}{}^i\,,\vphantom{\ds\int}\\
    (\tilde{M}_0)_\alpha{}^i&=&\Phi(\cF)_{\alpha}{}^i\,,\vphantom{\ds\int}\\
    (\tilde{N}_1)_a&=&-\Phi(C)_a\,,\vphantom{\ds\int}\\
    (\tilde{N}_0)_a&=&\beta_{b0i}\rhd_{\alpha a}{}^b\Phi(B)^{\alpha i}+ \Phi(\cG)_a\,,\vphantom{\ds\int}\\
    \end{array}
\end{equation}
and $\epsilon_{\mathfrak{g}}{}^{\alpha}$, $\epsilon_{\mathfrak{h}}{}^a{}_i$,  $\epsilon_{\mathfrak{l}}{}^A{}_{ij}$, $\epsilon_{\mathfrak{m}}{}^{\alpha}{}_i$, and $\epsilon_{\mathfrak{n}}{}^a$ are the independent parameters of the gauge transformations. 

The obtained gauge generator (\ref{eq:generator3BF}) is then employed to calculate the form variations of variables and their corresponding canonical momenta, denoted as $A(t,\vec{x})$, using the following equation,
\begin{equation}
    \delta_0 A(t,\vec{x}) =\{ A(t,\vec{x}), G\}\,.
\end{equation}
The form variations of all fields and canonical momenta are given in Appendix \ref{AppE}, equations (\ref{eq:formvar}), while the algebra of the generators (\ref{eq:gtilda}) is obtained in the Appendix \ref{AppB}, equations (\ref{eq:PBgen1})-(\ref{eq:PBgen4}). However, one must bear in mind that the gauge generator (\ref{eq:generator3BF}) is the generator of the symmetry transformations on a slice of spacetime, i.e., on a hypersurface $\Sigma_3$. 
Having in hand all these results, specifically the form variations of all variables and their canonical momenta (\ref{eq:formvar}), we can determine the full gauge symmetry of the theory, which will be done in the next section. 
%%%%%%%%%%%%%%%%%%%%%%%%%%%%%%%%%%%%%%%%%%%%%%%%%%%%%%%%%%%%%%%%%%%%%%%%%%%%%%%%%%%%%%%%%%%%%%%%%%%%%%%%%%%%%%%%%%%%%%%%%%%%%%%%%%%%%%%%%%%%%%%%%%%%%%%%%%%%%%%%%%%%%%%%%%%%%%%%%%%%%%%%%%%%%%%%
\section{Symmetries of the $3BF$ action}\label{secIV}

In order to systematically describe all gauge transformations of the $3BF$ action, we will discuss in turn each set of gauge parameters $\epsilon_{\mathfrak{g}}{}^{\alpha}$, $\epsilon_{\mathfrak{h}}{}^a{}_i$,  $\epsilon_{\mathfrak{l}}{}^A{}_{ij}$, $\epsilon_{\mathfrak{m}}{}^{\alpha}{}_i$, and $\epsilon_{\mathfrak{n}}{}^a$,  appearing in (\ref{eq:generator3BF}). The subsection \ref{secIVa} deals with the gauge group $G$, and the already familiar $G$-gauge transformations, which are already familiar from the ordinary $BF$ theory. In subsection \ref{secIVb} we discuss the gauge group $\tilde{H}_L$ which consists of the already familiar $H$-gauge and $L$-gauge transformations, familiar from the previous literature \cite{Wang2014}, while the subsection \ref{secIVc} examines the $M$-gauge and $N$-gauge transformations which are also present in the theory. Finally, the results of the subsections \ref{secIVa}, \ref{secIVb}, and \ref{secIVc} will be summarized in the subsection \ref{secIVd}, where we will present the complete structure of the gauge symmetry group.

\subsection{Gauge group $G$}\label{secIVa}

First, consider the infinitesimal transformation with the parameter $\epsilon{}_{\mathfrak{g}}{}^\alpha$, given by the form variations
\begin{equation}
\begin{array}{lcrclcr}
    \vphantom{\int\ds}\delta_0 \alpha^\alpha{}_{\mu}&=& \vphantom{\int\ds}\ds-\partial_\mu \epsilon_{\mathfrak{g}}{}^\alpha -f_{\beta\gamma}{}^\alpha \alpha^\beta{}_\mu \epsilon_{\mathfrak{g}}{}{}^\gamma\,,&\quad \quad& \delta_0 B^\alpha{}_{\mu\nu}&=& \vphantom{\int\ds}\ds f_{\beta\gamma}{}^\alpha \epsilon_{\mathfrak{g}}{}{}^\beta B^\gamma{}_{\mu\nu}\,,  \vphantom{\int\ds}\\ \vphantom{\int\ds}
    \delta_0 \beta^a{}_{\mu\nu}&=&\ds\rhd_{\alpha b}{}^a \epsilon_{\mathfrak{g}}{}^\alpha \beta^b{}_{\mu\nu}\,,&\quad \quad& \ds \delta_0 C^a{}_{\mu}&=&\ds \vphantom{\int\ds}\rhd_{\alpha b}{}^a \epsilon_{\mathfrak{g}}{}^\alpha C^b{}_{\mu}\,, \vphantom{\int\ds}\\ \vphantom{\int\ds}
    \delta_0 \gamma^A{}_{\mu\nu\rho}&=&\ds \vphantom{\int\ds}\rhd_{\alpha B}{}^A \epsilon_{\mathfrak{g}}{}^\alpha \gamma^B{}_{\mu\nu\rho}\,,&\quad \quad&\ds \delta_0 D^A&=&\ds \vphantom{\int\ds}\rhd_{\alpha B}{}^A \epsilon_{\mathfrak{g}}{}^\alpha D^B\,, \vphantom{\int\ds}
\end{array}
\end{equation}
which is analogous to writing the transformation as:
\begin{equation}
    \begin{array}{lcrclcr}
    \alpha&\to& \alpha'=\alpha-\nabla \epsilon_{\mathfrak{g}}{}\,,& \quad \quad &B&\to& B'=B-[B,\epsilon_{\mathfrak{g}}{}]\,,\vphantom{\int\ds}\\ \beta&\to&\beta'=\beta+\epsilon_{\mathfrak{g}}{}\rhd\beta\,,&\quad \quad &C&\to& C'=C+\epsilon_{\mathfrak{g}}{}\rhd C\,,\vphantom{\int\ds}\\\gamma&\to& \gamma'=\gamma+\epsilon_{\mathfrak{g}}{}\rhd\gamma\,,&\quad \quad &D&\to& D'=D+\epsilon_{\mathfrak{g}}{}\rhd D\,.\vphantom{\int\ds}
    \end{array}
\end{equation}
Based on these infinitesimal transformations, one can extrapolate the finite symmetry transformations, defined in the Theorem \ref{th:3bf2}.
\begin{Theorem}[$G$-gauge transformations]\label{th:3bf2}
In the $3BF$ theory for the $2$-crossed module $(L\stackrel{\delta}{\to}H \stackrel{\partial}{\to}G,\rhd, \{\_\,,\_\}_{\mathrm{pf}})$, the following transformation is a gauge symmetry,
\begin{equation}
\begin{array}{lcrclcr}
        \alpha&\to& \alpha'=\mathrm{Ad}_{g} \alpha + g\D g{}^{-1}\,,& \quad \quad &B&\to& B'=gBg^{-1}\,,\vphantom{\int\ds}\\ \beta&\to&\beta'=g\rhd\beta\,,&\quad \quad &C&\to& C'=g\rhd C\,,\vphantom{\int\ds}\\\gamma&\to& \gamma'=g\rhd\gamma\,,&\quad \quad &D&\to& D'=g\rhd D\,,\vphantom{\int\ds}
\end{array}
\end{equation}
where $g=\mathrm{exp} (\epsilon_{\mathfrak{g}}\cdot \hat{G})=\mathrm{exp} (\epsilon_{\mathfrak{g}}{}_\alpha \hat{G}^\alpha{})\in G$, and $\epsilon_{\mathfrak{g}}: \cM_4\to \mathfrak{g}$ is the parameter of the transformation.
\end{Theorem}
\begin{Proof}
Note that if one considers an element of the group, $g\in G$, the transformations of the Theorem \ref{th:3bf2} give rise to the following $3$-curvature transformation
\begin{equation}
\begin{array}{lcrclcrclcr}
    \cF&\to& \cF'=g\cF g^{-1}\,,&\quad \quad &\cG&\to& \cG'=g\rhd \cG\,,&\qquad&\cH&\to&\cH'=g\rhd \cH\,,\vphantom{\int\ds}
\end{array}
\end{equation}
and the invariance of the $3BF$ action under this transformation follows from the $G$-invariance of the symmetric bilinear forms on $\mathfrak{g}$, $\mathfrak{h }$, and $\mathfrak{l}$.
\end{Proof}

Let us consider two subsequent infinitesimal $G$-gauge transformations, determined by the small parameters $\epsilon_\mathfrak{g}{}^\alpha{}_1$ and $\epsilon_\mathfrak{g}{}^\beta{}_2$.
To calculate the commutator between the generators of the $G$-gauge transformations, we will make use of the Baker-Campbell-Hausdorff (BCH) formula in the case when the parameters of the transformations are small
\begin{equation}
    \mathrm{e}{}^{\epsilon_\mathfrak{g}{}^\alpha{}_1 \hat{G}_\alpha}\mathrm{e}{}^{\epsilon_\mathfrak{g}{}^\beta{}_2 \hat{G}_\beta}=\mathrm{e}^{{\epsilon_\mathfrak{g}{}^\alpha{}_1 \hat{G}_\alpha}+{\epsilon_\mathfrak{g}{}^\beta{}_2 \hat{G}_\beta}+\frac{1}{2}\epsilon_\mathfrak{g}{}^\alpha{}_1\,\epsilon_\mathfrak{g}{}^\beta{}_2\,[ \hat{G}_\alpha,{ \hat{G}_\beta}]+O(\epsilon_\mathfrak{g}{}^3)}\,,
\end{equation}
from which it follows:
\begin{equation}\label{eq:BCHsc}
    \mathrm{e}{}^{\epsilon_\mathfrak{g}{}^\alpha{}_1 \hat{G}_\alpha}\mathrm{e}{}^{\epsilon_\mathfrak{g}{}^\beta{}_2 \hat{G}_\beta}-\mathrm{e}{}^{\epsilon_\mathfrak{g}{}^\beta{}_2 \hat{G}_\beta}\mathrm{e}{}^{\epsilon_\mathfrak{g}{}^\alpha{}_1 \hat{G}_\alpha}=\epsilon_\mathfrak{g}{}^\alpha{}_1\,\epsilon_\mathfrak{g}{}^\beta{}_2\,[ \hat{G}_\alpha,{ \hat{G}_\beta}]+O(\epsilon_\mathfrak{g}{}^3)\,.
\end{equation}
Using the equation (\ref{eq:BCHsc}), we obtain that the generators of the $G$-gauge transformations defined in the Theorem \ref{th:3bf2} satisfy the following commutation relations:
\begin{equation}
    [\hat{G}_\alpha{}, \hat{G}_\beta]=f_{\alpha\beta}{}^\gamma \hat{G}_\gamma\,,
\end{equation}
where $f_{\alpha\beta}{}^\gamma$ are the structure constants of the algebra $\mathfrak{g}$. By noting that there exists an isomorphism between generators $\hat{G}_\alpha\cong \tau_\alpha$, one establishes that the group of the $G$-gauge transformations from the Theorem \ref{th:3bf2} is the same as the group $G$ of the $2$-crossed module $(L\stackrel{\delta}{\to}H \stackrel{\partial}{\to}G,\rhd, \{\_\,,\_\}{}_{\mathrm{pf}})$. This is an important result, which will not be true for the remaining symmetry transformations, as we shall see below.

\subsection{The gauge group $\tilde{H}_L$}\label{secIVb}
Let us now consider the form variations of the variables corresponding to the parameter $\epsilon{}_{\mathfrak{h}}{}^a{}_i$. For example, one can see from the equations (\ref{eq:formvar}) that the form-variation of the variables $\alpha^\alpha{}_{0}$ and $\alpha^\alpha{}_{i}$ are:
\begin{equation}
    \delta_0 \alpha^\alpha{}_o = 0\,,\quad \quad \delta_0\alpha^\alpha{}_i=-\partial_a{}^\alpha\epsilon_{\mathfrak{h}}{}^a{}_i\,.
\end{equation}
Taking into account that the action of the generator (\ref{eq:generator3BF}) gives the symmetry transformations on one hypersurface $\Sigma_3$ with the time component of the parameter equal to zero, $\epsilon{}_{\mathfrak{h}}{}^a{}_0=0$, one can extrapolate that for parameter of the spacetime gauge transformations $\epsilon{}_{\mathfrak{h}}{}^a{}_\mu$, the form-variation of the variable $\alpha^\alpha{}_\mu$ is given as:
\begin{equation}
    \delta_0\alpha^\alpha{}_\mu=-\partial_a{}^\alpha\epsilon_{\mathfrak{h}}{}^a{}_\mu\,,
\end{equation}
and similarly for the rest of the variables. Thus, the infinitesimal symmetry transformations in the whole spacetime corresponding to the parameter $\epsilon{}_{\mathfrak{h}}{}^a{}_\mu$ are given by the form variations:
\begin{equation}
    \begin{array}{lcrclcr}
         \vphantom{\int\ds}\ds\delta_0\alpha^\alpha{}_\mu&=&-\partial_a{}^\alpha\epsilon_{\mathfrak{h}}{}^a{}_\mu\,,&\quad \quad&\ds \delta_0B^\alpha{}_{\mu\nu}&=&\ds2C_{a[\mu|}\epsilon_{\mathfrak{h}}{}^b{}_{|\nu]} \rhd_{\beta  b}{}^a g^{\alpha\beta}\,,\vphantom{\int\ds}\\ \vphantom{\int\ds}\ds \delta_0\beta^a{}_{\mu\nu}&=&\ds-2\nabla_{[\mu|} \epsilon_{\mathfrak{h}}{}^a{}_{|\nu]}\,,&\quad\quad&\ds \delta_0C^a{}_\mu&=&\ds2 D_A X_{(ab)}{}^A\epsilon_{\mathfrak{h}}{}^b{}_\mu\,,\vphantom{\int\ds}\\ \vphantom{\int\ds}\ds \delta_0\gamma^A{}_{\mu\nu\rho}&=&\vphantom{\int\ds}\ds3!\beta^a{}_{[\mu\nu}\epsilon_{\mathfrak{h}}{}^b{}_{\rho]} X_{(ab)}{}^A\,,&\quad \quad&\ds \delta_0D&=&0\,.\vphantom{\int\ds}
    \end{array}
\end{equation}
For these infinitesimal transformations one obtains the finite symmetry transformations given in Theorem \ref{th:3BFeta}.

\begin{Theorem}[$H$-gauge transformations]\label{th:3BFeta}
In the $3BF$ theory for the $2$-crossed module $(L\stackrel{\delta}{\to}H \stackrel{\partial}{\to}G,\rhd, \{\_\,,\_\}{}_{\mathrm{pf}})$, the following transformation is a symmetry:
\begin{equation}
\begin{array}{lcrclcr}
    \alpha &\to& \alpha' = \alpha-\partial\epsilon_{\mathfrak{h}}{}\,,&\quad \quad&  B &\to& B'= B - C'\wedge^{\cT}\epsilon_{\mathfrak{h}}{}-\epsilon_{\mathfrak{h}}{}\wedge^{\cal D}\epsilon_{\mathfrak{h}}{}\wedge^{\cal D}D\,,\vphantom{\int\ds}\\\beta&\to&\beta'=\beta-\nabla'\epsilon_{\mathfrak{h}}{}-\epsilon_{\mathfrak{h}}{}\wedge\epsilon_{\mathfrak{h}}{}\,,& \quad \quad &C &\to& C'= C-D\wedge^{{\cal X}_1}\epsilon_{\mathfrak{h}}{}-D\wedge^{{\cal X}_2}\epsilon_{\mathfrak{h}}{}\,,\vphantom{\int\ds} \\\gamma&\to&\gamma'=\gamma+\{\beta',\epsilon_{\mathfrak{h}}{}\}{}_{\mathrm{pf}}+\{\epsilon_{\mathfrak{h}}{},\beta\}{}_{\mathrm{pf}}\,,
    &\quad \quad \quad &D &\to& D'= D\,,\vphantom{\int\ds}
\end{array}
\end{equation} 
where $\epsilon_{\mathfrak{h}}{} \in \cA^1(\cM_4,\mathfrak{h})$ is an arbitrary $\mathfrak{h}$-valued $1$-form, and $\nabla'$ denotes the covariant derivative with respect to the connection $\alpha'$. The maps $\cT$, $\cD$, $\mathcal{X}_1$, and $\mathcal{X}_2$ are defined in Appendix \ref{AppD}.
\end{Theorem}
\begin{Proof}
Note that the $3$-curvature transforms as \begin{equation}\label{hcurv}
    \begin{array}{lcr}
    \cF&\to& \cF'=\cF \,,\\ \cG&\to& \cG'=\cG-\cF\wedge^\rhd \epsilon_{\mathfrak{h}}{}\,,\\ \cH&\to&\cH'=\cH+\{\cG',\epsilon_{\mathfrak{h}}{}\}{}_{\mathrm{pf}}-\{\epsilon_{\mathfrak{h}}{}\,,\cG\}{}_{\mathrm{pf}}\,.
\end{array}
\end{equation}
Taking into account the transformations of the $3$-curvature (\ref{hcurv}) and the transformations of the Lagrange multipliers, the action $S_{3BF}$  transforms as:
\begin{equation}
\begin{aligned}
S'{}_{3BF}=&S_{3BF}+\int_{\cM_4}\Big(-\killing{C'\wedge^{\cT}\epsilon_{\mathfrak{h}}{}}{\cF}_{\mathfrak{g}}-\killing{\epsilon_{\mathfrak{h}}{}\wedge^{\cal D}\epsilon_{\mathfrak{h}}{}\wedge^{\cal D}D}{\cF}_{\mathfrak{g}}-\killing{C'}{\cF\wedge^\rhd \epsilon_{\mathfrak{h}}{}}_{\mathfrak{h}}-\killing{D\wedge^{{\cal X}_1}\epsilon_{\mathfrak{h}}{}}{\cG}_{\mathfrak{h}}\\&-\killing{D\wedge^{{\cal X}_2}\epsilon_{\mathfrak{h}}{}}{\cG}_{\mathfrak{h}}+\killing{D}{\{\cG,\epsilon_{\mathfrak{h}}{}\}{}_{\mathrm{pf}}}_{\mathfrak{l}}-\killing{D}{\{\cF\wedge^\rhd \epsilon_{\mathfrak{h}}{},\epsilon_{\mathfrak{h}}{}\}{}_{\mathrm{pf}}}_{\mathfrak{l}}-\killing{D}{\{\epsilon_{\mathfrak{h}}{}\,,\cG\}{}_{\mathrm{pf}}}_{\mathfrak{l}}\Big)\,.
\end{aligned}
\end{equation}
Using the definitions of the maps $\cT$,$\cD$, $\mathcal{X}_1$, and $\mathcal{X}_2$, given in Appendix \ref{AppD}, one sees that the terms in the parentheses cancel, specifically the first term with the third, second with seventh, fourth with eighth, and fifth with the sixth term.
\end{Proof}

The $H$-gauge transformations do not form a group. Namely, one can check that the two consecutive $H$-gauge transformations do not give a transformation of the same kind, i.e., the closure axiom of the group is not satisfied. This is analogous to the well-known structure of Lorentz group, where boost transformations are not closed, and thus do not form a group. Indeed, one must consider both rotations and boosts to obtain the set of transformations that forms the Lorentz group. In the case of the $H$-gauge transformations, we will show that together with the $H$-gauge transformations one needs to consider the transformations corresponding to the parameter $\epsilon_{\mathfrak{l}}{}^A{}_{ij}$. From the equations (\ref{eq:formvar}) one reads the form-variations on a space hypersurface $\Sigma_3$ corresponding to this parameter. Similarly as it is done in the case of the $H$-gauge trasformations, one extrapolates that the form-variations for all the variables corresponding to the parameter $\epsilon_{\mathfrak{l}}{}^A{}_{\mu\nu}$ are given as:
\begin{equation}
\begin{array}{lcrclcr}
   \vphantom{\int\ds}\ds\delta_0 \alpha^\alpha{}_{\mu}&=&0\,,&\quad \quad & \ds\delta_0 B^\alpha{}_{\mu\nu}&=&\ds -D_A\rhd_{\beta B}{}^A\epsilon_{\mathfrak{l}}{}^B{}_{\mu\nu}g^{\alpha\beta}\vphantom{\int\ds}\,, \\\vphantom{\int\ds}\ds\delta_0 \beta^a{}_{\mu\nu}&=&\ds\delta_A{}^a \epsilon_{\mathfrak{l}}{}^A{}_{\mu\nu}\,,&\quad \quad &\ds \delta_0 C^a{}_{\mu}&=&\ds0\,,\vphantom{\int\ds}\\\ds
    \delta_0 \gamma^A{}_{\mu\nu\rho}&=&\vphantom{\int\ds}\ds\nabla_\mu \epsilon_{\mathfrak{l}}{}^A{}_{\nu\rho}-\nabla_\nu \epsilon_{\mathfrak{l}}{}^A{}_{\mu\rho}+\nabla_\rho \epsilon_{\mathfrak{l}}{}^A{}_{\mu\nu}\,,&\quad \quad & \ds\delta_0 D^A&=&\ds0\,.\vphantom{\int\ds}
\end{array}
\end{equation}
These infinitesimal transformations correspond to the finite symmetry transformations defined in Theorem \ref{th:3BFteta}.
\begin{Theorem}[$L$-gauge transformations]\label{th:3BFteta}
In the $3BF$ theory for the $2$-crossed module $(L\stackrel{\delta}{\to}H \stackrel{\partial}{\to}G,\rhd, \{\_\,,\_\}{}_{\mathrm{pf}})$, the following transformation is a symmetry
\begin{equation}\label{l-gauge}
\begin{array}{lcrclcr}
      \alpha &\to& \alpha' = \alpha\,,&\quad \quad& B &\to& B' =B+D\wedge^\cS \epsilon_{\mathfrak{l}}{}\,,\\
     \beta&\to&\beta'=\beta+\delta\epsilon_{\mathfrak{l}}{}\,, &\quad \quad& C&\to& C'=C\,,\\\gamma&\to&\gamma'=\gamma+\nabla\epsilon_{\mathfrak{l}}{}\,,&\quad \quad &D&\to& D'=D\,,
\end{array}
\end{equation} 
where $\epsilon_{\mathfrak{l}}{}\in\cA^2(\cM_4,\mathfrak{l})$ is an arbitrary $\mathfrak{l}$-valued $2$-form, and the map $\cS$ is defined in Appendix \ref{AppD}.
\end{Theorem}
\begin{Proof}
Note that the $3$-curvature transforms as \begin{equation}\label{lcurv}
    \begin{array}{lcr}
    \cF&\to& \cF'=\cF \,,\\ \cG&\to& \cG'=\cG\,,\\ \cH&\to&\cH'=\cH+\cF\wedge^\rhd \epsilon_{\mathfrak{l}}{}\,.
\end{array}
\end{equation}
Taking into account the transformations (\ref{lcurv}) and the transformations of the Lagrange multipliers, the action transforms as:
\begin{equation}
S'{}_{3BF}=S_{3BF}+\int_{\cM_4}\Big(\killing{D\wedge^\cS \epsilon_{\mathfrak{l}}{}}{\cF}_{\mathfrak{g}}+\killing{D}{\cF\wedge^\rhd \epsilon_{\mathfrak{l}}{}}_{\mathfrak{l}}\Big)\,.
\end{equation}
According to the definition of the map $\cS$, the terms in the parentheses cancel.
\end{Proof}

Let us denote the generators of the $H$-gauge transformations given by the Theorem \ref{th:3BFeta} and the $L$-gauge transformations given by the Theorem \ref{th:3BFteta} as $\hat{H}_a{}^\mu$ and $\hat{L}_A{}^{\mu\nu}$, respectively. As we have commented above, one can now check that the transformations defined in the Theorem $\ref{th:3BFeta}$, i.e., the $H$-gauge transformations, do not form a group.
If one performs two consecutive $H$-gauge transformations, defined with parameters $\epsilon_\mathfrak{h}{}_1$ and $\epsilon_\mathfrak{h}{}_2$, one obtains
\begin{equation}
     \mathrm{e}{}^{\epsilon_\mathfrak{h}{}_1\cdot \hat{H}}\mathrm{e}{}^{\epsilon_\mathfrak{h}{}_2\cdot \hat{H}}-\mathrm{e}{}^{\epsilon_\mathfrak{h}{}_2\cdot \hat{H}}\mathrm{e}{}^{\epsilon_\mathfrak{h}{}_1\cdot \hat{H}}=2\,(\{{\epsilon_\mathfrak{h}{}_1}\wedge{\epsilon_\mathfrak{h}{}_2}\}{}_{\mathrm{pf}}-\{{\epsilon_\mathfrak{h}{}_2}\wedge{\epsilon_\mathfrak{h}{}_1}{}\}{}_{\mathrm{pf}})\cdot \hat{L} \,,
\end{equation}
where $\epsilon_\mathfrak{h}\cdot \hat{H}=\epsilon_\mathfrak{h}{}^a{}_{\mu}\hat{H}_a{}^\mu$ and $\epsilon_\mathfrak{l}\cdot \hat{L}=\frac{1}{2}\epsilon_\mathfrak{l}{}^A{}_{\mu\nu}\hat{L}_A{}^{\mu\nu}$. Using the equation analogous to BCH formula (\ref{eq:BCHsc}), one obtains that the commutator of the generators of two $H$-gauge transformations is the generator of an $L$-gauge transformation (see Appendix \ref{AppF} for the details of the calculation):
\begin{equation}\label{eq:HHcom}
    [\hat{H}_a{}^\mu, \hat{H}_b{}^\nu]=2X_{(ab)}{}^A \hat{L}_A{}^{\mu\nu}\,.
\end{equation}

Next, note that the transformations defined in Theorem $\ref{th:3BFteta}$ are the linear transformations, and the two subsequent $L$-gauge transformations give one $L$-gauge transformation with the parameter $\epsilon_\mathfrak{l}{}_{1}+\epsilon_\mathfrak{l}{}_{2}$. Formally, one can write the previous statement as
\begin{equation}
    \mathrm{e}{}^{\epsilon_\mathfrak{l}{}_{1}\cdot \hat{L}}\mathrm{e}{}^{\epsilon_\mathfrak{l}{}_{2}\cdot \hat{L}}=\mathrm{e}{}^{(\epsilon_\mathfrak{l}{}_{1}+\epsilon_\mathfrak{l}{}_{2})\cdot \hat{L}}\,,
\end{equation}
which leads to the conclusion that the generators of the $L$-gauge transformations are mutually commuting:
\begin{equation}\label{eq:comLL}
     [ \hat{L}_A{}^{\mu\nu},\hat{L}_B{}^{\rho\sigma}] = 0\,.
\end{equation}
Thus, the $L$-gauge transformations form an Abelian group, which will be denoted as $\tilde{L}$. According to the index structure of the parameters and generators, we can conclude that the group $\tilde L$ is isomorphic to $\realni^{6r}$, where $r$ is the dimension of the group $L$:
\begin{equation}
\tilde L \cong \realni^{6r}\,.
\end{equation}
Our analogy with the case of the Lorentz group can once again prove useful, since the closure of the $L$-gauge transformations resembles the fact that the composition of two rotations is a rotation. The Abelian group $\tilde{L}$ should not be confused with the non-Abelian group $L$ of the $2$-crossed module $(L\stackrel{\delta}{\to}H \stackrel{\partial}{\to}G,\rhd, \{\_\,,\_\}{}_{\mathrm{pf}})$.

Let us now examine the relationship between $H$-gauge transformations and $L$-gauge transformations. The following result,
\begin{equation}
    \begin{aligned}
    \mathrm{e}{}^{\epsilon_\mathfrak{h}\cdot \hat{H}}\mathrm{e}{}^{\epsilon_\mathfrak{l}\cdot \hat{L}}&=&\mathrm{e}{}^{\epsilon_\mathfrak{l}\cdot \hat{L}}\mathrm{e}{}^{\epsilon_\mathfrak{h}\cdot \hat{H}}\,,
    \end{aligned}
\end{equation}
leads to the conclusion that the commutator of generators of the $H$-gauge transformations and generators of the $L$-gauge transformations vanishes:
\begin{equation}\label{eq:comLH}
    [\hat{H}_a{}^\mu, \hat{L}_A{}^{\nu\rho}]=0\,.
\end{equation}
From the closure of the algebra (\ref{eq:HHcom}), (\ref{eq:comLL}), and (\ref{eq:comLH}), one can conclude that the $H$-gauge transformations together with the $L$-gauge transformations form a group, which will be denoted as $\Tilde{H}_L$. 
Lastly, the action of the group $G$ on the $H$-gauge and $L$-gauge transformations is examined by calculating the expressions:
\begin{equation}
\begin{array}{lcrclcr}
    [\epsilon_\mathfrak{g}\cdot \hat{G},\epsilon_\mathfrak{h}\cdot \hat{H}]&=&(\epsilon_\mathfrak{g}\rhd \epsilon_{\mathfrak{h}})\cdot \hat{H}\,,&\qquad&
    [\epsilon_\mathfrak{g}\cdot \hat{G},\epsilon_\mathfrak{l}\cdot \hat{L}]&=&(\epsilon_\mathfrak{g}\rhd \epsilon_{\mathfrak{l}})\cdot \hat{L}\,,\vphantom{\int\ds}
\end{array}
\end{equation}
which lead to the following commutators:
\begin{equation}
    \begin{array}{lcr}
    [{\hat{G}{}_\alpha},{\hat{H}{_a{}^\mu}}] & = & \rhd{{}_{\alpha a}}{}^b\,\hat{H}{_b{}^\mu}\,,\vphantom{\ds\int}\\
    
    [{\hat{G}_{\alpha}},{\hat{L}_A{}^{\mu\nu}}]&=&\rhd_{\alpha A}{}^B\,\hat{L}{}_B{}^{\mu\nu}\,.\vphantom{\ds\int}
    \end{array}
\end{equation}

Theorems \ref{th:3bf2}, \ref{th:3BFeta}, and \ref{th:3BFteta} represent the $G$-, $H$-, and $L$-gauge transformations, which are already familiar from the previous literature (see for example \cite{Wolf2014,Wang2014}).

\subsection{The gauge groups $M$ and $N$}\label{secIVc}

Next, consider the infinitesimal transformation with the parameter $\epsilon_{\mathfrak{m}}{}{}^\alpha{}_i$, given by the form variations in Appendix \ref{AppE}.
In a similar manner as done in the previous subsection, one establishes that the form variations obtained as a result of the Hamiltonian analysis are transformations on one hypersurface $\Sigma_3$, from which one can guess the symmetry in the whole spacetime. Keeping in mind that the variations on the hypersurface have the time component of the parameter set to $\epsilon_{\mathfrak{m}}{}^\alpha{}_0=0$, one extrapolates the form-variations of the whole spacetime for the parameter $\epsilon_{\mathfrak{m}}{}^\alpha{}_\mu$ to be:

\begin{equation}
\begin{array}{lcrclcr}
   \vphantom{\int\ds}\ds\delta_0 \alpha^\alpha{}_{\mu}&=&0\,,&\quad \quad & \ds\delta_0 B^\alpha{}_{\mu\nu}&=&\ds -2\nabla_{[\mu|} \epsilon_{\mathfrak{m}}{}{}^\alpha{}_{|\nu]}\vphantom{\int\ds}\,, \\\vphantom{\int\ds}\ds\delta_0 \beta^a{}_{\mu\nu}&=&\ds0\,,&\quad \quad &\ds \delta_0 C^a{}_{\mu}&=&\ds-\partial{}^a{}_\alpha \epsilon_{\mathfrak{m}}{}^\alpha{}_\mu\,,\vphantom{\int\ds}\\\ds
    \delta_0 \gamma^A{}_{\mu\nu\rho}&=&0\,,&\quad \quad & \ds\delta_0 D^A&=&\ds0\,.\vphantom{\int\ds}
\end{array}
\end{equation}
Based on this result, one obtains the finite symmetry transformations in the whole spacetime, as defined in Theorem \ref{th:3bf1}, which we will refer to as the $M$-gauge transformations.

\begin{Theorem}[$M$-gauge transformations]\label{th:3bf1}
In the $3BF$ theory for the $2$-crossed module $(L\stackrel{\delta}{\to}H \stackrel{\partial}{\to}G,\rhd, \{\_\,,\_\}{}_{\mathrm{pf}})$, the following transformation is a symmetry
\begin{equation}\label{mgauge}
\begin{array}{lcrclcr}
\alpha &\to& \alpha' = \alpha\,,&\quad \quad &B &\to& B' = B - \nabla \epsilon_{\mathfrak{m}}{}\,,\vphantom{\int\ds}\\
\beta &\to& \beta' = \beta\,,& \quad\quad &C{}^a&\to& C'{}^a{}=C^a-\partial{}^a{}_\alpha \epsilon_{\mathfrak{m}}{}^\alpha{}\,,\vphantom{\int\ds}\\
\gamma &\to &\gamma' =\gamma\,,&\quad\quad &D&\to&D'=D\,,\vphantom{\int\ds}
\end{array}
\end{equation} 
where $\epsilon_{\mathfrak{m}}{}\in \cA^1(\cM_4,\mathfrak{g})$ is an arbitrary $\mathfrak{g}$-valued $1$-form.
\end{Theorem}
\begin{Proof}
Consider the transformation of the $3BF$ action under the transformations of the variables defined in the Theorem \ref{th:3bf1}. One obtains:
\begin{equation}
    S_{3BF}'=S_{3BF}+\int_{\cM_4}\D^4 x\,\epsilon{}^{\mu\nu\rho\sigma}\left(-\frac{1}{2}(\nabla_\mu \epsilon_{\mathfrak{m}}{}^\alpha{}_\nu) \cF_\alpha{}_{\rho\sigma}-\frac{1}{3!}\partial^a{}_\alpha \epsilon_{\mathfrak{m}}{}^\alpha{}_\mu\cG_a{}_{\nu\rho\sigma} \right)\,.
\end{equation}
Using the definition of $3$-curvature, given by the expressions (\ref{eq:krivinekoeficijenti3BF}), one obtains:
\begin{equation}
    S_{3BF}'=S_{3BF}+\int_{\cM_4}\D^4 x\,\epsilon^{\mu\nu\rho\sigma}\Big( -\frac{1}{2}(\nabla_\mu \epsilon_{\mathfrak{m}}{}^\alpha{}_\nu)\left( F_\alpha{}_{\rho\sigma}-\partial^a{}_\alpha \beta_a{}_{\rho\sigma}\right) -\frac{1}{3!}\partial^a{}_\alpha\epsilon_{\mathfrak{m}}{}^\alpha{}_\mu\left(3\nabla_\nu \beta_a{}_{\rho\sigma}-\delta^A{}_a\gamma_A{}_{\nu\rho\sigma} \right)\Big)\,.
\end{equation}
Taking into account that the second and the third term cancel, while the last term is zero because of the identity (\ref{eq:deltapartial0}), the expression reduces to:
\begin{equation}
     S'{}_{3BF}=S_{3BF}-\frac{1}{2}\int_{\cM_4}\D^4 x\,\epsilon^{\mu\nu\rho\sigma}\epsilon_{\mathfrak{m}}{}^\alpha{}_\mu\nabla_\nu F_\alpha{}_{\rho\sigma}\,.
\end{equation}
Finally, the term $\epsilon^{\mu\nu\rho\sigma}\nabla_\nu F_\alpha{}_{\rho\sigma}=0$ is the BI (\ref{bia2BF}). One concludes that the action $S_{3BF}$ is invariant under the transformation defined in Theorem \ref{th:3bf1}.
\end{Proof}

Note that the transformations defined in Theorem $\ref{th:3bf1}$ are linear transformations, and the two subsequent $M$-gauge transformations give one $M$-gauge transformation with the parameter $\epsilon_\mathfrak{m}{}_{1}+\epsilon_\mathfrak{m}{}_{2}$. Denoting the generators of the $M$-gauge transformations as $\hat{M}_{\alpha}{}^{\mu}$, one can now write the previous statement formally as:
\begin{equation}
    \mathrm{e}{}^{\epsilon_\mathfrak{m}{}_{1}\cdot \hat{M}}\mathrm{e}{}^{\epsilon_\mathfrak{m}{}_{2}\cdot \hat{M}}=\mathrm{e}{}^{(\epsilon_\mathfrak{m}{}_{1}+\epsilon_\mathfrak{m}{}_{2})\cdot \hat{M}}\,, 
\end{equation}
where $\epsilon_\mathfrak{m}\cdot \hat{M}=\epsilon_\mathfrak{m}{}^\alpha{}_{\mu} \hat{M}_{\alpha}{}^{\mu}$, leading to the conclusion that:
\begin{equation}
     [ \hat{M}_\alpha{}^\mu,\hat{M}_\beta{}^\nu ] = 0\,.
\end{equation}
Thus, the $M$-gauge transformations form an Abelian group, which will be denoted as $\Tilde{M}$. According to the index structure of its parameters and generators, we see that this group is isomorphic to $\realni^{4p}$, where $p$ is the dimension of the group $G$:
\begin{equation}
\tilde M \cong \realni^{4p}\,.
\end{equation} 
Next, one can examine the relationship of $M$-gauge transformations with the $G$, $H$, and $L$-gauge transformations defined in the previous subsections. Specifically, considering the $G$-gauge symmetry generators, one finds
\begin{equation}
\begin{array}{lcrclcr}
    [\epsilon_\mathfrak{g}\cdot \hat{G},\epsilon_\mathfrak{m}\cdot \hat{M}]&=&(\epsilon_\mathfrak{g}\rhd \epsilon_{\mathfrak{m}})\cdot \hat{M}\,,
\end{array}
\end{equation}
obtaining the result:
\begin{equation}
    \begin{array}{lcr}
    [{\hat{G}_\alpha{}\,,\hat{M}_\beta{}^\mu}]&=&f_{\alpha\beta}{}^\gamma \hat{M}_\gamma{}^\mu\,.\vphantom{\ds\int}
    \end{array}
\end{equation}
Considering the $H$- and $L$-gauge transformations, one obtains
\begin{equation}
    \begin{aligned}
    \mathrm{e}{}^{\epsilon_\mathfrak{h}\cdot \hat{H}}\mathrm{e}{}^{\epsilon_\mathfrak{m}\cdot \hat{M}}&=&\mathrm{e}{}^{\epsilon_\mathfrak{m}\cdot \hat{M}}\mathrm{e}{}^{\epsilon_\mathfrak{h}\cdot\hat{ H}}\,,\\
    \mathrm{e}{}^{\epsilon_\mathfrak{l}\cdot \hat{L}}\mathrm{e}{}^{\epsilon_\mathfrak{m}\cdot \hat{M}}&=&\mathrm{e}{}^{\epsilon_\mathfrak{m}\cdot \hat{M}}\mathrm{e}{}^{\epsilon_\mathfrak{l}\cdot \hat{L}}\,,
    \end{aligned}
\end{equation}
leading to the conclusion that the generators of the $M$-gauge transformations commute with both the generators of $H$-gauge transformations and the generators of the $L$-gauge transformations:
\begin{equation}
    [\hat{H}_a, \hat{M}_\alpha{}^\mu]=0\,, \quad  [\hat{L}_A{}^{\mu\nu}, \hat{M}_\alpha{}^\rho] = 0\,.
\end{equation}

Finally, examining the infinitesimal transformation corresponding to the parameter $\epsilon_{\mathfrak{n}}{}^a$, given by the form-variations as calculated in (\ref{eq:formvar}),
\begin{equation}
\begin{array}{lcrclcr}
\vphantom{\int\ds}\delta_0 \alpha^\alpha{}_{\mu}&=&0\,,&\quad\quad&\delta_0B^\alpha{}_{\mu\nu} &=& \beta_{b\mu\nu}\rhd_{\alpha' a}{}^b\epsilon_{\mathfrak{n}}{}^a g^{\alpha\alpha'}\,,\\\vphantom{\int\ds}\delta_0\beta^a{}_{\mu\nu}&=&0\,, &\quad \quad&  \delta_0C^a{}_\mu&=&-\nabla_\mu\epsilon_{\mathfrak{n}}{}^a\,,\\\delta_0 \gamma^A{}_{\mu\nu\rho}&=&0\,,&\quad \quad&  \delta_0 D^A&=&\delta{}^A{}_a \epsilon_{\mathfrak{n}}{}^a\,.\vphantom{\int\ds}
\end{array}
\end{equation}
one obtains the Theorem \ref{th:3BFEa}, the symmetry transformations which will be referred to as $N$-gauge transformations. Note that the $N$-gauge transformations are simultaneously the transformations in the whole spacetime, since the parameter does not carry spacetime indices.
\begin{Theorem}[$N$-gauge transformations]
\label{th:3BFEa}
In the $3BF$ theory for the $2$-crossed module $(L\stackrel{\delta}{\to}H \stackrel{\partial}{\to}G,\rhd, \{\_\,,\_\}{}_{\mathrm{pf}})$, the following transformation is a symmetry
\begin{equation}\label{ngauge}
\begin{array}{lcrclcr}
   \vphantom{\int\ds}\alpha &\to& \alpha' = \alpha\,,&\quad \quad& B &\to& B'= B - \beta\wedge^\cT \epsilon_\mathfrak{n}\,,\vphantom{\int\ds}\\\vphantom{\int\ds}
    \beta&\to&\beta'=\beta\,,&\quad\quad& C{} &\to& C'=C - \nabla\epsilon_{\mathfrak{n}}\,,\vphantom{\int\ds}\\\vphantom{\int\ds} \gamma&\to&\gamma'=\gamma\,, &\quad \quad & D^A&\to& D'{}^A=D^A+\delta{}^A{}_a\epsilon_{\mathfrak{n}}{}^a\,,\vphantom{\int\ds}
\end{array}\end{equation} 
where  $\epsilon_\mathfrak{n}: \cM_4\to\mathfrak{h}$ is an arbitrary $\mathfrak{h}$-valued $0$-form.
\end{Theorem}
\begin{Proof}
Under the transformations defined in Theorem \ref{th:3BFEa}, the action is transformed as follows:
\begin{equation}
    S_{3BF}'=S_{3BF}+\int_{\cM_4}\D x^4\epsilon^{\mu\nu\rho\sigma}\left(\frac{1}{4} \beta_{b\mu\nu}\rhd_{\alpha a}{}^b\epsilon_{\mathfrak{n}}{}^a\cF^\alpha{}_{\rho\sigma}- \frac{1}{3!} (\nabla_\mu\epsilon_{\mathfrak{n}}{}^a  )  \cG_a{}_{\nu\rho\sigma}+\frac{1}{4!}\delta{}^A{}_a\epsilon_{\mathfrak{n}}{}^a\cH_{A\,\mu\nu\rho\sigma}\right)\,.
\end{equation}
Using the expressions for the $3$-curvature defined in (\ref{eq:3krivine}), one obtains
\begin{equation}\label{eq:tran}
\begin{aligned}
    S_{3BF}'&=&S_{3BF}+\int_{\cM_4}\D x^4\epsilon^{\mu\nu\rho\sigma}\Big(\frac{1}{4} \beta_{b\mu\nu}\rhd_{\alpha a}{}^b\epsilon_{\mathfrak{n}}{}^a\left(F^\alpha{}_{\rho\sigma}-\partial_c{}^\alpha \beta^c{}_{\rho\sigma}\right)- \frac{1}{3!}(\nabla_\mu \epsilon_{\mathfrak{n}}{}^a)\left(3\nabla_\nu\beta_a{}_{\rho\sigma}-\delta{}^A{}_a\gamma_{A\,\nu\rho\sigma}\right)\\&&
    \vphantom{\int\ds}+\frac{1}{4!}\delta{}^A{}_a\epsilon^a\left(4\nabla_\mu\gamma_{A\,\nu\rho\sigma}+6X_{(bc)}{}_A\beta^b{}_{\mu\nu}\beta^c{}_{\rho\sigma}\right)\Big)\vphantom{\int\ds}\,.
\end{aligned}
\end{equation}
Here, after one partial integration the last term in the first row of the equation (\ref{eq:tran}) cancels with the first term in the second row, while taking into account the identity 
\begin{equation}
    \frac{1}{2}\epsilon^{\mu\nu\rho\sigma}(\nabla_\nu\nabla_\mu\epsilon_{\mathfrak{n}}{}^a)\beta_a{}_{\rho\sigma}=-\frac{1}{4} \epsilon^{\mu\nu\rho\sigma} \beta_{b\rho\sigma}\rhd_{\alpha a}{}^b\epsilon_{\mathfrak{n}}{}^a F^\alpha{}_{\mu\nu}\,,
\end{equation}
the first term and the third term also cancel, leading to the following expression:
\begin{equation}
        \vphantom{\int\ds}S'{}_{3BF}=S_{3BF}+\int_{\cM_4}\D x^4\epsilon^{\mu\nu\rho\sigma}\Big(\frac{1}{4}\epsilon_{\mathfrak{n}}{}_a \rhd_{\alpha (b|}{}^{a}\partial_{|c)}{}^\alpha\beta^b{}_{\mu\nu}\beta^c{}_{\rho\sigma}+\frac{1}{4}\epsilon_{\mathfrak{n}}{}_a\delta{}_A{}^aX_{(bc)}{}^A\beta^b{}_{\mu\nu}\beta^c{}_{\rho\sigma}\Big)
        \,.
\end{equation}
Here, the remaining two terms vanish because of the symmetrized form of the identity (\ref{id:5}):
$$\rhd_{\alpha (b|}{}^a\partial_{|c)}{}^\alpha+\delta{}_A{}^aX_{(bc)}{}^A=f_{(bc)}{}^a=0\,,$$
as a consequence of the antisymmetry of the structure constants. One concludes that the $S_{3BF}$ action is invariant under the transformations defined in Theorem \ref{th:3BFEa}.
\end{Proof}

The $N$-gauge transformations defined in Theorem $\ref{th:3BFEa}$ define the group which will be denoted as $\Tilde{N}$. Note that these transformations are also linear, and the composition of two $N$-gauge transformations gives one $N$-gauge transformation with the parameter $\epsilon_\mathfrak{n}{}_{1}+\epsilon_\mathfrak{n}{}_{2}$. The generators of the group $\Tilde{N}$ will be denoted with $\hat{N}_a$, and one can write these results as:
\begin{equation}
\mathrm{e}{}^{\epsilon_\mathfrak{n}{}_{1}\cdot \hat{N}}\mathrm{e}{}^{\epsilon_\mathfrak{n}{}_{2}\cdot\hat{ N}}=\mathrm{e}{}^{(\epsilon_\mathfrak{n}{}_{1}+\epsilon_\mathfrak{n}{}_{2})\cdot \hat{N}}\,,
\end{equation}
where $\epsilon_\mathfrak{n}\cdot \hat{N}=\epsilon_\mathfrak{n}{}^a{} \hat{N}_{a}$, leading to the conclusion that:
\begin{equation}
  [ \hat{N}_a{},\hat{N}_b ] = 0\,.
\end{equation}
It follows that the group $\Tilde{N}$ is Abelian, and the index structure of parameters and generators indicates that it is isomorphic to $\realni^q$, where $q$ is the dimension of the group $H$. Therefore,
\begin{equation}
\tilde N \cong \realni^q\,.
\end{equation}

Next, one can examine the relationship of the $N$-gauge transformations with the $G$, $H$, $L$, and $M$-gauge transformations. First, considering the $G$-gauge transformations one obtains:
\begin{equation}
\begin{array}{lcrclcr}
    [\epsilon_\mathfrak{g}\cdot \hat{G},\epsilon_\mathfrak{n}\cdot \hat{N}]&=&(\epsilon_\mathfrak{g}\rhd \epsilon_{\mathfrak{n}})\cdot \hat{N}\,,\vphantom{\int\ds}
\end{array}
\end{equation}
from which it follows:
\begin{equation}
    \begin{array}{lcr}
    [{\hat{G}_{\alpha}},{\hat{N}_a}] & = & \rhd_{\alpha a}{}^b \, \hat{N}_{b}\,. \vphantom{\ds\int} \\
    \end{array}
\end{equation}

Let us now examine the relationship between $N$-gauge transformations and $H$-gauge transformations, calculating the following expression:
\begin{equation}
    \begin{aligned}
    \mathrm{e}{}^{\epsilon_\mathfrak{h}\cdot \hat{H}}\mathrm{e}{}^{\epsilon_\mathfrak{n}\cdot \hat{N}}-\mathrm{e}{}^{\epsilon_\mathfrak{n}\cdot \hat{N}}\mathrm{e}{}^{\epsilon_\mathfrak{h}\cdot \hat{H}}&=&-(\epsilon_\mathfrak{n}\wedge^\cT \epsilon_\mathfrak{h})\cdot \hat{M} \,,
    \end{aligned}
\end{equation}
where the proof is given in Appendix \ref{AppF}. One obtains that the commutator between the generators of $H$-gauge transformation and $N$-gauge transformation is the generator of $M$-gauge transformation:
\begin{equation}\label{eq:HNcom}
    [ \hat{H}_a{}^\mu,\hat{N}^b]= \rhd_{\alpha a}{}^b \hat{M}^\alpha{}^\mu\,.
\end{equation}
Analogously, one can check that the following is satisfied
\begin{equation}
     \mathrm{e}{}^{\epsilon_\mathfrak{l}\cdot \hat{L}}\mathrm{e}{}^{\epsilon_\mathfrak{n}\cdot \hat{N}}=\mathrm{e}{}^{\epsilon_\mathfrak{n}\cdot \hat{N}}\mathrm{e}{}^{\epsilon_\mathfrak{l}\cdot \hat{L}}\,,\quad \mathrm{e}{}^{\epsilon_\mathfrak{m}\cdot \hat{M}}\mathrm{e}{}^{\epsilon_\mathfrak{n}\cdot \hat{N}}=\mathrm{e}{}^{\epsilon_\mathfrak{n}\cdot \hat{N}}\mathrm{e}{}^{\epsilon_\mathfrak{m}\cdot \hat{M}}\,,
\end{equation}
leading to the conclusion that the generators of $L$-gauge, $M$-gauge, and $N$-gauge transformations mutually commute, i.e.,
\begin{equation}
    [ \hat{M}_\alpha{}^\mu,\hat{N}_a ] = 0\,,\qquad   [\hat{L}_A{}^{\mu\nu},\hat{N}_a] = 0\,.
\end{equation}
This concludes the calculation of the algebra of generators. 

\subsection{Structure of the symmetry group}\label{secIVd}

Summarizing the results of the previous subsections, one can write the algebra of the generators of the full gauge symmetry group as follows.
\begin{itemize}
    \item The algebra $\mathfrak{g}$ of the group $G$ of the $2$-crossed module $ (L\stackrel{\delta}{\to} H \stackrel{\partial}{\to}G\,, \rhd\,, \{\_\,,\_\}{}_{\mathrm{pf}})$: \begin{equation}\label{eq:completegaugesymmetry1}
 \begin{array}{lcr}
 [{\hat{G}_\alpha{} },{\hat{G}_\beta{}} ]&=&f_{\alpha\beta}{}^\gamma {\hat{G}_\gamma{} } \,.\vphantom{\ds\int}\\
 \end{array}
 \end{equation}
 \item The algebra of the group $\Tilde{H}_L$ consisting of the generators of $H$- and $L$-gauge transformations:
 \begin{equation}
 [\hat{H}_a{}^\mu ,\hat{H}_b{}^\nu  ]=2X_{(ab)}{}^A \hat{L}_A{}^{\mu\nu}  \,,\vphantom{\ds\int}\quad\quad
  [\hat{L}_A{}^{\mu\nu} ,\hat{L}_B{}^{\rho\sigma}   ] = 0\,,\vphantom{\ds\int}\quad\quad [\hat{H}_a{}^\mu ,\hat{L}_A{}^{\nu\rho}  ]= 0\,.\vphantom{\ds\int}
 \end{equation}
 \item The algebra of the generators of $M$-gauge transformations:
 \begin{equation}
\begin{array}{lcr}
 [\hat{M}_\alpha{}^\mu ,\hat{M}_\beta{}^\nu  ] &=& 0\,.\vphantom{\ds\int}
 \end{array}
\end{equation}
\item The algebra of the generators of $N$-gauge transformations:
 \begin{equation}
\begin{array}{lcr}
 [\hat{N}_a ,\hat{N}_b  ]&=& 0\,.\vphantom{\ds\int}
 \end{array}
\end{equation}
\item The commutators between the generators of the groups $\tilde{M}$ and $\tilde{N}$:
 \begin{equation}
\begin{array}{lcr}
 [\hat{M}_\alpha{}^\mu ,\hat{N}_a ] &=& 0\,.\vphantom{\ds\int}
 \end{array}
 \end{equation}
 \item The action of the generators of the group $\Tilde{H}_L$ on the generators of $M$- and $N$-gauge transformations:
 \begin{equation}
\begin{array}{lcr}
 
  [\hat{H}_a{}^\mu ,\hat{N}^b  ]&=& \rhd_{\alpha a}{}^b \hat{M}^\alpha{}^\mu  \,,\vphantom{\ds\int}\\
 
 [\hat{H}_a{}^\mu ,\hat{M}_\alpha{}^\nu  ]&=& 0\,,\vphantom{\ds\int}\\
 
  [ \hat{L}_A{}^{\nu\rho}  ,\hat{M}_\alpha{}^\mu ] &=& 0\,,\vphantom{\ds\int}\\
 
 [\hat{L}_A{}^{\mu\nu} ,\hat{N}_a  ] &=& 0\,.\vphantom{\ds\int}\\
 
     \end{array}
\end{equation}
\item The action of the generators of the group $G$ on the generators of $H$-, $L$-, $M$-, and $N$-gauge transformations:

\begin{equation}\label{eq:completegaugesymmetry4}
    \begin{array}{lcr}
     [{\hat{G}{}_\alpha} ,{\hat{H}{_a{}^\mu}} ] & = & \rhd{{}_{\alpha a}}{}^b\,\hat{H}{_b{}^\mu} \vphantom{\ds\int} \,,\\
 
 [{\hat{G}_{\alpha}} ,{\hat{L}_A{}^{\mu\nu}} ]&=&\rhd_{\alpha A}{}^B\hat{L}{}_B{}^{\mu\nu}  \,,\vphantom{\ds\int}\\
 
[{\hat{G}_\alpha{}} \,,\hat{M}_\beta{}^\mu ]&=&f_{\alpha\beta}{}^\gamma \hat{M}_\gamma{}^\mu  \,,\vphantom{\ds\int}\\
 
 [{\hat{G}_{\alpha}} ,{\hat{N}_a} ] & = & \rhd_{\alpha a}{}^b \, \hat{N}_{b}  \,. \vphantom{\ds\int} \\

\end{array}
\end{equation}
\end{itemize}

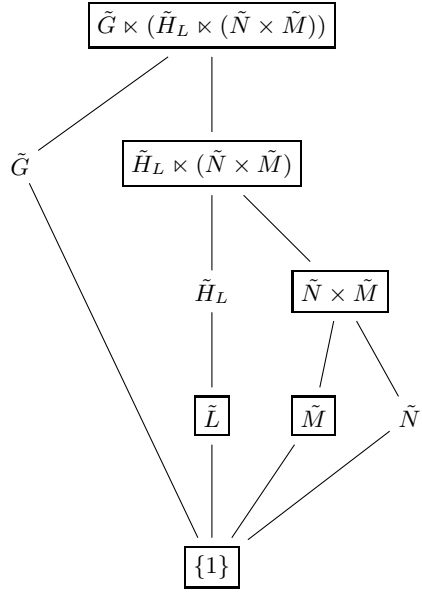
\begin{figure}[h!]
% center everything in the figure
\centering
% horizontal node distance
\newcommand{\mydistance}{.6cm}
\newcommand{\mydistancea}{1cm}
\begin{tikzpicture}[node distance=2cm]
\title{group $\cG$}
\node(tG)                           {$\boxed{\Tilde{G}\ltimes (\Tilde{{H}}_L\ltimes (\Tilde{{N}}\times\Tilde{ M}))}$};
\node(A)             [below=1cm of tG]              {$\boxed{\Tilde{{H}}_L\ltimes (\Tilde{N}\times \Tilde{M})}$};
\node(C)       [below=1cm of A] {${\Tilde{H}}_L$};
\node(B)       [right=\mydistance of C]
{$\boxed{\Tilde{N}\times \Tilde{M}}$};
\node(L)      [below=1cm of C] {$\boxed{\Tilde{L}}$};
\node(M)      [right=\mydistance of L]       {$\boxed{\Tilde{M}}$};
\node(N)      [right=\mydistance of M]       {$\Tilde{N}$};
\node(G)      [left=1cm of A]       {$\Tilde{G}$};
\node(1)       [below=3cm of C] {$\boxed{\{1\}}$};
\draw(1)      -- (G);
\draw(1)      -- (N);
\draw(1)      -- (M);
\draw(1)      -- (L);
\draw(L)      -- (C);
\draw(B)      -- (A);
\draw(tG)      -- (A);
\draw(M)      -- (B);
\draw(N)      -- (B);
\draw(C)      -- (A);
\draw(G) -- (tG);
\end{tikzpicture}
\caption{Relevant subgroups of the symmetry group ${\cG}_{3BF}$. The invariant subgroups are boxed. }\label{fig:01}
\end{figure}

Based on the equations (\ref{eq:completegaugesymmetry1})-(\ref{eq:completegaugesymmetry4}), one can investigate the symmetry group structure. 
On the Hesse-like diagram shown in Figure \ref{fig:01}, we have included only the relevant subgroups of the whole symmetry group ${\cG}_{3BF}$, where the invariant subgroups are boxed. 

Let us remember that the subgroup is an \emph{invariant subgroup}, or equivalently a \emph{normal subgroup}, if it is invariant under conjugation by members of the group of which it is a subgroup. Formally, one says the group $H$ is an invariant subgroup of the group $G$ if $H$ is a subgroup of $G$, i.e., $H \leq G$, and for all $ h\in H $ and $ g\in G$, the conjugation of the element of $H$ with the element of $G$ is an element of $H$, i.e., $\exists h'\in H$ such that $ghg^{-1}=h'$. On the level of algebra, the corresponding object is an \emph{ideal}. Formally written, an algebra $A$ is a subalgebra of an algebra $L$ with respect to the multiplication in $L$, i.e., $[A, A] \subset A$. Then, a subalgebra $A$ of $L$ is an \emph{ideal} in $L$ if its elements, multiplied with any element of the algebra, give again an element of the subalgebra, i.e., $[A, L] \subset A$.

With the above definitions in mind, note first that the groups $\Tilde{L}$, $\Tilde{M}$, and $\Tilde{N}$, are subgroups of the full symmetry group ${\cG}_{3BF}$. The groups $\Tilde{L}$ and $\Tilde{M}$ are invariant subgroups, since the only nontrivial commutators between the generators $\hat{L}_A{}^{\mu\nu}$, and $\hat{M}_\alpha{}^{\mu}$, are with the generators of the group $\tilde{G}$, and are equal to some linear combinations of the generators of $\tilde{L}$, and $\tilde{M}$, respectively. The group $\Tilde{N}$ is not an invariant subgroup, since the commutator between the generators $\hat{N}_a$ and $\hat{H}_a{}^\mu$ are linear combinations of the generators $\hat{M}_\alpha{}^\mu$. However, the generators of the groups $\tilde{N}$ and $\tilde{M}$ are mutually commuting, and the group $\tilde{N}$ is an invariant subgroup of the product of the groups $\tilde{M}$ and $\Tilde{N}$, which makes this product a direct product. The obtained group $\tilde{N}\times \tilde{M}$ is an invariant subgroup of the whole symmetry group.

On the other hand, we saw that the $H$-gauge transformations together with the $L$-gauge transformations form the group $\tilde{H}_L$. This group is not an invariant subgroup of the whole symmetry group ${\cG}_{3BF}$, because of the commutator of the generators $\hat{H}_a{}^\mu$ and $\hat{N}_b$. Similarly as before, one can join these two subgroups, of which one is invariant and one is not, using a semidirect product, to obtain a subgroup $\Tilde{{H}}_L\ltimes (\Tilde{N}\times \Tilde{M})$, that will as a result be an invariant subgroup of the complete symmetry group ${\cG}_{3BF}$. Here, the product is semidirect because the group $\tilde{H}_L$ is not an invariant subgroup of the group $\Tilde{{H}}_L\ltimes (\Tilde{N}\times \Tilde{M})$, due to the commutator between the generators $\hat{H}_a{}^\mu$ and $\hat{N}_b$.

Finally, following the same line of reasoning, one adds the $G$-gauge transformations and obtains the complete gauge symmetry group ${\cG}_{3BF}$ as:
\begin{equation}
    \cG_{3BF}={\Tilde{G}\ltimes (\Tilde{{H}}_L\ltimes (\Tilde{{N}}\times\Tilde{ M}))}\,.
\end{equation}
This concludes the analysis of the group of gauge symmetries for the $3BF$ action.

\section{Conclusions}\label{secV}
\subsection{Summary of the results}\label{secVa}
Let us summarize the results of the paper. In Section \ref{secII}, we have introduced a generalization of the $BF$ theory in the framework of higher category theory, the $3BF$ theory. Section \ref{secIII} contains the Hamiltonian analysis for the $3BF$ theory. In subsection \ref{secIIIa}, the basic canonical structure and the total Hamiltonian are obtained, while in subsection \ref{secIIIb} the complete Hamiltonian analysis of the $3BF$ theory is performed, resulting in the first-class and second-class constraints of the theory, as well as their Poisson brackets. In the subsection \ref{secIIIc} we have discussed the Bianchi identities and also the generalized Bianchi identities, since they enforce restrictions and reduce the number of independent first-class constraints present in the theory, and having those identities in mind, the counting of the dynamical degrees of freedom has been performed. As expected, it was established that the considered $3BF$ action is a topological theory. Finally, this section concludes with the subsection \ref{secIIId} where we have constructed the generator of the gauge symmetries for the topological theory, based on the calculations done in section \ref{secIIIb}, and we have found the form-variations for all the variables and their canonical momenta, listed in the Appendix \ref{AppE}, equations (\ref{eq:formvar}).

In section \ref{secIV}, the main results of our paper are presented. With the material of the subsection \ref{secIIIb} in hand, after obtaining the form variations of all variables and their canonical momenta, we proceeded to find all the gauge symmetries of the theory. The subsection \ref{secIVa} examined the gauge group $G$, and the already familiar $G$-gauge transformations. In subsection \ref{secIVb} we discussed the gauge group $\tilde{H}_L$ which gives the already familiar $H$-gauge and $L$-gauge transformations, while in the subsection \ref{secIVc} we analyzed the $M$-gauge and $N$-gauge transformations which represent a novel result. The results of the subsections \ref{secIVa}, \ref{secIVb}, and \ref{secIVc} are summarized in subsection \ref{secIVd}, where the complete structure of the symmetry group had been presented. The known $G$-, $H$-, and $L$-gauge transformations have been rigorously defined in Theorems \ref{th:3bf2}, \ref{th:3BFeta}, and \ref{th:3BFteta}, while the two novel $M$- and $N$-gauge transformations, have been defined in Theorems \ref{th:3bf1} and \ref{th:3BFEa}. The Lie algebra of the full gauge symmetry group $\cG_{3BF}$ has also been obtained. 

\subsection{Discussion}\label{secVb}
One of the most important consequences of our results is the relationship between a $2$-crossed module and a symmetry group of the corresponding $3BF$ action, which we denoted as a \emph{duality}. In particular, from the Lie algebra of the symmetry group $\cG_{3BF}$ one sees that the structure constants depend on the choices of groups $G$, $H$, and $L$ of the $2$-crossed module, on the action $\rhd$, and on the symmetric part of the Peiffer lifting. However, $\cG_{3BF}$ does not depend on the antisymmetric part of the Peiffer lifting, nor on the homomorphisms $\partial$ and $\delta$. This means that in principle one can have several different $2$-crossed modules dual to the same symmetry group. Therefore, the term "duality" is used in a loose sense, since there is no one-to-one correspondence between a $2$-crossed module and a symmetry group of the corresponding $3BF$ action. In addition, this result allows one to implement a strategy for the construction of a $2$-crossed module, by first specifying the choice of the group $\cG_{3BF}$, and then supplying the additional information about the homomorphisms and the antisymmetric part of the Peiffer lifting, in a way that satisfies all axioms in the definition of a $2$-crossed module.

Another important topic for discussion is the following. From the fact that the $3BF$ action is formulated in a manifestly covariant way, using differential forms, it should be obvious that the diffeomorphisms are a symmetry of the theory. However, by looking at the structure of the gauge group $\cG_{3BF}$, one does not immediately see whether $Dif\!f(\cM_4,\mathbb{R})$ is its subgroup. In fact, this issue is subtle, and it deserves some discussion. 

It is easy to see that every action, which depends on at least two fields $\phi_1(x)$ and $\phi_2(x)$, is invariant under the following transformation, determined by the Henneaux-Teitelboim (HT) parameter $\epsilon^{\footnotesize{\mathrm{HT}}}$ (see \cite{Horowitz} for details and naming),
\begin{equation}\label{eq:htgen}
    \delta_0{}^{\mathrm{HT}}\phi_1=\epsilon^{\mathrm{HT}}(x)\frac{\delta S}{\delta \phi_2}\,, \quad \quad \delta_0{}^{\mathrm{HT}}\phi_2=-\epsilon^{\mathrm{HT}}(x)\frac{\delta S}{\delta \phi_1}\,,
\end{equation}
which can be easily verified by calculating the variation of the action:
\begin{equation}
    \delta{}^{\mathrm{HT}} S [\phi_1,\phi_2]=\frac{\delta S}{\delta \phi_1}\delta_0{}^{\mathrm{HT}} \phi_1+\frac{\delta S}{\delta \phi_2}\delta_0{}^{\mathrm{HT}} \phi_2 =0\,.
\end{equation}
Since this invariance is present even in theories with no gauge symmetry, it
is not associated with constraints, and thus not present in the generator of gauge symmetries (\ref{eq:generator3BF}), see \cite{Horowitz} for details.

Now, let us consider the diffeomorphism transformation
\begin{equation}
    x^\mu\to x'{}^\mu=x^\mu+\xi^\mu(x)\,,
\end{equation}
where the parameter $\xi^\mu(x)$ is an arbitrary function, which we will consider to be infinitesimal. Also, let us denote all parameters of the gauge group collectively as $\epsilon_i(x)$. If diffeomorphisms are a symmetry of the action, then for every field $\phi(x)$ in the theory, and every parameter of the diffeomorphisms $\xi^\mu(x)$, there should exist a choice of the parameters $\epsilon_i(x)$ and  $\epsilon^{\mathrm{HT}}(x)$, such that:
\begin{equation}\label{eq:sumofvar}
    (\delta_0{}^{\mathrm{gauge}}+\delta_0{}^{\mathrm{HT}}+\delta_0{}^{\mathrm{diff}})\,\phi=0\,.
\end{equation}
In other words, if the diffeomorphisms are a symmetry of the theory, their form variations should be expressible as gauge form variations combined with HT form variations:
\begin{equation}\label{eq:diff}
    \delta_0{}^{\mathrm{diff}}\,\phi=-\delta_0{}^{\mathrm{gauge}}\phi-\delta_0{}^{\mathrm{HT}}\phi\,.
\end{equation}
In our case, the $3BF$ action depends on the fields $\alpha^\alpha{}_{\mu}$, $\beta^a{}_{\mu\nu}$, $\gamma^A{}_{\mu\nu\rho}$, $B^\alpha{}_{\mu\nu}$, $C^a{}_{\mu}$, and $D^A$\,. The HT parameters $\epsilon^{\mathrm{HT}}{}^{\alpha\beta}{}_{\mu\nu\rho}\,, \epsilon^{\mathrm{HT}}{}^{ab}{}_{\mu\nu\rho}\,,$ and $\epsilon^{\mathrm{HT}}{}^{AB}{}_{\mu\nu\rho}$ are defined via the following form variations, analogous to (\ref{eq:htgen}):
\begin{equation}
    \begin{array}{lcrclcr}
      \ds \delta_0{}^{\mathrm{HT}} \alpha^\alpha{}_\mu&=&\ds\frac{1}{2}\epsilon^{\mathrm{HT}}{}^{\alpha\beta}{}_{\mu\nu\rho}\frac{\delta S}{\delta B^\beta{}_{\nu\rho}}\,, &\quad\quad&\ds \delta_0{}^{\mathrm{HT}} B^\alpha{}_{\mu\nu}&=&\ds-\epsilon^{\mathrm{HT}}{}^{\alpha\beta}{}_{\rho\mu\nu}\frac{\delta S}{\delta \alpha^\beta{}_{\rho}}\,,\\
      \ds \delta_0{}^{\mathrm{HT}} \beta^a{}_{\mu\nu}&=&\ds\epsilon^{\mathrm{HT}}{}^{ab}{}_{\mu\nu\rho}\frac{\delta S}{\delta C^b{}_{\rho}}\,, &\quad\quad&\ds \delta_0{}^{\mathrm{HT}} C^a{}_{\mu}&=&\ds-\frac{1}{2}\epsilon^{\mathrm{HT}}{}^{ab}{}_{\nu\rho\mu}\frac{\delta S}{\delta \beta^b{}_{\nu\rho}}\,,\\
      \ds \delta_0{}^{\mathrm{HT}} \gamma^A{}_{\mu\nu\rho}&=&\ds\epsilon^{\mathrm{HT}}{}^{AB}{}_{\mu\nu\rho}\frac{\delta S}{\delta D^B}\,, &\quad\quad&\ds \delta_0{}^{\mathrm{HT}} D^A&=&\ds-\frac{1}{3!}\epsilon^{\mathrm{HT}}{}^{AB}{}_{\mu\nu\rho}\frac{\delta S}{\delta \gamma^B{}_{\mu\nu\rho}}\,,\\
           \end{array}
\end{equation}
while the gauge parameters $\epsilon_\mathfrak{g}{}^\alpha$, $ \epsilon_\mathfrak{h}{}^a{}_\mu$, $ \epsilon_\mathfrak{l}{}^A{}_{\mu\nu}$, $ \epsilon_\mathfrak{m}{}^\alpha{}_\mu$, and $\epsilon_\mathfrak{n}{}^a$ are defined in Theorems \ref{th:3bf2}--\ref{th:3BFEa}. Given these, there indeed exists a choice of these parameters, such that (\ref{eq:sumofvar}) is satisfied for all fields. 
Specifically, if one chooses the gauge parameters as
\begin{equation}
    \epsilon_{\mathfrak{g}}{}^\alpha=-\xi^\lambda \alpha^\alpha{}_{\lambda}\,, \quad \epsilon_{\mathfrak{h}}{}^a{}_\mu=\xi^\lambda \beta^a{}_{\mu\lambda}\,,\quad  \epsilon_{\mathfrak{l}}{}^A{}_{\mu\nu}=\xi^\lambda \gamma^A{}_{\mu\nu\lambda}\,,\quad \epsilon_{\mathfrak{m}}{}^\alpha{}_\mu=\xi^\lambda B^\alpha{}_{\mu\lambda}\,,\quad \epsilon_{\mathfrak{n}}{}^a=-\xi^\lambda C^a{}_\lambda\,,
\end{equation}
and the HT parameters as
\begin{equation}
    \epsilon^{\mathrm{HT}}{}^{\alpha\beta}{}_{\mu\nu\rho}=\xi^\lambda g^{\alpha\beta}\epsilon_{ \mu\nu\rho\lambda}\,,\quad \quad \epsilon^{\mathrm{HT}}{}^{ab}{}_{\mu\nu\rho}=\xi^\lambda g^{ab}\epsilon_{\lambda \mu\nu\rho}\,, \quad \quad \epsilon^{\mathrm{HT}}{}^{AB}{}_{\mu\nu\rho}=\xi^\lambda g^{AB}\epsilon_{\mu\nu\rho\lambda }\,,
\end{equation}
one can obtain, using (\ref{eq:diff}), precisely the standard form variations corresponding to diffeomorphisms:
\begin{equation}
    \begin{array}{lcr}
    \delta_0{}^{\mathrm{diff}}\alpha^\alpha{}_\mu&=&-\partial_\mu\xi^\lambda\alpha^\alpha{}_\lambda-\xi^\lambda\partial_\lambda \alpha^\alpha{}_\mu\,,\\
    \delta_0{}^{\mathrm{diff}}\beta^a{}_{\mu\nu}&=&-\partial_\mu\xi^\lambda \beta^a{}_{\lambda\nu}-\partial_\nu\xi^\lambda \beta^a{}_{\mu\lambda}-\xi^\lambda \partial_\lambda\beta^a{}_{\mu\nu}\,,\\
    \delta_0{}^{\mathrm{diff}}\gamma^A{}_{\mu\nu\rho}&=&-\partial_\mu\xi^\lambda \gamma^A{}_{\lambda\nu\rho}-\partial_\nu\xi^\lambda \gamma^A{}_{\mu\lambda\rho}-\partial_\rho\xi^\lambda \gamma^A{}_{\mu\nu\lambda}-\xi^\lambda \partial_\lambda\gamma^A{}_{\mu\nu\rho}\,,\\
    \delta_0{}^{\mathrm{diff}}B^\alpha{}_{\mu\nu}&=&-\partial_\mu\xi^\lambda B^\alpha{}_{\lambda\nu}-\partial_\nu\xi^\lambda B^\alpha{}_{\mu\lambda}-\xi^\lambda \partial_\lambda B^\alpha{}_{\mu\nu}\,,\\
    \delta_0{}^{\mathrm{diff}}C^a{}_\mu&=&-\partial_\mu\xi^\lambda C^a{}_\lambda-\xi^\lambda\partial_\lambda C^a{}_\mu\,,\\
    \delta_0{}^{\mathrm{diff}} D^A&=&-\xi^\lambda\partial_\lambda  D^A\,.
    \end{array}
\end{equation}
This establishes that diffeomorphisms are indeed contained in the full gauge symmetry group $\cG_{3BF}$, up to the HT transformations, which are always a symmetry of the theory.

\subsection{Future lines of investigation}\label{secVc}
Based on the results obtained in this work, one can imagine various additional topics for further research.

First, since we have obtained that the pure $3BF$ theory is a topological theory, it does not describe a realistic physical theory which ought to contain local propagating degrees of freedom. To build a realistic physical theory, one introduces the degrees of freedom by imposing the simplicity constraints on the topological action. In our previous work \cite{Radenkovic2019}, we have formulated the classical actions that manifestly distinguish the topological sector from the simplicity constraints, for all the fields present in the Standard Model coupled to Einstein-Cartan gravity. Specifically, we have defined the constrained $2BF$ actions describing the Yang-Mills field and Einstein-Cartan gravity, and also the constrained $3BF$ actions describing the Klein-Gordon, Dirac, Weyl and Majorana fields coupled to gravity in the standard way. The natural continuation of this line of research would be the Hamiltonian analysis of all such constrained $3BF$ models of gravity coupled to various matter fields, and the study of their canonical quantization.

On the other hand, as an alternative to the canonical quantization, one may choose the spinfoam quantization approach, and define the path integral of the theory as the state sum for the Regge-discretized $3BF$ action. The topological nature of the $3BF$ action, together with the structure of the gauge $3$-group, should ensure that such a sum is a topological invariant, i.e., that it is triangulation independent. This construction was recently carried out in \cite{RadVoj2022}, where the $3BF$ state sum for a general $2$-crossed module and a closed and orientable $4$-dimensional manifold $\cM_4$ is defined. Unfortunately, in order to rigorously define this state sum, one needs the higher category generalizations of the Peter-Weyl and Plancherel theorems, from ordinary groups to the cases of $2$-groups and $3$-groups. These theorems ought to determine the domains of various labels living on simplices of the triangulation, as a consequence of the representation theory of $3$-groups. Until these mathematical results are obtained, one can try to guess the appropriate structure of the irreducible representations of a $3$-group and construct the topological invariant $Z$ for the $3BF$ topological action, in analogy with what was done in the case of $2BF$ theory, see \cite{GirelliPfeifferPopescu2008, MikovicVojinovic2012}. Once the topological state sum is obtained, one can proceed to impose the simplicity constraints, and thus construct the state sum corresponding to the tentative quantum theory of gravity with matter. The classical action for gravity and matter is formulated in \cite{Radenkovic2019} in a way that explicitly distinguishes between the topological sector and the simplicity constraints sector of the action, making the procedure of imposing the constraints straightforward.

Next, it would be useful to investigate in more depth the mathematical structure and properties of the simplicity constraints, in particular their role as the gauge fixing conditions for the symmetry group $\cG_{3BF}$. The simplicity constraints should explicitly break the symmetry group $\cG_{3BF}$ to the subgroup corresponding to the constrained $3BF$ theory, which may then be further spontaneously broken by the Higgs mechanism.

One of the results obtained in this work is a duality between the gauge symmetry group of the $3BF$ action, $\cG_{3BF}$, and the underlining $3$-group, i.e., the $2$-crossed module $(L\stackrel{\delta}{\to} H \stackrel{\partial}{\to}G\,, \rhd\,, \{\_\,,\_\}{}_{\mathrm{pf}})$. This duality should be better understood. On one hand, the group $\cG_{3BF}$ can provide further insight into the construction of the TQFT state sum, i.e., a topological invariant corresponding to the underlining $3$-group structure. On the other hand, this duality is interesting from the perspective of pure mathematics, since it can provide deeper insight in the structure of $3$-groups. In addition, one could expect that the $3BF$ theory would have a $3$-group of higher gauge symmetries, but it is not obvious if the five types of gauge transformations can form a 3-group structure or not. This is an important topic for future research.

Finally, in \cite{MV2020} it was pointed out that it may be useful to make one more step in the categorical generalization, and consider a $4BF$ theory as a description of a quantum gravity model with matter fields. One could then calculate the gauge group of the $4BF$ action, and compare the results with the results obtained for the $3BF$ theory.

The list is not conclusive, and there may be many other interesting topics to study.
\acknowledgments
This research was supported by the Ministry of Education, Science and Technological Development of the Republic of Serbia (MPNTR), and by the Science Fund of the Republic of
Serbia, Program DIASPORA, No. 6427195, SQ2020. The contents of this
publication are the sole responsibility of the authors and can in no way
be taken to reflect the views of the Science Fund of the Republic of
Serbia.
\appendix
\section{$2$-crossed module}\label{AppA}
\begin{theorem-non}[Differential $2$-crossed module] A differential $2$-crossed module is given by an exact sequence of Lie algebras:
\begin{displaymath}
    \mathfrak{l}\stackrel{\delta}{\to} \mathfrak{h} \stackrel{\partial}{\to}\mathfrak{g}\,,
\end{displaymath}
together with left action $\rhd$ of $\mathfrak{g}$ on $\mathfrak{g}$, $\mathfrak{h}$, and $\mathfrak{l}$, by derivations, and on itself via adjoint representation, and a $\mathfrak{g}$-equivariant bilinear map called the {\bf Peiffer lifting}:
$$\{\_\,,\,\_\}{}_{\mathrm{pf}} : \mathfrak{h} \times \mathfrak{h} \to \mathfrak{l}\,.$$
Fixing the basis in the algebras as $T_A \in \mathfrak{l}$, $t_a \in \mathfrak{h}$ and $\tau_\alpha \in \mathfrak{g}$:
$$
    [T_A,T_B]={f_{AB}}^C \, T_C\,, \quad \quad [t_a,t_b]={f_{ab}}^c \, t_c\,, \quad \quad  [\tau_\alpha,\tau_\beta]={f_{\alpha\beta}}^\gamma \, \tau_\gamma\,,
$$
one defines the maps $\partial$ and $\delta$ as:
$$
    \partial (t_a)={\partial_a}^\alpha \, \tau_\alpha\,, \quad \quad \quad \delta (T_A)={\delta_A}^a \, t_a\,,
$$
and the action of $\mathfrak{g}$ on the generators of $\mathfrak{l}$, $\mathfrak{h}$, and $\mathfrak{g}$ is, respectively:
$$
    \tau_\alpha \rhd T_A={\rhd_{\alpha A}}^B \,T_B\,, \quad \quad \tau_\alpha \rhd t_a={\rhd_{\alpha a}}^b\, t_b\,, \quad \quad \tau_\alpha \rhd \tau_\beta = {\rhd_{\alpha\beta}}^\gamma\, \tau_\gamma \,.
$$
The coefficients ${X_{ab}}^A$ are introduced as:
$$
\{t_a,\,t_b\}{}_{\mathrm{pf}}={X_{ab}}^A T_A\,.
$$
The maps $\partial$ and $\delta$ satisfy the following identity:
\begin{equation}\label{eq:deltapartial0}
    {\partial_a}^\alpha \,{\delta_A}^a = 0\,.
\end{equation}
Note that when $\eta$ is a $\mathfrak{g}$-valued differential form and $\omega$ is $\mathfrak{l}$-, $\mathfrak{h}$-, or $\mathfrak{g}$-valued differential form, the previous action is defined as:
$$
\eta \wedge^\rhd \omega = \eta^\alpha \wedge \omega^A \rhd_{\alpha A }{}^B \, T_B\,, \quad \quad \eta \wedge^\rhd \omega = \eta^\alpha \wedge \omega^a \rhd_{\alpha a }{}^b \, t_b\,, \quad \quad \eta \wedge^\rhd \omega = \eta^\alpha \wedge \omega^\beta f_{\alpha \beta}{}^\gamma \, \tau_\gamma\,,
$$
where the forms are multiplied via the wedge product $\wedge$, while the generators of $G$ act on the generators of the three groups via the action $\rhd$.

The following identities are satisfied:
\begin{enumerate}
\item In the differential crossed module $(L \stackrel{\delta}{\to}H \,, \rhd')$ the action $\rhd'$ of $\mathfrak{h}$ on $\mathfrak{l}$ is defined for each $\underline{h} \in \mathfrak{h}$ and $\underline{l} \in \mathfrak{l}$ as:
$$\underline{h} \rhd' \underline{l} =-\{\delta(\underline{l}),\,\underline{h}\}{}_{\mathrm{pf}}\,,$$
or written in the basis where $t_{a} \rhd' T_A = \rhd'{}_{aA}{}^B T_B$ the previous identity becomes:
\begin{equation}
    {{\rhd'}_{a A}}^B=-{\delta_A}^b {X_{ba}}^B\,;
\end{equation}
\item The action of $\mathfrak{g}$ on itself is via adjoint representation:
\begin{equation}
       {\rhd_{\alpha\beta}}^\gamma={f_{\alpha\beta}}^\gamma\,;
\end{equation} 
\item The action of $\mathfrak{g}$ on $\mathfrak{h}$ and $\mathfrak{l}$ is equivariant, i.e., the following identities are satisfied:
\begin{equation}
    \partial_a{}^\beta f_{\alpha \beta}{}^\gamma = \rhd_{\alpha a}{}^b \partial_b{}^\gamma\,, \quad \quad \delta_A{}^a\rhd_{\alpha a}{}^b=\rhd_{\alpha A}{}^B \delta_B{}^b\,;
\end{equation}
\item The Peiffer lifting is $\mathfrak{g}$-equivariant, i.e., for each $\underline{h}_1,\underline{h}_2\in\mathfrak{h}$ and $\underline{g}\in\mathfrak{g}$:
$$\underline{g}\rhd\{\underline{h}_1,\underline{h}_2\}{}_{\mathrm{pf}}=\{\underline{g}\rhd \underline{h}_1,\underline{h}_2\}{}_{\mathrm{pf}}+\{\underline{h}_1,\,\underline{g}\rhd \underline{h}_2\}{}_{\mathrm{pf}}\,,$$
or written in the basis:
\begin{equation}
    {X_{ab}}^B {\rhd_{\alpha B}}^A={\rhd_{\alpha a}}^c {X_{cb}}^A+{\rhd_{\alpha b}}^c {X_{ac}}^A\,;
\end{equation}
\item $\delta(\left\{\underline{h}_1,\,\underline{h}_2  \right \}{}_{\mathrm{pf}})=\left \langle\underline{h}_1,\,\underline{h}_2\right \rangle{}_{\mathrm{p}}\,, \qquad \forall \underline{h}_1,\,\underline{h}_2 \in \mathfrak{h}$.

The map $(\underline{h}{}_1, \, \underline{h}{}_2) \in \mathfrak{h} \times \mathfrak{h} \to \langle\underline{h}{}_1, \, \underline{h}{}_2\rangle{}_{\mathrm{p}}\in \mathfrak{h}$ is bilinear $\mathfrak{g}$-equivariant map called the {\bf Peiffer paring}, i.e., all $\underline{h}{}_1 \,, \underline{h}{}_2 \in \mathfrak{h}$ and $\underline{g} \in \mathfrak{g}$ satisfy the following identity:
$$ \underline{g} \rhd \langle \underline{h}{}_1 \,, \underline{h}{}_2 \rangle{}_{\mathrm{p}} = \langle \underline{g} \rhd \underline{h}{}_1 \,, \underline{h}{}_2\rangle{}_{\mathrm{p}} + \langle \underline{h}{}_1 \,, \underline{g} \rhd\underline{h}{}_2 \rangle{}_{\mathrm{p}}\,. $$
Fixing the basis the identity becomes:
\begin{equation}\label{id:5}
    {X_{ab}}^A {\delta_A}^c=f_{ab}{}^c-{\partial_a}^\alpha{\rhd_{\alpha b}}^c\,;
\end{equation}
\item $[\underline{l}_1,\,\underline{l}_2]=\left\{\delta(\underline{l}_1),\,\delta(\underline{l}_2)\right \}{}_{\mathrm{pf}}\,, \qquad \forall \underline{l}_1,\,\underline{l}_2 \in \mathfrak{l}$, i.e.,
\begin{equation}
    {f_{AB}}^C={\delta_A}^a{\delta_B}^b {X_{ab}}^C\,;
\end{equation}
\item $\left\{[\underline{h}_1,\,\underline{h}_2],\,\underline{h}_3\right \}{}_{\mathrm{pf}}=\partial(\underline{h}_1)\rhd\left\{\underline{h}_2,\,\underline{h}_3\right \}{}_{\mathrm{pf}}+\left\{\underline{h}_1,\,[\underline{h}_2,\,\underline{h}_3]\right \}{}_{\mathrm{pf}}-\partial(\underline{h}_2)\rhd\left\{\underline{h}_1,\,\underline{h}_3\right \}{}_{\mathrm{pf}}-\left\{\underline{h}_2,\,[\underline{h}_1,\,\underline{h}_3]\right \}{}_{\mathrm{pf}}\,, \qquad \forall \underline{h}_1,\,\underline{h}_2,\,\underline{h}_3 \in \mathfrak{h}$, i.e.,
$$\left\{[\underline{h}_1,\,\underline{h}_2],\underline{h}_3\right \}{}_{\mathrm{pf}}=\{ \partial(\underline{h}_1) \rhd \underline{h}_2,\,\underline{h}_3\}{}_{\mathrm{pf}}-\{ \partial(\underline{h}_2) \rhd \underline{h}_1\,,\underline{h}_3\}{}_{\mathrm{pf}}-\{\underline{h}_1,\delta\{\underline{h}_2,\,\underline{h}_3\}{}_{\mathrm{pf}}\}{}_{\mathrm{pf}}+\{\underline{h}_2,\,\delta\{\underline{h}_1,\underline{h}_3\}{}_{\mathrm{pf}}\}{}_{\mathrm{pf}}\,,$$
\begin{equation}
    {f_{ab}}^d {X_{dc}}^B={\partial_a}^\alpha {X_{bc}}^A {\rhd_{\alpha A}}^B+{X_{ad}}^B {f_{bc}}^d-{\partial_b}^\alpha{\rhd_{\alpha A}}^B {X_{ac}}^A-{X_{bd}}^B{f_{ac}}^d\,;
\end{equation}
\item $\left\{\underline{h}_1,\,[\underline{h}_2,\,\underline{h}_3]\right \}{}_{\mathrm{pf}}=\left\{\delta\left\{\underline{h}_1,\,\underline{h}_2\right \}{}_{\mathrm{pf}},\underline{h}_3\right \}{}_{\mathrm{pf}}-\left\{\delta\left\{\underline{h}_1,\,\underline{h}_3\right \}{}_{\mathrm{pf}},\,\underline{h}_2\right \}{}_{\mathrm{pf}}\,, \qquad \forall \underline{h}_1,\,\underline{h}_2,\,\underline{h}_3 \in \mathfrak{h}$, i.e.,
\begin{equation}
    {X_{ad}}^A {f_{bc}}^d={X_{ab}}^B{\delta_B}^d {X_{dc}}^A-{X_{ac}}^B{\delta_B}^d {X_{db}}^A\,;
\end{equation}
\item$\left\{\delta(\underline{l}),\,\underline{h}\right \}{}_{\mathrm{pf}}+\left\{\underline{h},\,\delta(\underline{l})\right \}{}_{\mathrm{pf}}=-\partial(\underline{h}) \rhd \underline{l}\,, \qquad \forall \underline{l} \in \mathfrak{l}\,, \quad \forall \underline{h} \in \mathfrak{h}\,,$ i.e.,
\begin{equation}
    {\delta_A}^a {X_{ab}}^B+{\delta_A}^a {X_{ba}}^B=-{\partial_b}^\alpha {\rhd_{\alpha A}}^B\,.
\end{equation}
\end{enumerate}
\end{theorem-non}

A reader interested in more details about $3$-groups is referred to \cite{Wolf2014,Wang2014}.

The structure constants satisfy the Jacobi identities
\begin{equation}
f_{\alpha\gamma}{}^\delta \,f_{\beta\epsilon}{}^\gamma = 2 f_{\alpha [\beta|}{}^\gamma \,f{}_{\gamma|\epsilon]}{}^\delta\,, \qquad f{}_{ad}{}^{c} \,f{}_{be}{}^{d} = 2f{}_{a [b|}{}^{d} \,f{}_{d|e]}{}^{c}\,, \qquad f{}_{AD}{}^{C} \,f{}_{BE}{}^{D} = 2f{}_{A [B|}{}^{D} \,f{}_{D|E]}{}^{C} \,.\label{jac}
\end{equation}
Also, the following relations are useful: 
\begin{equation}  
f{}_{\beta\gamma}{}^\alpha \rhd_{\alpha b}{}^a =2 \rhd_{[\beta| c}{}^a \rhd{}_{|\gamma] b}{}^c\,, \quad \quad f{}_{\beta\gamma}{}^\alpha \rhd_{\alpha B}{}^A = 2\rhd_{[\beta| C}{}^A \rhd{}_{|\gamma] B}{}^C\,.
\end{equation}
\section{Additional relations of the constraint algebra}\label{AppB}
In this Appendix the useful technical results used in the subsection \ref{secIIIb} are given.
First, since the secondary constraints, given by the equations (\ref{SekundarneVeze3BF}), must be preserved during the evolution of the system, the consistency conditions of secondary constraints must be enforced. However, no tertiary constraints arise from these conditions, since one obtains the following PB:
\begin{equation}\label{eq:secondclassconstwithHamiltonian}
    \begin{array}{lcl}
    \{\cS(\cF)^{\alpha}{}^i\,, \,H_T\}&=&f_{\beta\gamma}{}^\alpha \cS(\cF){}^\beta{}^i \alpha^\gamma{}_0\,,\vphantom{\ds\int}\\
    \{\cS(\nabla B)_\alpha{}\,, \,H_T\}&=&f_{\beta\gamma}{}_\alpha B^\gamma{}_{0k}\cS(\cF)^\beta{}^k+f_{\beta}{}_\alpha{}^\gamma \alpha^\beta{}_0\cS(\nabla B)_\gamma+C_a{}_0\rhd_{\alpha b}{}^a\cS(\cG){}^b{}\vphantom{\ds\int}\\&&\ds-\rhd_{\alpha a}{}^b \beta^a{}_{0k}\cS(\nabla C)_b{}^k\vphantom{\ds\int}+\ds\frac{1}{2}\rhd_{\alpha}{}^B{}_A\gamma^A{}_{0jk}S(\nabla D){}_B{}^{jk}\,,\vphantom{\ds\int}\\
    \{\cS(\cG)^a\,,\,H_T\}&=&\ds\rhd_{\alpha b}{}^a \beta^b{}_{0k} \cS(\cF)^\alpha{}^k-\alpha^\alpha{}_0\rhd_{\alpha b}{}^a \cS(\cG)^b\,,\vphantom{\ds\int}\\
    \{\cS(\nabla C)_a{}^i\,,\,H_T\}&=&\ds C_b{}_0\rhd_{\alpha}{}^b{}_a\cS(\cF)^\alpha{}^i+\rhd_{\alpha}{}_a{}^b \alpha^\alpha{}_0 \cS(\nabla C)_b{}^i+2X_{(ab)}{}^A\beta^b{}_{0j}\cS(\nabla D)_A{}^{ij}\,,\vphantom{\ds\int}\\
    \{\cS(\nabla D){}_A{}^{ij}\,,\,H_T\}&=&\alpha^\alpha{}_0 \rhd_{\alpha A}{}^B \cS(\nabla D){}_B{}^{ij}\,.\vphantom{\ds\int}
    \end{array}
\end{equation}

The PB between the first-class constraints, given by the equations (\ref{eq:VezePrveKlase}), and the second-class constraints, given by the equations (\ref{eq:VezeDrugeKlase}), are given by:
\begin{equation}\label{eq:jednacinaPB1i2}
\begin{array}{lcl}
\pb{\Phi(\cF)^{\alpha i}(\vec{x})}{\chi(\alpha)_{\beta}{}^j(\vec{y})} & = & -f{{}_{\beta\gamma}}{}^\alpha \,\chi(B)^{\gamma}{}{}^{ij}(\vec{x})\, \delta^{(3)}(\vec{x}-\vec{y}) \,, \vphantom{\ds\int} \\
\pb{\Phi(\cG)^a (\vec{x})}{\chi(\alpha)_{\alpha}{}^i(\vec{y})} & = & \rhd{_{\alpha b}}{}^a \,\chi(C)^{b}{}^i (\vec{x})\,\delta^{(3)}(\vec{x}-\vec{y}) \,, \vphantom{\ds\int} \\
\pb{\Phi(\cG)^a(\vec{x})}{\chi(\beta)_b{}^{ij}(\vec{y})} & = & -\rhd{{}_{\alpha b}}{}^a\, \chi(B)^{\alpha}{}{}^{ij}(\vec{x})\, \delta^{(3)} (x-y)\,, \vphantom{\ds\int} \\
\pb{\Phi(\nabla C)^{a i}(\vec{x})}{\chi(\alpha)_{\alpha}{}^j(\vec{y})} & = & -\rhd{{}_{\alpha b}{}^a}\, \chi(\beta)^{b}{}^{ij}(\vec{x})\, \delta^{(3)}(\vec{x}-\vec{y}) \,, \vphantom{\ds\int} \\
\pb{\Phi(\nabla C)^{a i}(\vec{x})}{\chi(\beta)_{b}{}^{jk}(\vec{y})} & = & 2X{^{(ac)}}{}^A g_{bc}\, \chi(\gamma){}_A{}^{ijk}(\vec{x}) \, \delta^{(3)}(\vec{x}-\vec{y}) \,, \vphantom{\ds\int}\\
\pb{\Phi(\nabla C)^{a i}(\vec{x})}{\chi(C)_b{}^j(\vec{y})} & = & \rhd{{}_{\alpha b}{}^{a}}\,\chi(B)^\alpha{}^{ij}(\vec{x}) \, \delta^{(3)}(\vec{x}-\vec{y}) \,, \vphantom{\ds\int} \\
\pb{\Phi(\nabla C)^{a i}(\vec{x})}{\chi(D)_A(\vec{y})} & = & 2X^{(ab)}{}_A \, \chi(C){}_b{}^i(\vec{x}) \, \delta^{(3)}(\vec{x}-\vec{y}) \,, \vphantom{\ds\int}\\
\pb{\Phi(\nabla B)^{\alpha}(\vec{x})}{\chi(\alpha)_{\beta}{}^i(\vec{y})} & = & f{{}_{\beta\gamma}}{}^\alpha\, \chi(\alpha)^{\gamma}{}{}^i(\vec{x})\, \delta^{(3)}(\vec{x}-\vec{y}) \,, \vphantom{\ds\int} \\
\pb{\Phi(\nabla B)^{\alpha}(\vec{x})}{\chi(\beta)_a{}^{ij}(\vec{y})} & = &g^{\alpha\beta} \rhd{_{\beta a }{}^{b}}\, \chi(\beta)_{b}{}^{ij}(\vec{x})\, \delta^{(3)} (\vec{x}-\vec{y})\,, \vphantom{\ds\int} \\
\pb{\Phi(\nabla B)^{\alpha}(\vec{x})}{\chi(\gamma)_A{}^{ijk}(\vec{y})} & = & g^{\alpha\beta}\rhd{_{\beta A}{}^{B}}\, \chi(\gamma)_{B}{}^{ijk}(\vec{x})\, \delta^{(3)}(\vec{x}-\vec{y}) \,, \vphantom{\ds\int} \\
\pb{\Phi(\nabla B)^{\alpha}(\vec{x})}{\chi(B)_{\beta}{}^{ij}(\vec{y})} & = & f{_{\beta \gamma}}{}^\alpha\, \chi(B){}^{\gamma ij}(\vec{x})\, \delta^{(3)}(\vec{x}-\vec{y}) \,. \vphantom{\ds\int} \\
\pb{\Phi(\nabla B)^{\alpha}(\vec{x})}{\chi(C)_{a}{}^{i}(\vec{y})} & = & -\rhd{^{\alpha}{}^b{}_a}\, \chi(C){}_{b}{}^i(\vec{x})\, \delta^{(3)}(\vec{x}-\vec{y}) \,. \vphantom{\ds\int} \\
\pb{\Phi(\nabla B)^{\alpha}(\vec{x})}{\chi(D)_{A}{}(\vec{y})} & = & g^{\alpha\beta}\rhd{_{\beta A}{}^B}\, \chi(D){}_{B}(\vec{x})\, \delta^{(3)}(\vec{x}-\vec{y}) \,, \vphantom{\ds\int} \\
\pb{\Phi(\nabla D){^A{}^{ij}}(\vec{x})}{\chi(\alpha)_\alpha{}^{k}(\vec{y})}&=&\rhd_{\alpha B}{}^A \chi(\gamma)^B{}^{ijk}(\vec{x})\, \delta^{(3)}(\vec{x}-\vec{y})\,,\vphantom{\ds\int}\\
\pb{\Phi(\nabla D){^A{}^{ij}}(\vec{x})}{\chi(D)_B{(\vec{y})}}&=&-\rhd_{\alpha B}{}^A \chi(B)^\alpha{}^{ij}(\vec{x})\, \delta^{(3)}(\vec{x}-\vec{y})\,.\vphantom{\ds\int}\\
\end{array}
\end{equation}
Finally, it is useful to calculate PB between the first-class constraints, given by the equations (\ref{eq:VezePrveKlase}), and the total Hamiltonian, given by the equation (\ref{totalhamiltonian}):
\begin{equation}
   \begin{array}{lcl}\label{eq:komutatorisaH}
    \{\Phi(\cF)^{\alpha}{}^i\,, \,H_T\}&=&f_{\beta\gamma}{}^\alpha \Phi(\cF){}^\beta{}^i \alpha^\gamma{}_0\,,\vphantom{\ds\int}\\
    \{\Phi(\nabla B)_\alpha{}\,, \,H_T\}&=&f_{\beta\gamma}{}_\alpha B^\gamma{}_{0k}\Phi(\cF)^\beta{}^k+f_{\beta}{}_\alpha{}^\gamma \alpha^\beta{}_0\Phi(\nabla B)_\gamma+C_a{}_0\rhd_{\alpha b}{}^a\Phi(\cG){}^b{}\vphantom{\ds\int}\\&&-\rhd_{\alpha a}{}^b \beta^a{}_{0k}\Phi(\nabla C)_b{}^k\vphantom{\ds\int}+\ds\frac{1}{2}\rhd_{\alpha}{}^B{}_A\gamma^A{}_{0jk}\Phi(\nabla D){}_B{}^{jk}\,,\\
    \{\Phi(\cG)^a\,,\,H_T\}&=&\rhd_{\alpha b}{}^a \beta^b{}_{0k} \Phi(\cF)^\alpha{}^k-\alpha^\alpha{}_0\rhd_{\alpha b}{}^a \Phi(\cG)^b\,,\vphantom{\ds\int}\\
    \{\Phi(\nabla C)_a{}^i\,,\,H_T\}&=&C_b{}_0\rhd_{\alpha}{}^b{}_a\Phi(\cF)^\alpha{}^i+\rhd_{\alpha}{}_a{}^b \alpha^\alpha{}_0 \Phi(\nabla C)_b{}^i+2X_{(ab)}{}^A\beta^b{}_{0j}\Phi(\nabla D)_A{}^{ij}\,,\vphantom{\ds\int}\\
    \{\Phi(\nabla D){}_A{}^{ij}\,,\,H_T\}&=&\alpha^\alpha{}_0 \rhd_{\alpha A}{}^B \Phi(\nabla D){}_B{}^{ij}\,.\vphantom{\ds\int}
    \end{array}
\end{equation}
The calculated PB brackets given by the equation (\ref{eq:komutatorisaH}) will be useful for calculation of the generator of gauge symmetries (\ref{eq:generator3BF}).
With these results one can proceed to the construction of the gauge symmetry generator on one hypersurface $\Sigma_3$ given in the equation (\ref{eq:generator3BF}), and ultimately obtain the finite gauge symmetry of the whole spacetime.

The PB algebra of gauge symmetry generators $(\tilde{M}_0)_\alpha{}^i$, $(\tilde{M}_1)_\alpha{}^i$, $(\tilde{G}_0)_\alpha$, $(\tilde{G}_1)_\alpha$,  $(\tilde{H}_0)_a{}^i$, $(\tilde{H}_1)_a{}^i$, $(\tilde{N}_0)_a$, $(\tilde{N}_1)_a$, $(\tilde{L}_0)_A{}^{ij}$, and $(\tilde{L}_1)_A{}^{ij}$, as defined in (\ref{eq:gtilda}), is:
\begin{equation}\label{eq:PBgen1}
\vphantom{\int\ds}
        \{(\tilde{G}_0)_\alpha(\vec{x})\,,(\tilde{G}_0)_\beta(\vec{y})\}=f_{\alpha\beta}{}^\gamma (\tilde{G}_0)_\gamma\, \delta^{(3)}(\vec{x}-\vec{y})\,,
\end{equation}

\begin{equation}\label{eq:PBgen2}
    \begin{aligned}
        \vphantom{\int\ds}
        \{(\tilde{H}_0)_a{}^i(\vec{x})\,,(\tilde{H}_0)_b{}^j(\vec{y})\}&=&2X_{(ab)}{}^A(\tilde{L}_0){}_A{}^{ij}\, \delta^{(3)}(\vec{x}-\vec{y})\,,\vphantom{\int\ds}\\
         \vphantom{\int\ds}\{(\tilde{H}_0)_a{}^i{}(\vec{x})\,,(\tilde{H}_1)_b{}^j(\vec{y})\}&=&2X_{(ab)}{}^A(\tilde{L}_1){}_A{}^{ij}\, \delta^{(3)}(\vec{x}-\vec{y})\,,\\
    \end{aligned}
\end{equation}
\begin{equation}\label{eq:PBgen3}
    \begin{aligned}
        \vphantom{\int\ds}
        \{(\tilde{H}_0)_a{}^i(\vec{x})\,,(\tilde{N}_0)^b(\vec{y})\}&=&\rhd_{\alpha a}{}^b(\tilde{M}_0)^\alpha{}^i\, \delta^{(3)}(\vec{x}-\vec{y})\,,\vphantom{\int\ds}\\
        \vphantom{\int\ds}\{(\tilde{H}_1)_a{}^i(\vec{x})\,,(\tilde{N}_0)^b(\vec{y})\}&=&\rhd_{\alpha a}{}^b(\tilde{M}_1)^\alpha{}^i\, \delta^{(3)}(\vec{x}-\vec{y})\,,\\
        \vphantom{\int\ds}\{(\tilde{H}_0)_a{}(\vec{x})\,,(\tilde{N}_1)^b{}^i(\vec{y})\}&=&\rhd_{\alpha a}{}^b(\tilde{M}_1)^\alpha{}^i\, \delta^{(3)}(\vec{x}-\vec{y})\,,\\
    \end{aligned}
\end{equation}

\begin{equation}\label{eq:PBgen4}
    \begin{aligned}
         \vphantom{\int\ds}\{(\tilde{G}_0)_\alpha{}(\vec{x})\,,(\tilde{M}_0)_\beta{}^i(\vec{y})\}&=&f_{\alpha\beta}{}^\gamma(\tilde{M}_0)_\gamma{}^i\, \delta^{(3)}(\vec{x}-\vec{y})\,,\\
        \vphantom{\int\ds}
        \{(\tilde{G}_0)_\alpha{}(\vec{x})\,,(\tilde{M}_1)_\beta{}^i(\vec{y})\}&=&f_{\alpha\beta}{}^\gamma(\tilde{M}_1)_\gamma{}^i\, \delta^{(3)}(\vec{x}-\vec{y})\,,\\
    \pb{(\tilde{G}_0){}_\alpha(\vec{x})}{(\tilde{H}_1){_a{}^i}(\vec{y})} & = & \rhd{{}_{\alpha a}}{}^b\,(\tilde{H}_1){_b{}^i}(\vec{x})\,\delta^{(3)}(\vec{x}-\vec{y})\,,\vphantom{\ds\int}\\
    \pb{(\tilde{G}_0){}_\alpha(\vec{x})}{(\tilde{H}_0){_a{}^i}(\vec{y})} & = &\rhd{{}_{\alpha a}}{}^b\,(\tilde{H}_0){_b{}^i}(\vec{x})\,\delta^{(3)}(\vec{x}-\vec{y})\,,\vphantom{\ds\int}\\
    \pb{(\tilde{G}_0)_{\alpha} (\vec{x})}{(\tilde{N}_1)_a(\vec{y})} & = & \rhd_{\alpha a}{}^b \, (\tilde{N}_1)_{b}(\vec{x})\, \delta^{(3)}(\vec{x}-\vec{y}) \,, \vphantom{\ds\int} \\
    \pb{(\tilde{G}_0){}_\alpha(\vec{x})}{(\tilde{N}_0){_a{}}(\vec{y})} & = &\rhd{{}_{\alpha a}}{}^b\,(\tilde{N}_0){_b{}}(\vec{x})\,\delta^{(3)}(\vec{x}-\vec{y})\,,\vphantom{\ds\int}\\
    \pb{(\tilde{G}_0)_{\alpha}(\vec{x})}{(\tilde{L}_0)_A{}^{ij}(\vec{y})}&=&\rhd_{\alpha A}{}^B(\tilde{L}_0){}_B{}^{ij}(\vec{x})\,\delta^{(3)}(\vec{x}-\vec{y})\,.\vphantom{\ds\int}
    \end{aligned}
\end{equation}

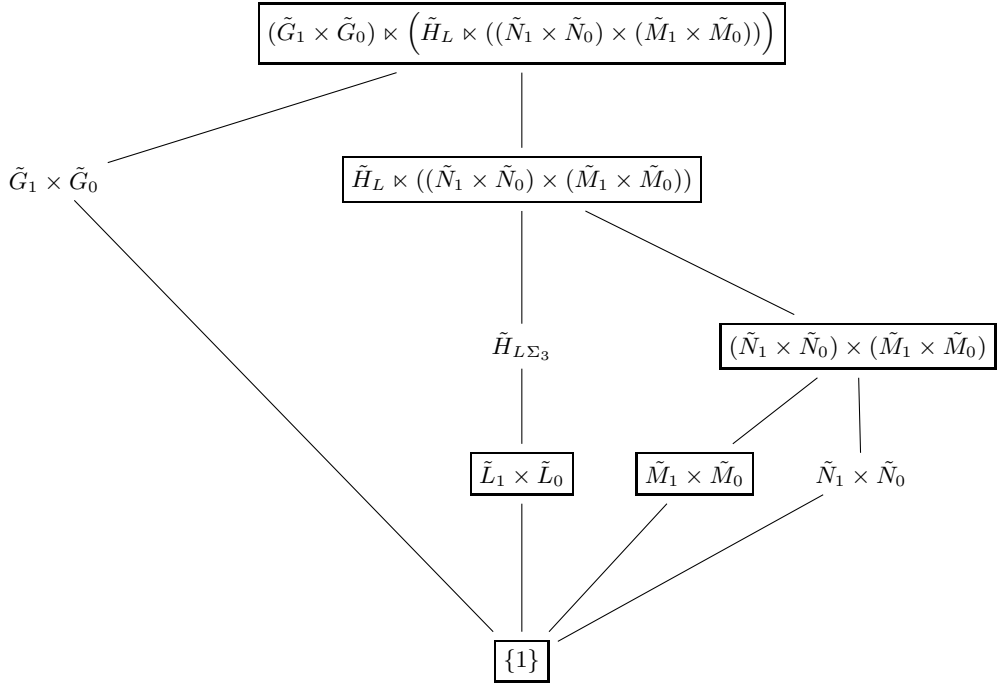
\begin{figure}\label{fig:2}
% center everything in the figure
\centering
% horizontal node distance
\newcommand{\mydistance}{.6cm}
\newcommand{\mydistancea}{1cm}
\begin{tikzpicture}[node distance=2cm]
\title{group $\Tilde{\cG}$}
\node(G)                           {$\boxed{(\Tilde{G}_1\times \Tilde{G}_0)\ltimes\left(\Tilde{H}_L\ltimes((\Tilde{N}_1\times \Tilde{N}_0)\times (\Tilde{M}_1\times \Tilde{M}_0))\right)}$};
\node(B)       [below=1cm of G] {$\boxed{\Tilde{H}_L\ltimes((\Tilde{N}_1\times \Tilde{N}_0)\times (\Tilde{M}_1\times \Tilde{M}_0))}$};
\node(F)       [below=1.5cm of B]       {$ \Tilde{H}_L{}_{\Sigma_3}$};
\node(MN1)       [right=2cm of F] {$\boxed{(\Tilde{N}_1\times \Tilde{N}_0)\times (\Tilde{M}_1\times \Tilde{M}_0)}$};
\node(L1)      [below=1cm of F] {$\boxed{\Tilde{L}_1\times \Tilde{L}_0}$};
\node(M1)      [right=\mydistance of L1]       {$\boxed{\Tilde{M}_1\times \Tilde{M}_0}$};
\node(N1)      [right=\mydistance of M1]       {$\Tilde{N}_1\times \Tilde{N}_0$};
\node(G1)      [left=3cm of B]       {$\Tilde{G}_1\times \Tilde{G}_0$};
\node(1)   [below=3.5cm of F] {$\boxed{\{1\}}$};
\draw(G)      -- (B);
\draw(F)      -- (B);
\draw(L1)     -- (F);
\draw(MN1)     -- (B);
\draw(G1)     -- (G);
\draw(1)     -- (G1);
\draw(1)     -- (L1);
\draw(1)     -- (M1);
\draw(1)     -- (N1);
\draw(M1)-- (MN1);
\draw(N1)-- (MN1);
\end{tikzpicture}
\caption{The symmetry group $\cG_{\Sigma_3}$ of the Poisson bracket algebra in the phase space. The invariant subgroups are boxed.}
\end{figure}
The gauge symmetry group has the following structure. 
First, the groups $\Tilde{M}_1\times \Tilde{M}_0$, $\Tilde{N}_1\times \Tilde{N}_0$ and $\Tilde{L}_1\times \Tilde{L}_0$ with the corresponding algebras $\mathfrak{a}_1$, $\mathfrak{a}_2$ and $\mathfrak{a}_3$, respectively, where:
\begin{equation}
    \mathfrak{a}_1=\mathrm{span}\{ (\tilde{M}_1)_\alpha{}^i\}\,\oplus\,\mathrm{span}\{ (\tilde{M}_0)_\alpha{}^i\}\,, \quad \mathfrak{a}_2=\mathrm{span}\{ (\tilde{N}_1)_a \}\,\oplus\, \mathrm{span}\{(\tilde{N}_0)_a\}\,, \quad \mathfrak{a}_3=\mathrm{span}\{ (\tilde{L}_1)_A{}^{ij}\}\,\oplus\,\mathrm{span}\{ (\tilde{L}_0)_A{}^{ij}\}\,,
\end{equation}
are the subgroups of the full symmetry group $\Tilde{\cG}_{\Sigma_3}$. Besides, the subgroups $\Tilde{L}_1\times \Tilde{L}_0$ and ${\Tilde{M}_1\times \Tilde{M}_0}$ are the invariant subgroups. The group $\Tilde{N}_1\times \Tilde{N}_0$ is not an invariant subgroup of the whole symmetry group, since the Poisson brackets $\{(\tilde{H}_0)_a{}^i(\vec{x})\,,(\tilde{N}_0)^b(\vec{y})\}$ and $\{(\tilde{H}_1)_a{}^i(\vec{x})\,,(\tilde{N}_0)^b(\vec{y})\}$ are equal to some linear combinations of the generators of $\Tilde{M}_1\times \Tilde{M}_0$. Nevertheless, one can form a direct product $(\Tilde{N}_1\times \Tilde{N}_0)\times (\Tilde{M}_1\times \Tilde{M}_0)$, since the generators of these groups are mutually commuting, giving a group which is an invariant subgroup of the complete symmetry group.

Next, consider a subgroup $\tilde{H}_L{}_{\Sigma_3}$ determined by the algebra spanned by the generators $(\Tilde{L}_1){}_A{}^{ij}$, $(\Tilde{L}_0){}_A{}^{ij}$, $(\tilde{H}_1){}_{a}{}^{i}$, and $(\tilde{H}_0){}_{a}{}^{i}$. This group is not invariant subgroup of the whole symmetry group, because of the PB $\{(\tilde{H}_0)_a{}^i(\vec{x})\,,(\tilde{N}_0)^b(\vec{y})\}$ and $\{(\tilde{H}_1)_a{}^i(\vec{x})\,,(\tilde{N}_0)^b(\vec{y})\}$, due to the same argument as before. Now, one can join these two subgroups, of which one is invariant and one is not, using a semidirect product into an invariant subgroup $H_L\ltimes((N_1\times N_0)\times (M_1\times M_0))$, determined by the algebra $\mathfrak{a}_4$: $$\mathfrak{a}_4=\mathrm{span}\{(\tilde{M}_0)_\alpha{}^i, (\tilde{M}_1)_\alpha{}^i,    (\tilde{H}_0)_a{}^i, (\tilde{H}_1)_a{}^i, (\tilde{N}_0)_a, (\tilde{N}_1)_a, (\tilde{L}_0)_A{}^{ij}, (\tilde{L}_1)_A{}^{ij}\}\,.$$.

Finally, following the same line of reasoning, one adds the group $\Tilde{G}_1\times \Tilde{G}_0$ and obtains the full gauge symmetry group $\tilde{\cG}{}_{\Sigma_3}$ to be equal to:
$$\tilde{\cG}{}_{\Sigma_3}=(\Tilde{G}_1\times \Tilde{G}_0)\ltimes\left(\Tilde{H}_L\ltimes((\Tilde{N}_1\times \Tilde{N}_0)\times (\Tilde{M}_1\times \Tilde{M}_0))\right)\,.$$

The complete symmetry group structure is shown in the Figure \ref{fig:2}. Here, the invariant subgroups of the whole symmetry group are boxed.

\section{Construction of the symmetry generator}\label{AppC}

When one substitutes the generators (\ref{eq:gtilda}) into the equation (\ref{eq:generator3BF}), one obtains the gauge generator of the $3BF$ theory in the following form
\begin{equation}\label{eq:generator3BF*}
\begin{array}{ccl}
    G&=&-\ds \int_{\Sigma_3}\D^3 \vec{x} \bigg( (\nabla_0 \epsilon_{\mathfrak{m}{}}{}^{\alpha}{}_i)\Phi(B){}_{\alpha}{}^i-\epsilon_{\mathfrak{m}{}}{}^{\alpha}{}_i\Phi(\cF)_{\alpha}{}^i+(\nabla_0 \epsilon_{\mathfrak{g}{}}{}^{\alpha})\Phi(\alpha){}_{\alpha}+\epsilon_{\mathfrak{g}{}}{}^\alpha \big( f_{\alpha\gamma}{}^\beta B_{\beta 0i}\Phi(B)^\gamma{}^i\\ && \ds \qquad \vphantom{\ds\int}+C_{a0}\rhd_{\alpha b}{}^a \Phi(C)^{b0}+\beta_{a0i}\rhd_{\alpha b}{}^a \Phi(\beta)^{b0i}-\frac{1}{2}\gamma^A{}_{0ij}\rhd_{ \alpha A}{}^B\Phi(\gamma)_B{}^{ij}-\Phi(\nabla B){}_{\alpha}\big)\\\ds& &\qquad\vphantom{\ds\int}+(\nabla_0\epsilon_{\mathfrak{n}{}}{}^{a})\Phi(C)_a-\epsilon_{\mathfrak{n}{}}{}^a\left(\beta_{b0i}\rhd_{\alpha a}{}^b\Phi(B)^{\alpha i}+ \Phi(\cG)_a\right) \\   \ds&&\qquad\vphantom{\ds\int}+( \nabla_0 \epsilon{}_{\mathfrak{h}{}}{}^a{}_i)\Phi(\beta){}_a{}^i-\epsilon{}_{\mathfrak{h}{}}{}^a{}_i\,\left(C_{b0}\rhd_{\alpha a}{}^b\Phi(B)^{\alpha i}-2\beta^{b}{}_{0j}X_{(ab)}{}^A\Phi(\gamma)_A{}^{ij}+ \Phi(\nabla C)_a{}^i\right)\\ \ds &&\vphantom{\ds\int} \ds\qquad\ds-\frac{1}{2}(\nabla_0\epsilon_{\mathfrak{l}}{}
   ^A{}_{ij})\Phi(\gamma)_A{}^{ij}+ \frac{1}{2}\epsilon{}_{\mathfrak{l}}{}^A{}_{ij}\Phi(\nabla D)_A{}^{ij}\bigg)\,,
\end{array}
\end{equation}
where $\epsilon_{\mathfrak{g}}{}^{\alpha}$, $\epsilon_{\mathfrak{h}}{}^a{}_i$,  $\epsilon_{\mathfrak{l}}{}^A{}_{ij}$, $\epsilon_{\mathfrak{m}}{}^{\alpha}{}_i$, and $\epsilon_{\mathfrak{n}}{}^a$ are the independent parameters of the gauge transformations.

The generator of gauge transformations (\ref{eq:generator3BF*}) in $ 3BF $ theory given by the action (\ref{eq:bfcgdh}), is obtained by the Castellani's procedure, requiring the following requirements to be met
\begin{equation}
\begin{aligned}
G_1&=&C_{PFC}\,,\\
    G_0+\{G_1,H_T\}&=&C_{PFC}\,,\\
    \{G_0,H_T\}&=&C_{PFC}\,,
\end{aligned}
\end{equation}
where $C_{PFC}$ denotes some first-class constraints, and assuming that the generator has the following structure:
\begin{equation}
\begin{array}{ccl}
    G&=&\vphantom{\ds\int}\ds\int_{\Sigma_3}\D^3\, \vec{x}\,\Big( \dot{\epsilon}_{\mathfrak{m}{}}{}^\alpha{}_i (G_1)_{\mathfrak{m}{}}{}_\alpha{}^i+\epsilon_{\mathfrak{m}{}}{}^\alpha{}_i(G_0)_{\mathfrak{m}{}}{}_\alpha{}^i+\dot{\epsilon}_{\mathfrak{g}{}}{}^\alpha (G_1)_{\mathfrak{g}{}}{}_\alpha+\epsilon_{\mathfrak{g}{}}{}^\alpha(G_0)_{\mathfrak{g}{}}{}_\alpha\\\vphantom{\ds\int}&&+\dot{\epsilon}_{\mathfrak{h}{}}{}^a{}_i (G_1)_{\mathfrak{h}{}}{}_a{}^i+\epsilon_{\mathfrak{h}{}}{}^a{}_i(G_0)_{\mathfrak{h}{}}{}_a{}^i+\dot{\epsilon}{}_{\mathfrak{n}}{}^a (G_1)_{\mathfrak{n}{}}{}_a+\epsilon_{\mathfrak{n}{}}{}^a(G_0)_{\mathfrak{n}{}}{}_a \vphantom{\int \ds}\\\vphantom{\ds\int}
    && \ds+\frac{1}{2}\dot{\epsilon}_{\mathfrak{l}{}}{}^A{}_{ij}(G_1)_{\mathfrak{l}{}}{}_A{}^{ij}+\frac{1}{2}\epsilon_{\mathfrak{l}{}}{}^A{}_{ij}(G_0)_{\mathfrak{l}{}}{}_A{}^{ij}\Big)\,.\vphantom{\int \ds}
\end{array}
\end{equation}
The first step of Castellani's procedure, imposing the set of conditions
\begin{equation}
       \vphantom{\int \ds} (G_1)_{\mathfrak{m}{}}{}_\alpha{}^i=C_{PFC}\,,\quad \vphantom{\int \ds}(G_1)_{\mathfrak{g}{}}{}_\alpha=C_{PFC}\,,\quad(G_1)_{\mathfrak{h}{}}{}_a{}^i=C_{PFC}\,,\quad (G_1)_{\mathfrak{n}{}}{}_a=C_{PFC}\,, \quad (G_1)_{\mathfrak{l}{}}{}_A{}^{ij}=C_{PFC}\,, 
\end{equation}
is satisfied with a natural choice:
\begin{equation}
   \vphantom{\int \ds} (G_1)_{\mathfrak{m}{}}{}_\alpha{}^i=-\Phi(B)_\alpha{}^i\,,  \quad (G_1)_{\mathfrak{g}{}}{}_\alpha=-\Phi(\alpha)_\alpha\,,\quad (G_1)_{\mathfrak{h}{}}{}_a{}^i=-\Phi(C)_\alpha{}^i\,, \quad (G_1)_{\mathfrak{n}{}}{}_a=-\Phi(\beta)_a\,,\quad (G_1)_{\mathfrak{l}{}}{}_A{}^{ij}=\Phi(\gamma)_A{}^{ij}\,.
\end{equation}
It remains to determine the five generators $G_0$.

The Castellani's second condition for the generator $(G_0)_{\mathfrak{m}{}}{}_\alpha{}^i$ gives:
\begin{equation}
    \begin{array}{l}
    \ds\vphantom{\int\ds}(G_0)_{\mathfrak{m}{}}{}_\alpha{}^i-\{\Phi(B)_\alpha{}^i\,,H_T\}=(C_{PFC}){}_\alpha{}^i\,,\\\vphantom{\int\ds}
    \ds (G_0)_{\mathfrak{m}{}}{}_\alpha{}^i-\Phi(\cF)_{\alpha}{}^ i=(C_{PFC}){}_\alpha{}^i\,,
    \end{array}
\end{equation}
that is $(G_0)_{\mathfrak{m}{}}{}_\alpha{}^i=(C_{PFC}){}_\alpha{}^i+\Phi(\cF)_{\alpha}{}^i$. Subsequently, from the Castellani's third condition it follows
\begin{equation}
\begin{array}{l}
\vphantom{\int\ds}\ds\{(G_0)_{\mathfrak{m}{}}{}_\alpha{}^i\,,H_T\}=(C_{PFC1})_\alpha{}^i\,,\vphantom{\int\ds}\\\vphantom{\int\ds}\ds
\{(C_{PFC}){}_\alpha{}^i+\Phi(\cF)_{\alpha}{}^i\,,H_T\}=(C_{PFC1})_\alpha{}^i\,,\vphantom{\int\ds}\\\vphantom{\int\ds}\ds
\{(C_{PFC}){}_\alpha{}^i\,,H_T\}-f_{\beta\gamma\alpha}\alpha^{\beta}{}_{0}\Phi(\cF)^{\gamma i}=(C_{PFC1})_\alpha{}^i\,,\vphantom{\int\ds}\\
\end{array}
\end{equation}
which gives $$(C_{PFC}){}_\alpha{}^i=f_{\beta\gamma\alpha}\alpha^{\beta}{}_{0}\Phi(B)^{\gamma i}\,.$$ 
It follows that the generator is:
\begin{equation}
    (G_0)_{\mathfrak{m}{}}{}_\alpha{}^i=f_{\beta\gamma\alpha}\alpha^{\beta}{}_{0}\Phi(B)^{\gamma i}+\Phi(\cF)_{\alpha}{}^i\,.
\end{equation}

The Castellani's second condition for the generator $(G_0)_{\mathfrak{g}{}}{}_\alpha$ gives:
\begin{equation}
\begin{array}{l}
    \vphantom{\int\ds}\ds(G_0)_{\mathfrak{g}{}}{}_\alpha - \{\Phi(\alpha)_\alpha\,,H_T\}=(C_{PFC}){}_\alpha\,,\vphantom{\int\ds}\\\vphantom{\int\ds}\ds(G_0)_{\mathfrak{g}{}}{}_\alpha - \Phi(\nabla B)_\alpha=(C_{PFC}){}_\alpha\,,\vphantom{\int\ds}
\end{array}
\end{equation}
that is $(G_0)_{\mathfrak{g}{}}{}_\alpha=(C_{PFC}){}_\alpha+\Phi(\nabla B)_\alpha$. Subsequently, from the Castellani's third condition it follows
\begin{equation}
\begin{array}{l}
\ds\{(G_0)_{\mathfrak{g}{}}{}_\alpha{}\,,H_T\}=(C_{PFC1})_\alpha\,,\vphantom{\int\ds}\\\ds
\{(C_{PFC}){}_\alpha{}+\Phi(\nabla B)_{\alpha}\,,H_T\}=(C_{PFC1})_\alpha\,,\vphantom{\int\ds}\\\ds
\{(C_{PFC}){}_\alpha\,,H_T\}+B_{\beta 0i} f_{\alpha\gamma}{}^\beta \Phi(\cF)^{\gamma i}-\alpha^\beta{}_0 f_{ \alpha\beta}{}^\gamma \Phi(\nabla B)_\gamma\vphantom{\int\ds}\\\ds+C_{a0}\rhd_{\alpha b}{}^a\Phi(\cG)^b+\beta_{a0i}\rhd_{\alpha b }{}^a\Phi(\nabla C)^b{}^i-\frac{1}{2}\gamma^A{}_{0ij}\rhd_{ \alpha A}{}^B\Phi(\nabla D)_B{}^{ij}=(C_{PFC1})_\alpha\,,\vphantom{\int\ds}\\
\end{array}
\end{equation}
which gives $$(C_{PFC}){}_\alpha= -B_{\beta 0i} f_{\alpha\gamma}{}^\beta \Phi(B)^{\gamma i}+\alpha^\beta{}_0 f_{ \alpha\beta}{}^\gamma \Phi(\alpha)_\gamma-C_{a0}\rhd_{\alpha b}{}^a\Phi(C)^b-\beta_{a0i}\rhd_{\alpha b }{}^a\Phi(\beta)^b{}^i+\frac{1}{2}\gamma^A{}_{0ij}\rhd_{ \alpha A}{}^B\Phi(\gamma)_B{}^{ij}\,.$$ 
It follows that the generator is:
\begin{equation}
    (G_0)_{\mathfrak{g}{}}{}_\alpha=-B_{\beta 0i} f_{\alpha\gamma}{}^\beta \Phi(B)^{\gamma i}+\alpha^\beta{}_0 f_{ \alpha\beta}{}^\gamma \Phi(\alpha)_\gamma-C_{a0}\rhd_{\alpha b}{}^a\Phi(C)^b-\beta_{a0i}\rhd_{\alpha b }{}^a\Phi(\beta)^b{}^i+\frac{1}{2}\gamma^A{}_{0ij}\rhd_{ \alpha A}{}^B\Phi(\gamma)_B{}^{ij}+\Phi(\nabla B)_\alpha\,.
\end{equation}

The Castellani's second condition for the generator $(G_0)_{\mathfrak{n}{}}{}_a$ gives
\begin{equation}
\begin{array}{l}
    \phantom{\int\ds}\ds(G_0)_{\mathfrak{n}{}}{}_a - \{\Phi(C)_a\,,H_T\}=(C_{PFC}){}_a{}\,,\phantom{\int\ds}\\\phantom{\int\ds}\ds(G_0)_{\mathfrak{n}{}}{}_a - \Phi(\cG)_a{}=(C_{PFC}){}_a{}\,,\phantom{\int\ds}
\end{array}
\end{equation}
that is $(G_0)_{\mathfrak{n}{}}{}_a{}=(C_{PFC}){}_a{}+\Phi(\cG)_a$. Subsequently, from the Castellani's third condition it follows
\begin{equation}
\begin{array}{l}
\phantom{\int\ds}\ds\{(G_0)_{\mathfrak{n}{}}{}_a\,,H_T\}=(C_{PFC1})_a\,,\phantom{\int\ds}\\\phantom{\int\ds}\ds
\{(C_{PFC}){}_a+\Phi(\cG)_{a}\,,H_T\}=(C_{PFC1})_a{}\,,\phantom{\int\ds}\\\phantom{\int\ds}\ds \{(C_{PFC}){}_a{},H_T\}+\alpha^\alpha{}_0\rhd_{\alpha a}{}^b \Phi(\cG)_b-\beta_b{}_{0i}\rhd_{\alpha a}{}^b \Phi(\cF)^{\alpha i}=(C_{PFC1})_a{}\,,\phantom{\int\ds}\\
\end{array}
\end{equation}
which gives $$(C_{PFC})_a{}=-\alpha^\alpha{}_0\rhd_{\alpha a}{}^b \Phi(C)_b+\beta_b{}_{0i}\rhd_{\alpha a}{}^b \Phi(B)^{\alpha i}\,.$$
It follows that the generator is:
$$(G_0)_{\mathfrak{n}{}}{}_a{}=-\alpha^\alpha{}_0\rhd_{\alpha a}{}^b \Phi(C)_b+\beta_b{}_{0i}\rhd_{\alpha a}{}^b \Phi(B)^{\alpha i}+\Phi(\cG)_a\,.$$

The Castellani's second condition for the generator $(G_0)_{\mathfrak{h}{}}{}_a{}^i$ gives:
\begin{equation}
\begin{array}{l}
    \phantom{\int\ds}\ds(G_0)_{\mathfrak{h}{}}{}_a{}^i - \{\Phi(\beta)_a{}^i\,,H_T\}=(C_{PFC}){}_a{}^i\,,\phantom{\int\ds}\\\phantom{\int\ds}\ds(G_0)_{\mathfrak{h}{}}{}_a{}^i - \Phi(\nabla C)_a{}^i=(C_{PFC}){}_a{}^i\,,\phantom{\int\ds}
\end{array}
\end{equation}
that is $(G_0)_{\mathfrak{h}{}}{}_a{}^i=(C_{PFC}){}_a{}^i+\Phi(\nabla C)_a{}^i$. Subsequently, from the Castellani's third condition it follows
\begin{equation}
\begin{array}{l}
\phantom{\int\ds}\ds\{(G_0)_{\mathfrak{h}{}}{}_a{}^i\,,H_T\}=(C_{PFC1})_a{}^i\,,\phantom{\int\ds}\\\phantom{\int\ds}\ds
\{(C_{PFC}){}_a{}^i+\Phi(\nabla C)_{a}{}^i\,,H_T\}=(C_{PFC1})_a{}^i\,,\phantom{\int\ds}\\\phantom{\int\ds} \ds\{(C_{PFC}){}_a{}^i,H_T\}+\alpha^\alpha{}_0 \rhd_{\alpha a}{}^b\Phi(\nabla C)_b{}^i-C_{b0}\rhd_{\alpha a}{}^b\Phi(\cF)^{\alpha i}+2\beta^{b}{}_{0j}X_{(ab)}{}^A\Phi(\nabla D)_A{}^{ij}=(C_{PFC1})_a{}^i\,,\phantom{\int\ds}\\
\end{array}
\end{equation}
which gives $$(C_{PFC})_a{}^i=-\alpha^\alpha{}_0 \rhd_{\alpha a}{}^b\Phi(\beta)_b{}^i+C_{b0}\rhd_{\alpha a}{}^b\Phi(B)^{\alpha i}-2\beta^{b}{}_{0j}X_{(ab)}{}^A\Phi(\gamma)_A{}^{ij}\,.$$
It follows that the generator is:
$$(G_0)_{\mathfrak{h}{}}{}_a{}^i=-\alpha^\alpha{}_0 \rhd_{\alpha a}{}^b\Phi(\beta)_b{}^i+C_{b0}\rhd_{\alpha a}{}^b\Phi(B)^{\alpha i}-2\beta^{b}{}_{0j}X_{(ab)}{}^A\Phi(\gamma)_A{}^{ij}+\Phi(\nabla C)_a{}^i\,.$$

The Castellani's second condition for the generator $(G_0)_{\mathfrak{l}{}}{}_A{}^{ij}$ gives:
\begin{equation}
    \begin{array}{l}
    \phantom{\int\ds}\ds(G_0)_{\mathfrak{l}{}}{}_A{}^{ij}+\{\Phi(\gamma)_A{}^{ij}\,,H_T\}=(C_{PFC}){}_A{}^{ij}\,,\phantom{\int\ds}\\\phantom{\int\ds}    \ds
    (G_0)_{\mathfrak{l}{}}{}_A{}^{ij}+\Phi(\nabla D)_{A}{}^{ij}=(C_{PFC}){}_A{}^{ij}\,,\phantom{\int\ds}
    \end{array}
\end{equation}
that is $(G_0)_{\mathfrak{l}{}}{}_A{}^{ij}=(C_{PFC}){}_A{}^{ij}-\Phi(\nabla D)_{A}{}^{ij}$.
Subsequently, from the Castellani's third condition it follows:
\begin{equation}
\begin{array}{l}
\phantom{\int\ds}\ds\{(G_0)_{\mathfrak{l}{}}{}_A{}^{ij}\,,H_T\}=(C_{PFC1})_A{}^{ij}\,,\phantom{\int\ds}\\\phantom{\int\ds}\ds
\{(C_{PFC}){}_A{}^{ij}-\Phi(\nabla D)_{A}{}^{ij},H_T\}=(C_{PFC1})_A{}^{ij}\,,\phantom{\int\ds}\\\phantom{\int\ds}\ds
\{(C_{PFC}){}_A{}^{ij}\,,H_T\}-\alpha^\alpha{}_0\rhd_{\alpha A}{}^B \Phi(\nabla D)_B{}^{ij}=(C_{PFC1})_A{}^{ij }\,,\phantom{\int\ds}
\end{array}
\end{equation}
which gives $$(C_{PFC}){}_A{}^{ij}=\alpha^\alpha{}_0\rhd_{\alpha A}{}^B \Phi(\gamma)_B{}^{ij}\,.$$ 
It follows that the generator is:
\begin{equation}
    (G_0)_{\mathfrak{l}{}}{}_A{}^{ij}=\alpha^\alpha{}_0\rhd_{\alpha A}{}^B \Phi(\gamma)_B{}^{ij}-\Phi(\nabla D)_{A}{}^{ij}\,.
\end{equation}

At this point, it is useful to summarize the results, and introduce the new notation:
\begin{equation}
\phantom{\int\ds}\dot{\epsilon}_{\mathfrak{m}{}}{}^\alpha{}_i (G_1)_{\mathfrak{m}{}}{}_\alpha{}^i+\epsilon_{\mathfrak{m}{}}{}^\alpha{}_i(G_0)_{\mathfrak{m}{}}{}_\alpha{}^i=-\nabla_0 {\epsilon}_{\mathfrak{m}{}}{}^\alpha{}_i\Phi(B)_\alpha{}^i+\epsilon_{\mathfrak{m}{}}{}^\alpha{}_i\Phi(\cF)_{\alpha}{}^i=\nabla_0 {\epsilon}_{\mathfrak{m}{}}{}^\alpha{}_i(\tilde{M}_1)_\alpha{}^i+\epsilon_{\mathfrak{m}{}}{}^\alpha{}_i(\tilde{M}_0)_\alpha{}^i\,.
\end{equation}
Note that the time derivative of the parameter combines with some of the other terms into a covariant derivative in the time directions. 

For the second part of the total generator one obtains:
\begin{equation}
    \begin{array}{lcl}
    \phantom{\int\ds}\ds\dot{\epsilon}_{\mathfrak{g}{}}{}^\alpha (G_1)_{\mathfrak{g}{}}{}_\alpha+\epsilon_{\mathfrak{g}{}}{}^\alpha(G_0)_{\mathfrak{g}{}}{}_\alpha&=&\ds-\dot{\epsilon}_{\mathfrak{g}{}}{}^\alpha\Phi(\alpha)_\alpha-\epsilon_{\mathfrak{g}{}}{}^\alpha\big( B_{\beta 0i} f_{\alpha\gamma}{}^\beta \Phi(B)^{\gamma i}-\alpha^\beta{}_0 f_{ \alpha\beta}{}^\gamma \Phi(\alpha)_\gamma \phantom{\int\ds}\\ \phantom{\int\ds}\ds&&\ds+C_{a0}\rhd_{\alpha b}{}^a\Phi(C)^b+\beta_{a0i}\rhd_{\alpha b }{}^a\Phi(\beta)_b{}^i-\frac{1}{2}\gamma^A{}_{0ij}\rhd_{ \alpha A}{}^B\Phi(\gamma)_B{}^{ij}-\Phi(\nabla B)_\alpha\big)  \phantom{\int\ds}\\\phantom{\int\ds}&=& \ds-\nabla_0 {\epsilon}_{\mathfrak{g}{}}{}^\alpha\Phi(\alpha)_\alpha-\epsilon_{\mathfrak{g}{}}{}^\alpha\big(B_{\beta 0i} f_{\alpha\gamma}{}^\beta \Phi(B)^{\gamma i} \phantom{\int\ds}\\\phantom{\int\ds}\ds&&+\ds C_{a0}\rhd_{\alpha b}{}^a\Phi(C)^b+\beta_{a0i}\rhd_{\alpha b }{}^a\Phi(\beta)_b{}^i-\frac{1}{2}\gamma^A{}_{0ij}\rhd_{ \alpha A}{}^B\Phi(\gamma)_B{}^{ij}-\Phi(\nabla B)_\alpha\big) \phantom{\int\ds}\\ \phantom{\int\ds}\ds&=&\ds\nabla_0{\epsilon}_{\mathfrak{g}{}}{}^\alpha (\tilde{G}_1)_\alpha+\epsilon_{\mathfrak{g}{}}{}^\alpha(\tilde{G}_0)_\alpha\,. \phantom{\int\ds}
    \end{array}
\end{equation}
Furthermore, it follows:
\begin{equation}
\begin{array}{lcl}
\phantom{\int\ds}\dot{\epsilon}_{\mathfrak{h}{}}{}^a{}_i (G_1)_{\mathfrak{h}{}}{}_a{}^i+\epsilon_{\mathfrak{h}{}}{}^a{}_i(G_0)_{\mathfrak{h}{}}{}_a{}^i&=&-\ds\nabla_0{\epsilon}_{\mathfrak{h}{}}{}^a{}_i\Phi(\beta)_\alpha{}^i+\epsilon_{\mathfrak{h}{}}{}^a{}_i\big(C_{b0}\rhd_{\alpha a}{}^b\Phi(B)^{\alpha i}-2\beta^{b}{}_{0j}X_{(ab)}{}^A\Phi(\gamma)_A{}^{ij}+\Phi(\nabla C)_a{}^i\big)\phantom{\int\ds}\\\phantom{\int\ds}&=&\ds\nabla_0{\epsilon}_{\mathfrak{h}{}}{}^a{}_i (\tilde{H}_1)_a{}^i+\epsilon_{\mathfrak{h}{}}{}^a{}_i(\tilde{H}_0)_a{}^i\,,\phantom{\int\ds}
\end{array}
\end{equation}
\begin{equation}
\begin{array}{lcl}
        \phantom{\int\ds}\dot{\epsilon}_{\mathfrak{n}{}}{}^a (G_1)_{\mathfrak{n}{}}{}_a+\epsilon_{\mathfrak{n}{}}{}^a(G_0)_{\mathfrak{n}{}}{}_a&=&\ds-\nabla_0{\epsilon}_{\mathfrak{n}{}}{}^a\Phi(C)_a+\epsilon_{\mathfrak{n}{}}{}^a(\beta_b{}_{0i}\rhd_{\alpha a}{}^b \Phi(B)^{\alpha i}+\Phi(\cG)_a) \phantom{\int\ds}\\ \phantom{\int\ds}&=&\ds\nabla_0{\epsilon}_{\mathfrak{n}{}}{}^a (\tilde{N}_1)_a+\epsilon_{\mathfrak{n}{}}{}^a(\tilde{N}_0)_a\,. \phantom{\int\ds}
        \end{array}
\end{equation}
Finally, one gets:
\begin{equation}
    \begin{array}{lcl}
          \phantom{\int\ds}\ds\frac{1}{2}\dot\epsilon_{\mathfrak{l}{}}{}^A{}_{ij}(G_1)_{\mathfrak{l}{}}{}_A{}^{ij}+\frac{1}{2}\epsilon_{\mathfrak{l}{}}{}^A{}_{ij}(G_0)_{\mathfrak{l}{}}{}_A{}^{ij}&=&\ds\frac{1}{2}\dot\epsilon_{\mathfrak{l}{}}{}^A{}_{ij}\Phi(\gamma)_A{}^{ij}+\frac{1}{2}\epsilon_{\mathfrak{l}{}}{}^A{}_{ij}\alpha^\alpha{}_0\rhd_{\alpha A}{}^B \Phi(\gamma)_B{}^{ij}-\frac{1}{2}\epsilon_{\mathfrak{l}{}}{}^A{}_{ij}\Phi(\nabla D)_{A}{}^{ij} \phantom{\int\ds}\\ \phantom{\int\ds}&=&\ds\frac{1}{2}\nabla_0\epsilon_{\mathfrak{l}{}}{}^A{}_{ij}\Phi(\gamma)_A{}^{ij}-\frac{1}{2}\epsilon_{\mathfrak{l}{}}{}^A{}_{ij}\Phi(\nabla D)_{A}{}^{ij} \phantom{\int\ds}\\ \phantom{\int\ds}&=&\ds\frac{1}{2}\nabla_0\epsilon_{\mathfrak{l}{}}{}^A{}_{ij}(\tilde{L}_1)_A{}^{ij}+\frac{1}{2}\epsilon_{\mathfrak{l}{}}{}^A{}_{ij}(\tilde{L}_0)_A{}^{ij}\,. \phantom{\int\ds}
    \end{array}
\end{equation}

\section{Definitions of maps $\cT$, $\cS$, $\cD$, ${\cal X}_1$, and ${\cal X}_2$}\label{AppD}
Given $G$-invariant symmetric non-degenerate bilinear forms in $\mathfrak{g}$ and $\mathfrak{h}$, one can define a bilinear antisymmetric map ${\cal T}: \mathfrak{h} \times \mathfrak{h} \to \mathfrak{g}$ by the rule:
$$
    \langle {\cal T}(\underline{h}_1,\,\underline{h}_2)\,, \underline{g}\rangle{}_\mathfrak{g}=-\langle \underline{h}_1,\,\underline{g}\rhd \underline{h}_2\rangle{}_\mathfrak{h}, \quad \quad \forall \underline{h}_1,\,\underline{h}_2 \in \mathfrak{h}\,, \quad \forall \underline{g} \in \mathfrak{g}\,.
$$
Written in basis:
$$
    \cT(t_a,t_b)=\cT_{ab}{}^\alpha \tau_\alpha\,,
$$
where the components of the map $\cT$ are:
$$
    \cT_{ab}{}^\alpha =-g_{ac}\rhd_{\beta b}{}^cg^{\alpha\beta}\,.
$$
See \cite{FariaMartinsMikovic2011} for more properties and the construction of $2BF$ invariant topological action using this map. 

The transformations of the Lagrange multipliers and the $3BF$ invariant topological action is defined via maps 
$$\cS  : \mathfrak{l} \times \mathfrak{l} \to \mathfrak{g}\,, \qquad {\cal X}_1: \mathfrak{l} \times \mathfrak{h} \to \mathfrak{h}\,,\qquad  {\cal X}_2:\mathfrak{l} \times\mathfrak{h} \to\mathfrak{h}\,, \qquad{\cal D}: \mathfrak{h} \times \mathfrak{h} \times \mathfrak{l} \to \mathfrak{g}\,., $$ 
as it is defined in \cite{Radenkovic2019}. 
The map $\cS  : \mathfrak{l} \times \mathfrak{l} \to \mathfrak{g}$ is defined by the rule:
$$
    \langle {\cal S} (\underline{l}_1,\,\underline{l}_2),\, \underline{g}\rangle{}_{\mathfrak{g}}=-\langle \underline{l}_1,\,\underline{g}\rhd \underline{l}_2\rangle{}_{\mathfrak{l}}\,, \quad \quad \forall \underline{l}_1, \forall \underline{l}_2 \in \mathfrak{l}\,,\quad \forall \underline{g} \in \mathfrak{g}\,.
$$

Written in the basis:
$${\cal S}(T_A,T_B)={\cal S}{}_{AB}{}^\alpha \tau_\alpha\,,$$
the defining relation for ${\cal S}$ becomes:
$${\cal S}_{AB}{}^\alpha =-\rhd_{\beta [B}{}^C g_{A]C}g^{\alpha\beta}\,.$$
Given two $\mathfrak{l}$-valued forms $\eta$ and $\omega$, one can define a $\mathfrak{g}$-valued form:
$$\omega \wedge^{\cal S} \eta = \omega^A \wedge \eta^B {\cal S}{}_{AB}{}^\alpha \tau_\alpha\,.$$
Using this map, the transformations of the Lagrange multipliers under $L$-gauge are defined in \cite{Radenkovic2019}.

Further, to define the transformations of the Lagrange multipliers under $H$-gauge transformations the bilinear map ${\cal X}_1: \mathfrak{l} \times \mathfrak{h} \to \mathfrak{h}$ is defined:
$$\langle {\cal X}_1(\underline{l},\,\underline{h}_1),\,\underline{h}_2 \rangle{}_{\mathfrak{h}}=-\langle \underline{l},\,\{\underline{h}_1,\,\underline{h}_2\}\rangle{}_{\mathfrak{l}}\,,\quad \quad \forall \underline{h}_1,\,\underline{h}_2 \in \mathfrak{h}\,,\quad \forall \underline{l}\in\mathfrak{l}\,, $$
and bilinear map ${\cal X}_2:\mathfrak{l} \times\mathfrak{h} \to\mathfrak{h} $ by the rule:
$$\langle {\cal X}_2(\underline{l},\,\underline{h}_2),\,\underline{h}_1 \rangle{}_{\mathfrak{h}}=-\langle \underline{l},\,\{\underline{h}_1,\,\underline{h}_2\}\rangle{}_{\mathfrak{l}}\,,\quad \quad \forall \underline{h}_1,\,\underline{h}_2 \in \mathfrak{h}\,,\quad \forall \underline{l}\in\mathfrak{l}\,. $$
As far as the bilinear maps ${\cal X}_1$ and ${\cal X}_2$ one can define the coefficients in the basis as:
$${\cal X}_1(T_A,t_a)={\cal X}_1{}_{Aa}{}^b \,t_b\,, \quad \quad {\cal X}_2(T_A,t_a)={\cal X}_2{}_{Aa}{}^b \,t_b\,.$$
When written in the basis the defining relations for the maps ${\cal X}_1$ and ${\cal X}_2$ become:
$$ {\cal X}_1{}_{Ab}{}^c=-X_{ba}{}^Bg_{AB}g^{ac}\,,\quad \quad \quad {\cal X}_2{}_{Ab}{}^c=-X_{ab}{}^Bg_{AB}g^{ac}\,.$$
Given $\mathfrak{l}$-valued differential form $\omega$ and $\mathfrak{h}$-valued differential form $\eta$, one defines a $\mathfrak{h}$-valued form as:
$$\omega\wedge^{{\cal X}_1}\eta=\omega^A\wedge\eta^a{{\cal X}_1}_{Aa}{}^{b}t_b\,, \quad \quad \omega\wedge^{{\cal X}_2}\eta=\omega^A\wedge\eta^a{{\cal X}_2}_{Aa}{}^{b}t_b\,. $$

Finally, a trilinear map ${\cal D}: \mathfrak{h} \times \mathfrak{h} \times \mathfrak{l} \to \mathfrak{g} $ is needed:
$$\langle {\cal D}(\underline{h}_1,\,\underline{h}_2,\,\underline{l}),\,\underline{g}\rangle{}_{\mathfrak{g}}=-\langle \underline{l},\,\{\underline{g}\rhd\underline{h}_1,\,\underline{h}_2\}\rangle{}_{\mathfrak{l}}\,,\quad \quad \forall \underline{h}_1,\,\underline{h}_2 \in \mathfrak{h}\,, \quad \forall \underline{l} \in \mathfrak{l},\,\quad \forall \underline{g}\in \mathfrak{g}\,, $$
One can define the coefficients of the trilinear map as:
$${\cal D}(t_a,\,t_b,\,T_A)= {\cal D}_{abA}{}^\alpha \tau_\alpha\,, $$
and the defining relation for the map ${\cal D}$ expressed in terms of coefficients becomes:
$${\cal D}{}_{abA}{}^\beta =-\rhd_{\alpha a}{}^c X_{cb}{}^B g_{AB}g^{\alpha\beta}\,.$$
Given two $\mathfrak{h}$-valued forms $\omega$ and $\eta$, and $\mathfrak{l}$-valued form $\xi$, the $g$-valued form is given by the formula:
$$\omega\wedge^{\cal D}\eta\wedge^{\cal D}\xi=\omega^a\wedge\eta^b\wedge\xi^A {\cal D}{}_{abA}{}^\beta \tau_\beta\,.$$

With these maps in hand, the transformations of the Lagrange multipliers under $H$-gauge transformations are defined, see \cite{Radenkovic2019}.

\section{Form-variations of all fields and momenta}\label{AppE}
The obtained gauge generator (\ref{eq:generator3BF}) is employed to calculate the form variations of variables and their corresponding canonical momenta, denoted as $A(t,\vec{x})$, using the following equation,
\begin{equation}
    \delta_0 A(t,\vec{x}) =\{ A(t,\vec{x}), G\}\,.
\end{equation}
The computed form variations are given as follows:
{\thickmuskip=0mu%
\medmuskip=0mu%
\thinmuskip=0mu%
\begin{equation}\label{eq:formvar}
\resizebox{.9\hsize}{!}{$
    \begin{array}{lclclclc}
       \delta_0 B^\alpha{}_{0i}&=&-\nabla_0\epsilon_{\mathfrak{m}}{}^\alpha{}_i+f_{\beta\gamma}{}^\alpha\epsilon_{\mathfrak{g}}{}^\beta B^\gamma{}_{0i}
       \quad&\quad& \delta_0 \pi(B)_\alpha{}^{0i}&=&f_{\alpha\beta}{}^\gamma\epsilon_{\mathfrak{g}}{}^\beta\pi(B)_\gamma{}^{0i}\,,\phantom{\ds\int}\\
       &&+\epsilon_{\mathfrak{n}}{}^a\rhd_{\alpha a}{}^b\beta_{b0i}+\epsilon_{\mathfrak{h}}{}^a{}_i \rhd_{\alpha a}{}^b  C_{b0}\,, &\quad& &&\phantom{\ds\int}\\
       \delta_0 B^\alpha{}_{ij}&=&-2\nabla_{[i|}\epsilon_{\mathfrak{m}}{}^\alpha{}_{|j]}+f_{\beta\gamma}{}^\alpha\epsilon_{\mathfrak{g}}{}^\beta B^{\gamma}{}_{ij} -\epsilon_{\mathfrak{l}}{}^A{}_{ij}\rhd_{\alpha A}{}^BD_B &\quad& \delta_0 \pi(B)_\alpha{}^{ij}&=&f_{\alpha\beta}{}^\gamma\epsilon_{\mathfrak{g}}{}^\beta\pi(B)_\gamma{}^{ij}\,,\phantom{\ds\int}\\
       &&+\epsilon_{\mathfrak{n}}{}^a\rhd_{\alpha a}{}^b\beta_{bij}+2\epsilon_{\mathfrak{h}}{}^a{}_{[j|} \rhd_{\alpha a}{}^b  C_{b|i]}\,, &\quad& &&\phantom{\ds\int}\\
       \delta_0  \alpha^\alpha{}_{0}&=&-\nabla_0 \epsilon_{\mathfrak{g}}{}^\alpha\,,  \quad&\quad& \delta_0 \pi(\alpha)_\alpha{}^{0}&=&f_{\alpha\beta}{}^\gamma \epsilon_{\mathfrak{m}}{}^\beta{}_i \pi(B)_\gamma{}^{0i}+f_{\alpha\beta }{}^\gamma \epsilon_{\mathfrak{g}}{}^\beta \pi(\alpha)_\gamma{}^0\phantom{\ds\int}\\
       &&&&&&+\rhd_{\alpha b}{}^a\epsilon_{\mathfrak{n}}{}^b\pi(C)_a{}^0+\rhd_{\alpha b}{}^a\epsilon_{\mathfrak{h}}{}^b{}_i\pi(\beta)_a{}^i\,\phantom{\ds\int}\\&&&&&&\ds-\frac{1}{2}\rhd_{\alpha B}{}^A\epsilon_{\mathfrak{l}}{}^B{}_{ij}\pi(\gamma)_A{}^{0ij}\,,\phantom{\int\ds}\\
       \delta_0 \alpha^\alpha{}_{i}&=&-\nabla_i\epsilon_{\mathfrak{g}}{}^\alpha-\partial_a{}^\alpha \epsilon_{\mathfrak{h}}{}^a{}_i\,,  &\quad& \delta_0 \pi(\alpha)_\alpha{}^{i}&=&f_{\alpha\beta}{}^\gamma\epsilon_{\mathfrak{m}}{}^\beta{}_{j}\pi(B)_\gamma{}^{ij}+f_{\alpha\beta}{}^\gamma\epsilon_{\mathfrak{g}}{}^\beta\pi(\alpha)_\gamma{}^i\phantom{\ds\int}\\
       &&&&&&+\rhd_{\alpha b}{}^\alpha \epsilon_{\mathfrak{n}}{}^b \pi(C)_a{}^i+\rhd_{\alpha b}{}^\alpha \epsilon_{\mathfrak{h}}{}^b{}_j \pi(\beta)_a{}^{ij}\,\phantom{\ds\int}\\
       &&&&&&\ds-\frac{1}{2}\rhd_{\alpha B}{}^A\epsilon_{\mathfrak{l}}{}^B{}_{jk}\pi(\gamma)_A{}^{ijk}-\epsilon^{0ijk}\nabla_j\epsilon_{\mathfrak{m}}{}_{\alpha k}\,,\phantom{\ds\int}\\
       &&&&&&\ds -\frac{1}{2}\epsilon^{0ijk}\epsilon_{\mathfrak{n}}{}^a\rhd_{\alpha b}{}^a\beta^b{}_{jk}\,,\phantom{\ds\int}\\
       \ds\delta_0 C^a{}_{0}&=&-\nabla_0\epsilon_{\mathfrak{n}}{}^a+\epsilon_{\mathfrak{g}}{}^\alpha \rhd_{\alpha b}{}^a C^b{}_0\ds\,,  &\quad& \delta_0 \ds\pi(C)_a{}^{0}&=&-\epsilon_{\mathfrak{g}}{}^\alpha\rhd_{\alpha a}{}^b\pi(C)_b{}^0+\epsilon_{\mathfrak{h}}{}_b{}_i\rhd_{\alpha a}{}^b\pi(B)^{\alpha 0i}\ds\,,\phantom{\ds\int}\\
       \ds\delta_0 C^a{}_{i}&=&-\nabla_i\epsilon_{\mathfrak{n}}{}^a+\epsilon_{\mathfrak{g}}{}^\alpha \rhd_{\alpha b}{}^a C^b{}_i\ds  &\quad& \delta_0 \ds\pi(C)_a{}^{i}&=&-\epsilon_{\mathfrak{g}}{}^\alpha\rhd_{\alpha a}{}^b\pi(C)_b{}^i+\epsilon_{\mathfrak{h}}{}_b{}_j\rhd_{\alpha a}{}^b\pi(B)^{\alpha ij}\ds\,,\phantom{\ds\int}\\
       &&-\epsilon_{\mathfrak{m}}{}^\alpha{}_i\partial^a{}_\alpha+2\epsilon_{\mathfrak{h}}{}^b{}_i\, D_A\,X_{(bc)}{}^Ag^{ac}\,,\ds&&&&\,\phantom{\ds\int}\\
       \ds\delta_0 \beta^a{}_{0i}&=&-\nabla_0\epsilon_{\mathfrak{h}}{}^a{}_i+\epsilon_{\mathfrak{g}}{}^\alpha \rhd_{\alpha b}{}^a \beta_{b0i}\ds\,,  &\quad& \ds\delta_0 \pi(\beta)_a{}^{0i}&=&-\epsilon_{\mathfrak{g}}{}^\alpha\rhd_{\alpha a}{}^b\pi(\beta)_b{}^{0i}+\epsilon_{\mathfrak{n}}{}_b\rhd_{\alpha a}{}^b\pi(B)^{\alpha0i}\ds\phantom{\ds\int}\\
       &&&&&&-2\epsilon_{\mathfrak{h}}{}^b{}_{j}X_{(ab)}{}^A\pi(\gamma)_A{}^{0ij}\,,\phantom{\ds\int}\\
       \ds\delta_0 \beta^a{}_{ij}&=&-2\nabla_{[i|}\epsilon_{\mathfrak{h}}{}^a{}_{|j]}+\epsilon_{\mathfrak{g}}{}^\alpha \rhd_{\alpha b}{}^a \beta^{b}{}_{ij}+\epsilon_{\mathfrak{l}}{}^A{}_{ij}\delta_A{}^a\ds\,,  \quad&\quad& \ds\delta_0 \pi(\beta)_a{}^{ij}&=&-\epsilon_{\mathfrak{g}}{}^\alpha\rhd_{\alpha a}{}^b\pi(\beta)_b{}^{ij}+\epsilon_{\mathfrak{n}}{}_b\rhd_{\alpha a}{}^b\pi(B)^{\alpha ij}\ds\,\phantom{\ds\int}\\
       &&&&&&-2\epsilon_{\mathfrak{h}}{}^b{}_{k}X_{(ab)}{}^A\pi(\gamma)_A{}^{ijk}\phantom{\ds\int}\\
       &&&&&&+\epsilon^{0ijk}\nabla_k\epsilon_{\mathfrak{n}}{}_a+\epsilon^{0ijk}\epsilon_{\mathfrak{h}}{}^\alpha{}_k\partial{}_a{}_\alpha\ds\,,\phantom{\ds\int}\\
       \ds\delta_0\gamma^A{}_{0ij}&=&\epsilon_{\mathfrak{g}}{}^\alpha \gamma{}^B{}_{0ij}\rhd_{\alpha B}{}^A+\nabla_0\epsilon_{\mathfrak{l}}{}^A{}_{ij} \,&\quad&\ds\delta_0\pi(\gamma)_A{}^{0ij}&=&-\epsilon_{\mathfrak{g}}{}^\alpha\,\rhd_{\alpha A}{}^B\,\pi(\gamma)_B{}^{0ij}\,,\phantom{\ds\int}\\
       &&-4\epsilon_{\mathfrak{h}}{}^a{}_{[i|}\beta^b{}_{0|j]}\,X_{(ab)}{}^A\,,\phantom{\ds\int}&&&&\\
       \ds\delta_0\gamma^A{}_{ijk}&=&\epsilon_{\mathfrak{g}}{}^\alpha \gamma{}^B{}_{ijk}\rhd_{\alpha B}{}^A+\nabla_{i}\epsilon_{\mathfrak{l}}{}^A{}_{jk}&\quad&\ds\delta_0\pi(\gamma)_A{}^{ijk}&=&-\epsilon_{\mathfrak{g}}{}^\alpha\,\rhd_{\alpha A}{}^B\,\pi(\gamma)_B{}^{ijk}+\epsilon^{oijk}\delta_a{}_A\epsilon_{\mathfrak{n}}{}^a\,,\phantom{\ds\int}\\
       &&-\nabla_j\epsilon_{\mathfrak{l}}{}^A{}_{ik}+\nabla_k\epsilon_{\mathfrak{l}}{}^A{}_{ij}+3!\epsilon_{\mathfrak{h}}{}^a{}_{[i}\,\beta^b{}_{jk]}\,X_{(ab)}{}^A\,,\quad&\quad&&&\phantom{\ds\int}\\
       \ds \delta_0 D^A&=&\epsilon_{\mathfrak{n}}{}^a\delta_a{}^A+\epsilon_{\mathfrak{g}}{}^\alpha D^B\rhd_{\alpha B}{}^A\,,&\quad&\ds \delta_0\pi(D)_A&=&\ds-2\epsilon_{\mathfrak{h}}{}^a{}_iX_{(ab)}{}_A\pi(C)^{bi}\phantom{\ds\int}\\
       &&&&&&\ds-\frac{1}{2}\epsilon_{\mathfrak{l}}{}_B{}^{ij}\rhd_{\alpha A}{}^B\pi(B)^\alpha{}_{0ij}\phantom{\ds\int}\\
       &&&&&&-\epsilon_{\mathfrak{g}}{}^\alpha\rhd_{\alpha A}{}^B\pi(D)_B \phantom{\ds\int}
    \end{array}$}
\end{equation}}

\section{Symmetry algebra calculations}\label{AppF}
To obtain the structure of the symmetry group of the $3BF$ action, as presented in the subsection \ref{secIVd}, one has to calculate the commutators between the generators of all the symmetries, i.e., the $G$-, $H$-, $L$-, $M$-, and $N$-gauge symmetries. This process is described in the subsections \ref{secIVa}, \ref{secIVb}, and \ref{secIVc}, while details of the calculation which are not straightforward will be given in the following.
\subsection{Commutator $[H,H]$}
Let us derive the commutator of the generators of the $H$-gauge transformations, i.e., the equation (\ref{eq:HHcom}). 
After transforming the variables under $H$-gauge transformations for the parameter $\epsilon_{\mathfrak{h}}{}_1$ one obtains the following
\begin{equation}
\begin{aligned}
    \alpha'& = &\alpha-\partial\epsilon_{\mathfrak{h}}{}_1\,,\vphantom{\int\ds}\\\beta'&=&\beta-\overset{\alpha-\partial\epsilon_{\mathfrak{h}}{}_1}{\nabla}\epsilon_{\mathfrak{h}}{}_1-\epsilon_{\mathfrak{h}}{}_1\wedge\epsilon_{\mathfrak{h}}{}_1\,,\vphantom{\int\ds} \\\gamma'&=&\gamma+\{\beta-\overset{\alpha-\partial\epsilon_{\mathfrak{h}}{}_1}{\nabla}\epsilon_{\mathfrak{h}}{}_1-\epsilon_{\mathfrak{h}}{}_1\wedge\epsilon_{\mathfrak{h}}{}_1,\epsilon_{\mathfrak{h}}{}_1\}{}_{\mathrm{pf}}+\{\epsilon_{\mathfrak{h}}{}_1,\beta\}{}_{\mathrm{pf}}\,,
    \vphantom{\int\ds}\\
    B'&=& B - (C-D\wedge^{{\cal X}_1}\epsilon_{\mathfrak{h}}{}_1-D\wedge^{{\cal X}_2}\epsilon_{\mathfrak{h}}{}_1)\wedge^{\cT}\epsilon_{\mathfrak{h}}{}_1-\epsilon_{\mathfrak{h}}{}_1\wedge^{\cal D}\epsilon_{\mathfrak{h}}{}_1\wedge^{\cal D}D\,,\vphantom{\int\ds}\\
    C'&=& C-D\wedge^{{\cal X}_1}\epsilon_{\mathfrak{h}}{}_1-D\wedge^{{\cal X}_2}\epsilon_{\mathfrak{h}}{}_1\,,\vphantom{\int\ds}\\
    D'&=& D\,,\vphantom{\int\ds}
    \end{aligned}
\end{equation} 
and transforming the variables once more for the parameter $\epsilon_{\mathfrak{h}}{}_2$ one obtains:
\begin{equation}
\begin{aligned}
    \alpha''& = &\alpha-\partial\epsilon_{\mathfrak{h}}{}_1-\partial\epsilon_{\mathfrak{h}}{}_2\,,\vphantom{\int\ds}\\
    \beta''&=&\beta-\overset{\alpha-\partial\epsilon_{\mathfrak{h}}{}_1}{\nabla}\epsilon_{\mathfrak{h}}{}_1-\epsilon_{\mathfrak{h}}{}_1\wedge\epsilon_{\mathfrak{h}}{}_1-\overset{\alpha-\partial\epsilon_{\mathfrak{h}}{}_1-\partial\epsilon_{\mathfrak{h}}{}_2}{\nabla}\epsilon_{\mathfrak{h}}{}_2-\epsilon_{\mathfrak{h}}{}_2\wedge\epsilon_{\mathfrak{h}}{}_2\,,\vphantom{\int\ds} \\
    \gamma''&=&\gamma+\{\beta-\overset{\alpha-\partial\epsilon_{\mathfrak{h}}{}_1}{\nabla}\epsilon_{\mathfrak{h}}{}_1-\epsilon_{\mathfrak{h}}{}_1\wedge\epsilon_{\mathfrak{h}}{}_1,\epsilon_{\mathfrak{h}}{}_1\}{}_{\mathrm{pf}}+\{\epsilon_{\mathfrak{h}}{}_1,\beta\}{}_{\mathrm{pf}}
    \vphantom{\int\ds}\\
    &&+\{\beta-\overset{\alpha-\partial\epsilon_{\mathfrak{h}}{}_1}{\nabla}\epsilon_{\mathfrak{h}}{}_1-\epsilon_{\mathfrak{h}}{}_1\wedge\epsilon_{\mathfrak{h}}{}_1-\overset{\alpha-\partial\epsilon_{\mathfrak{h}}{}_1-\partial\epsilon_{\mathfrak{h}}{}_2}{\nabla}\epsilon_{\mathfrak{h}}{}_2-\epsilon_{\mathfrak{h}}{}_2\wedge\epsilon_{\mathfrak{h}}{}_2,\epsilon_{\mathfrak{h}}{}_2\}{}_{\mathrm{pf}}+\{\epsilon_{\mathfrak{h}}{}_2,\beta-\overset{\alpha-\partial\epsilon_{\mathfrak{h}}{}_1}{\nabla}\epsilon_{\mathfrak{h}}{}_1-\epsilon_{\mathfrak{h}}{}_1\wedge\epsilon_{\mathfrak{h}}{}_1\}{}_{\mathrm{pf}}\,,
    \vphantom{\int\ds}\\
    B''&=& B - (C-D\wedge^{{\cal X}_1}\epsilon_{\mathfrak{h}}{}_1-D\wedge^{{\cal X}_2}\epsilon_{\mathfrak{h}}{}_1)\wedge^{\cT}\epsilon_{\mathfrak{h}}{}_1-\epsilon_{\mathfrak{h}}{}_1\wedge^{\cal D}\epsilon_{\mathfrak{h}}{}_1\wedge^{\cal D}D\vphantom{\int\ds}\\
    && - (C-D\wedge^{{\cal X}_1}\epsilon_{\mathfrak{h}}{}_1-D\wedge^{{\cal X}_2}\epsilon_{\mathfrak{h}}{}_1-D\wedge^{{\cal X}_1}\epsilon_{\mathfrak{h}}{}_2-D\wedge^{{\cal X}_2}\epsilon_{\mathfrak{h}}{}_2)\wedge^{\cT}\epsilon_{\mathfrak{h}}{}_2-\epsilon_{\mathfrak{h}}{}_2\wedge^{\cal D}\epsilon_{\mathfrak{h}}{}_2\wedge^{\cal D}D\,,\vphantom{\int\ds}\\
    C''&=& C-D\wedge^{{\cal X}_1}\epsilon_{\mathfrak{h}}{}_1-D\wedge^{{\cal X}_2}\epsilon_{\mathfrak{h}}{}_1-D\wedge^{{\cal X}_1}\epsilon_{\mathfrak{h}}{}_2-D\wedge^{{\cal X}_2}\epsilon_{\mathfrak{h}}{}_2\,,\vphantom{\int\ds}\\
    D''&=& D\,.\vphantom{\int\ds}
    \end{aligned}
\end{equation} \label{e1e2}
It is easy to see that for variables $\alpha^\alpha{}_\mu$, $C^a{}_\mu$ and $D^A$ the following is obtained:
\begin{equation}\label{eq:F3}
\begin{array}{lcr}
      \mathrm{e}{}^{\epsilon_\mathfrak{h}{}_1\cdot H}\mathrm{e}{}^{\epsilon_\mathfrak{h}{}_2\cdot H}\alpha^\alpha{}_\mu&=&\mathrm{e}{}^{\epsilon_\mathfrak{h}{}_2\cdot H}\mathrm{e}{}^{\epsilon_\mathfrak{h}{}_1\cdot H}\alpha^\alpha{}_\mu\,, \phantom{\ds\int}\\
      \mathrm{e}{}^{\epsilon_\mathfrak{h}{}_1\cdot H}\mathrm{e}{}^{\epsilon_\mathfrak{h}{}_2\cdot H}C^a{}_{\mu}&=&\mathrm{e}{}^{\epsilon_\mathfrak{h}{}_2\cdot H}\mathrm{e}{}^{\epsilon_\mathfrak{h}{}_1\cdot H}C^a{}_{\mu}\,, \phantom{\ds\int}\\
      \mathrm{e}{}^{\epsilon_\mathfrak{h}{}_1\cdot H}\mathrm{e}{}^{\epsilon_\mathfrak{h}{}_2\cdot H}D^A&=&\mathrm{e}{}^{\epsilon_\mathfrak{h}{}_2\cdot H}\mathrm{e}{}^{\epsilon_\mathfrak{h}{}_1\cdot H}D^A\,. \phantom{\ds\int}
      \end{array}
\end{equation}
For the remaining variables, $\beta^a{}_{\mu\nu}$, $\gamma^A{}_{\mu\nu\rho}$ and $B^\alpha{}_{\mu\nu}$, after subtracting (\ref{e1e2}) and the corresponding equation where ${\epsilon_\mathfrak{h}{}_1}\leftrightarrow{\epsilon_\mathfrak{h}{}_2}$, one obtains:
\begin{equation}\label{eq:F4}
    \begin{array}{lcl}
    \big(\mathrm{e}{}^{\epsilon_\mathfrak{h}{}_1\cdot H}\mathrm{e}{}^{\epsilon_\mathfrak{h}{}_2\cdot H}-\mathrm{e}{}^{\epsilon_\mathfrak{h}{}_2\cdot H}\mathrm{e}{}^{\epsilon_\mathfrak{h}{}_1\cdot H}\big)\ds\frac{1}{2}\beta^a{}_{\mu\nu}&=&\partial{}_b{}^\alpha{\epsilon_\mathfrak{h}{}_2}{}^b{}_{[\mu|}{\epsilon_\mathfrak{h}{}_1}{}^c{}_{|\nu]}\rhd{}_{\alpha c}{}^a-\partial{}_b{}^\alpha{\epsilon_\mathfrak{h}{}_1}{}^b{}_{[\mu|}{\epsilon_\mathfrak{h}{}_2}{}^c{}_{|\nu]}\rhd{}_{\alpha c}{}^a\,\phantom{\ds\int}\\&=&2\delta{}_{A}{}^a\,X_{(bc)}{}^A{\epsilon_\mathfrak{h}{}_1}{}^b{}_{[\mu|}{\epsilon_\mathfrak{h}{}_2}{}^c{}_{|\nu]}\,\phantom{\ds\int}\\&=&\delta{}_{A}{}^a(\{{\epsilon_\mathfrak{h}{}_1}\wedge{\epsilon_\mathfrak{h}{}_2}{}\}{}_{\mathrm{pf}}-\{{\epsilon_\mathfrak{h}{}_2}\wedge{\epsilon_\mathfrak{h}{}_1}{}\}{}_{\mathrm{pf}}){}^A{}_{\mu\nu}\,,\phantom{\ds\int}\\
    \big(\mathrm{e}{}^{\epsilon_\mathfrak{h}{}_1\cdot H}\mathrm{e}{}^{\epsilon_\mathfrak{h}{}_2\cdot H}-\mathrm{e}{}^{\epsilon_\mathfrak{h}{}_2\cdot H}\mathrm{e}{}^{\epsilon_\mathfrak{h}{}_1\cdot H}\big)\ds\frac{1}{3!}\gamma^A{}_{\mu\nu\rho}&=&2(\partial_{[\mu} \epsilon_{\mathfrak{h}_1}{}^a{}_\nu)\epsilon_{\mathfrak{h}_{2}}{}^b{}_{\rho]} X_{(ab)}{}^A+2 \epsilon_{\mathfrak{h}_1}{}^a{}_{[\nu}(\partial_\mu\epsilon{}_{\mathfrak{h}_{2}}{}^b{}_{\rho]}) X_{(ab)}{}^A\\&&\ds+2\alpha^\alpha{}_{[\mu}\epsilon_{\mathfrak{h}_1}{}^a{}_\nu \epsilon{}_{\mathfrak{h}_{2}}{}^b{}_{\rho]} X_{(ab)}{}^B\rhd_{\alpha B}{}^A\phantom{\ds\int}\\
    &=&\ds\nabla_{[\mu} (\{{\epsilon_\mathfrak{h}{}_1}\wedge{\epsilon_\mathfrak{h}{}_2}{}\}{}_{\mathrm{pf}}-\{{\epsilon_\mathfrak{h}{}_2}\wedge{\epsilon_\mathfrak{h}{}_1}\}{}_{\mathrm{pf}}){}^A{}_{\nu\rho]}\,,\phantom{\ds\int}\\
   \big(\mathrm{e}{}^{\epsilon_\mathfrak{h}{}_1\cdot H}\mathrm{e}{}^{\epsilon_\mathfrak{h}{}_2\cdot H}-\mathrm{e}{}^{\epsilon_\mathfrak{h}{}_2\cdot H}\mathrm{e}{}^{\epsilon_\mathfrak{h}{}_1\cdot H}\big)\ds\frac{1}{2}B^\alpha{}_{\mu\nu}&=&D^A{\epsilon_\mathfrak{h}{}_2}{}^a{}_{[\mu|}{\epsilon_\mathfrak{h}{}_1}{}^b{}_{|\nu]}(X_1{}_{Aa}{}^c+X_2{}_{Aa}{}^c)\cT{}_{cb}{}^\alpha\phantom{\ds\int}\\&&-D^A{\epsilon_\mathfrak{h}{}_1}{}^b{}_{[\mu|}{\epsilon_\mathfrak{h}{}_2}{}^a{}_{|\nu]}(X_1{}_{Ab}{}^c+X_2{}_{Ab}{}^c)\cT{}_{ca}{}^\alpha \,\\&=&-2D_A{\epsilon_\mathfrak{h}{}_1}{}^a{}_{[\mu|}{\epsilon_\mathfrak{h}{}_2}{}^b{}_{|\nu]}(X_{(ac)}{}^A\rhd_{\alpha b}{}^c+X_{(bc)}{}^A\rhd_{\alpha a}{}^c)\phantom{\ds\int}\\ &=&-2D_A{\epsilon_\mathfrak{h}{}_1}{}^a{}_{[\mu|}{\epsilon_\mathfrak{h}{}_2}{}^b{}_{|\nu]}X_{(ab)}{}^B\rhd_{\alpha B}{}^A\phantom{\ds\int}\\&=&(D\wedge^\cS (\{{\epsilon_\mathfrak{h}{}_1}\wedge{\epsilon_\mathfrak{h}{}_2}\}{}_{\mathrm{pf}}-\{{\epsilon_\mathfrak{h}{}_2}\wedge{\epsilon_\mathfrak{h}{}_1}\}{}_{\mathrm{pf}})^\alpha{}_{\mu\nu}.
    \end{array}
\end{equation}
Comparing (\ref{eq:F3}) and (\ref{eq:F4}) with (\ref{l-gauge}), one concludes that the commutator of two $H$-gauge transformations is the $L$-gauge transformation with the parameter $\epsilon_\mathfrak{l}{}^A{}_{\mu\nu}=4{\epsilon_\mathfrak{h}{}_1}{}^a{}_{[\mu|}{\epsilon_\mathfrak{h}{}_2}{}^b{}_{|\nu]}X_{(ac)}{}^A$:
\begin{equation}
    \mathrm{e}{}^{\epsilon_\mathfrak{h}{}_1\cdot H}\mathrm{e}{}^{\epsilon_\mathfrak{h}{}_2\cdot H}-\mathrm{e}{}^{\epsilon_\mathfrak{h}{}_2\cdot H}\mathrm{e}{}^{\epsilon_\mathfrak{h}{}_1\cdot H}=2\,(\{{\epsilon_\mathfrak{h}{}_1}\wedge{\epsilon_\mathfrak{h}{}_2}\}{}_{\mathrm{pf}}-\{{\epsilon_\mathfrak{h}{}_2}\wedge{\epsilon_\mathfrak{h}{}_1}{}\}{}_{\mathrm{pf}})\cdot \hat{L}\,.
\end{equation}

\subsection{Commutator $[H,N]$}
Let us calculate the commutator between the generators of $H$-gauge transformation and $N$-gauge transformation, i.e., derive the equation (\ref{eq:HNcom}). This is done by calculating the expressions 
\begin{equation}
      \big(\mathrm{e}{}^{\epsilon_\mathfrak{h}\cdot H}\mathrm{e}{}^{\epsilon_\mathfrak{n}\cdot N}-\mathrm{e}{}^{\epsilon_\mathfrak{n}\cdot N}\mathrm{e}{}^{\epsilon_\mathfrak{h}\cdot H}\big)A\,,\\
\end{equation}
for all variables $A$ present in the theory. 
It is easy to see that for variables $\alpha^\alpha{}_\mu$, $\beta^a{}_{\mu\nu}$, $\gamma^A{}_{\mu\nu\rho}$, and $D^A$ the following is obtained:
\begin{equation}\label{eq:F7}
\begin{array}{lcr}
      \mathrm{e}{}^{\epsilon_\mathfrak{h}\cdot H}\mathrm{e}{}^{\epsilon_\mathfrak{n}\cdot N}\alpha^\alpha{}_\mu&=&\mathrm{e}{}^{\epsilon_\mathfrak{n}\cdot N}\mathrm{e}{}^{\epsilon_\mathfrak{h}\cdot H}\alpha^\alpha{}_\mu\,, \phantom{\ds\int}\\
      \mathrm{e}{}^{\epsilon_\mathfrak{h}\cdot H}\mathrm{e}{}^{\epsilon_\mathfrak{n}\cdot N}\beta^a{}_{\mu\nu}&=&\mathrm{e}{}^{\epsilon_\mathfrak{n}\cdot N}\mathrm{e}{}^{\epsilon_\mathfrak{h}\cdot H}\beta^a{}_{\mu\nu}\,, \phantom{\ds\int}\\
      \mathrm{e}{}^{\epsilon_\mathfrak{h}\cdot H}\mathrm{e}{}^{\epsilon_\mathfrak{n}\cdot N}\gamma^A{}_{\mu\nu\rho}&=&\mathrm{e}{}^{\epsilon_\mathfrak{n}\cdot N}\mathrm{e}{}^{\epsilon_\mathfrak{h}\cdot H}\gamma^A{}_{\mu\nu\rho}\,, \phantom{\ds\int}\\
      \mathrm{e}{}^{\epsilon_\mathfrak{h}\cdot H}\mathrm{e}{}^{\epsilon_\mathfrak{n}\cdot N}D^A&=&\mathrm{e}{}^{\epsilon_\mathfrak{n}\cdot N}\mathrm{e}{}^{\epsilon_\mathfrak{h}\cdot H}D^A\,. \phantom{\ds\int}
      \end{array}
\end{equation}
For the remaining variables, $B^\alpha{}_{\mu\nu}$ and $C^a{}_\mu$, after the $H$-gauge transformation one obtains the following:
\begin{equation}
    \begin{aligned}
   &B'{}=B-(C-D\wedge^{\chi_1} \epsilon_\mathfrak{h}-D\wedge^{\chi_2} \epsilon_\mathfrak{h})\wedge^\tau \epsilon_\mathfrak{h}- \epsilon_\mathfrak{h}\wedge^\cD \epsilon_\mathfrak{h}\wedge^\cD D\,,\\
   &C'=C-D\wedge^{\chi_1} \epsilon_\mathfrak{h}-D\wedge^{\chi_2} \epsilon_\mathfrak{h}\,.
    \end{aligned}
\end{equation}
Next, transforming those variables with $N$-gauge transformation one obtains:
\begin{equation}\label{bchn}
    \begin{array}{lcl}
    B''&=&B'-\beta'\wedge^\cT \epsilon_\mathfrak{n}\\&=&B-(C-D\wedge^{\chi_1} \epsilon_\mathfrak{h}-D\wedge^{\chi_2} \epsilon_\mathfrak{h})\wedge^\tau \epsilon_\mathfrak{h}- \epsilon_\mathfrak{h}\wedge^\cD \epsilon_\mathfrak{h}\wedge^\cD D-(\beta-\overset{\{\alpha^\alpha-\partial_a{}^\alpha\epsilon_\mathfrak{h}{}^a\}}{\nabla \epsilon_\mathfrak{h}}-\epsilon_\mathfrak{h}\wedge\epsilon_\mathfrak{h})\wedge^\cT \epsilon_\mathfrak{n}\,,\\
    C''&=&C'-\overset{\{\alpha^\alpha-\partial_a{}^\alpha\epsilon_\mathfrak{h}{}^a\}}{\nabla} \epsilon_\mathfrak{n}\\&=&C-D\wedge^{\chi_1} \epsilon_\mathfrak{h}-D\wedge^{\chi_2} \epsilon_\mathfrak{h}-\overset{\{\alpha^\alpha-\partial_a{}^\alpha\epsilon_\mathfrak{h}{}^a\}}{\nabla} \epsilon_\mathfrak{n}\,.
    \end{array}
\end{equation}
Let us now exchange the order of transformations, and first transform the variables with $N$-gauge transformation, 
\begin{equation}
    \begin{aligned}
   &B^\cdot{}=B-\beta\wedge^\cT \epsilon_\mathfrak{n}\,,\\
   &C^\cdot=C-\nabla \epsilon_\mathfrak{n}\,,
    \end{aligned}
\end{equation}
and then with $H$-gauge transformation:
\begin{equation}\label{bcnh}
    \begin{array}{lcl}
    B^{\cdot\cdot}&=&B^{\cdot}-(C^{\cdot}-D^{\cdot}\wedge^{\chi_1} \epsilon_\mathfrak{h}-D^{\cdot}\wedge^{\chi_2} \epsilon_\mathfrak{h})\wedge^\tau \epsilon_\mathfrak{h}- \epsilon_\mathfrak{h}\wedge^\cD \epsilon_\mathfrak{h}\wedge^\cD D^{\cdot}\\&=&B-\beta\wedge^\cT \epsilon_\mathfrak{n}-(C-\nabla \epsilon_\mathfrak{n}-(D+\delta \epsilon_\mathfrak{n})\wedge^{\chi_1} \epsilon_\mathfrak{h}-(D+\delta \epsilon_\mathfrak{n})\wedge^{\chi_2} \epsilon_\mathfrak{h})\wedge^\tau \epsilon_\mathfrak{h}- \epsilon_\mathfrak{h}\wedge^\cD \epsilon_\mathfrak{h}\wedge^\cD (D+\delta \epsilon_\mathfrak{n})\,,\\
    C^{\cdot\cdot}&=&C^{\cdot}-D^{\cdot}\wedge^{\chi_1} \epsilon_\mathfrak{h}-D^{\cdot}\wedge^{\chi_2} \epsilon_\mathfrak{h}\\&=&C-\nabla \epsilon_\mathfrak{n}-(D+\delta \epsilon_\mathfrak{n})\wedge^{\chi_1} \epsilon_\mathfrak{h}-(D+\delta \epsilon_\mathfrak{n})\wedge^{\chi_2} \epsilon_\mathfrak{h}\,.
    \end{array}
\end{equation}
After subtracting (\ref{bchn}) and (\ref{bcnh}) one obtains:
\begin{equation}
    \begin{array}{lcl}
   \big(\mathrm{e}{}^{\epsilon_\mathfrak{h}\cdot H}\mathrm{e}{}^{\epsilon_\mathfrak{n}\cdot N}-\mathrm{e}{}^{\epsilon_\mathfrak{n}\cdot N}\mathrm{e}{}^{\epsilon_\mathfrak{h}\cdot H}\big)B^\alpha&=&\nabla \epsilon_\mathfrak{n}{}^a \wedge \epsilon_{\mathfrak{h}}{}^b \cT_{ab}{}^\alpha+\delta{}^A{}_a\epsilon_\mathfrak{n}{}^a\epsilon_{\mathfrak{h}}{}^b \wedge \epsilon_{\mathfrak{h}}{}^d X_1{}_{Ab}{}^c \cT_{cd}{}^\alpha\phantom{\int\ds}\\&&+\delta{}^A{}_a\epsilon_\mathfrak{n}{}^a\epsilon_{\mathfrak{h}}{}^b \wedge \epsilon_{\mathfrak{h}}{}^d X_2{}_{Ab}{}^c \cT_{cd}{}^\alpha-\epsilon_{\mathfrak{h}}{}^a \wedge \epsilon_{\mathfrak{h}}{}^b \delta_A{}^c\epsilon_{\mathfrak{n}}{}^c D_{Aab}{}^\alpha \,,\phantom{\int\ds}\\&&-\nabla\epsilon_\mathfrak{h}{}^a\wedge\epsilon_\mathfrak{n}{}^b \cT_{ab}{}^\alpha+ \partial_a{}^\beta\epsilon_\mathfrak{h}{}^a\rhd_{\beta c}{}^b \epsilon_\mathfrak{h}{}^c \epsilon_\mathfrak{n}{}^d\cT_{bd}{}^\alpha- \epsilon_\mathfrak{h}{}^a \wedge \epsilon_\mathfrak{h}{}^b f_{ab}{}^c \epsilon_\mathfrak{n}{}^d \cT_{cd}{}^\alpha\,,\phantom{\int\ds}\\ 
   \big(\mathrm{e}{}^{\epsilon_\mathfrak{h}\cdot H}\mathrm{e}{}^{\epsilon_\mathfrak{n}\cdot N}-\mathrm{e}{}^{\epsilon_\mathfrak{n}\cdot N}\mathrm{e}{}^{\epsilon_\mathfrak{h}\cdot H}\big)C^c&=&-(\delta{}^A{}_a\epsilon_\mathfrak{n}^a)\wedge\epsilon_\mathfrak{h}{}^b{X_1}_{Ab}{}^c-(\delta{}^A{}_a\epsilon_\mathfrak{n}^a)\wedge\epsilon_\mathfrak{h}{}^b{X_2}_{Ab}{}^c-\partial_a{}^\beta\epsilon_\mathfrak{h}{}^a\rhd_{\beta b}{}^c \epsilon_\mathfrak{n}{}^b\,,\phantom{\int\ds}
    \end{array}
\end{equation}
where after using the definitions of the maps $\cT$, $\cD$, $\chi_1$, and $\chi_2$ one obtains the result
\begin{equation}\label{eq:F13}
 \begin{array}{lcl}
   \big(\mathrm{e}{}^{\epsilon_\mathfrak{h}\cdot H}\mathrm{e}{}^{\epsilon_\mathfrak{n}\cdot N}-\mathrm{e}{}^{\epsilon_\mathfrak{n}\cdot N}\mathrm{e}{}^{\epsilon_\mathfrak{h}\cdot H}\big)B^\alpha&=&\nabla \epsilon_\mathfrak{n}{}^a \wedge \epsilon_{\mathfrak{h}}{}^b \cT_{ab}{}^\alpha-\nabla\epsilon_\mathfrak{h}{}^a\wedge\epsilon_\mathfrak{n}{}^b \cT_{ab}{}^\alpha=\nabla ( \epsilon_\mathfrak{n}{} \wedge^{\cT} \epsilon_{\mathfrak{h}}){}^\alpha\,,\phantom{\int\ds}\\ 
   \big(\mathrm{e}{}^{\epsilon_\mathfrak{h}\cdot H}\mathrm{e}{}^{\epsilon_\mathfrak{n}\cdot N}-\mathrm{e}{}^{\epsilon_\mathfrak{n}\cdot N}\mathrm{e}{}^{\epsilon_\mathfrak{h}\cdot H}\big)C^c&=&\partial^c{}_\alpha (\epsilon_\mathfrak{n}\wedge^\cT \epsilon_\mathfrak{h})^{\alpha}\,,\phantom{\int\ds}\\
    \end{array}
\end{equation}
Comparing (\ref{eq:F7}) and (\ref{eq:F13}) with (\ref{mgauge}), one obtains that:
\begin{equation}
      \big(\mathrm{e}{}^{\epsilon_\mathfrak{h}\cdot H}\mathrm{e}{}^{\epsilon_\mathfrak{n}\cdot N}-\mathrm{e}{}^{\epsilon_\mathfrak{n}\cdot N}\mathrm{e}{}^{\epsilon_\mathfrak{h}\cdot H}\big)=-(\epsilon_\mathfrak{n}\wedge^\cT \epsilon_\mathfrak{h})\cdot M\,.\\
\end{equation}

\end{document}